\documentclass[12pt,preprint]{iopart}
\usepackage{iopams}  
\usepackage{mlmodern}

\usepackage{bm,bbm}
\usepackage{verbatim}
\expandafter\let\csname equation*\endcsname\relax

\expandafter\let\csname endequation*\endcsname\relax
\usepackage{amsmath}

\usepackage[capitalise]{cleveref}
\usepackage{dsfont}
\usepackage{lipsum}
\usepackage{xcolor}
\usepackage{soul}
\usepackage{caption}
\usepackage{subcaption}
\usepackage{graphicx}

\usepackage{cite}

\usepackage{tocloft}

\usepackage{amsthm}

\newtheoremstyle{resultStyle}
  {6pt}{6pt}{\itshape}{}{\bfseries}{.}{.5em}{}

\newtheoremstyle{remarkStyle}
  {6pt}{6pt}{\normalfont}{}{\bfseries}{.}{.5em}{}

\theoremstyle{resultStyle}
\newtheorem{result}{Result}[section]

\theoremstyle{remarkStyle}
\newtheorem{remark}{Remark}[section]

\newcommand{\bX}{{\bm X}}

\begin{document}

\title{Rank-One Signal Recovery in Sparse Wishart Noise}

\author{Preben Forer$^*$, Urte Adomaityte$^\dagger$, Pierpaolo Vivo$^*$}

\address{$^*$Department of Mathematics, King’s College London, Strand, London, WC2R 2LS, United Kingdom}
\address{$^\dagger$Université Paris 1 Panthéon-Sorbonne, CNRS, Centre d’Économie de la Sorbonne, Paris, France}

\begin{abstract}
We study the high-dimensional recovery of a signal vector $\bm{x}$ in the presence of sparse Wishart-like noise. We define an $N \times N$ matrix $\bm A = \bm J+(\theta/N)\bm{xx}^{\top}$, where $\bm{xx}^{\top}$ is the rank-one deformation of the random noise matrix $\bm J$. We consider a Wishart-like matrix $\bm J=\bm{X}^{\top} \bm X$, where $\bm X$ is a sparse $M \times N$ random matrix with entries $X_{ij} = c_{ij}W_{ij}$, with $c_{ij}$ regulating the density of non-zero elements, and $W_{ij}$ the bond weights. Using the replica method, we compute analytically the top eigenpair statistics of $\bm A$, and their dependence on the signal strength $\theta$, the rectangularity ratio $\alpha=\sqrt{M/N}$, and the average connectivity of the noise. The spectral observables are expressed in terms of a system of Recursive Distributional Equations, which are efficiently solved via a Population Dynamics algorithm. They allow us to compute the average largest eigenvalue $\langle\lambda_1\rangle_{\bm A}$, the average top eigenvector component density, and the average overlap between the top eigenvector of $\bm A$ and $\bm{x}$. We identify a critical threshold $\theta_{\mathrm{crit}}$--depending on the average connectivity of the noise--that marks a BBP-like phase transition: below this value, $\langle\lambda_1\rangle_{\bm A}$ is unaffected by the signal, and the overlap vanishes. Thus, the signal is not recoverable from the top eigenvector of $\bm A$. For $\theta>\theta_{\mathrm{crit}}$, the signal-related outlier eigenvalue becomes $\langle\lambda_1\rangle_{\bm A}$ and the overlap is nonzero, allowing for recovery of the signal. The results are in excellent agreement with numerical diagonalisation. We show that in the dense limit, the recovery threshold and eigen-statistics converge to the results predicted by the classical BBP transition for additive rank-one deformations of dense Wishart matrices.
\end{abstract}
\tableofcontents

\section{Introduction} \label{sec:Introduction}

The unprecedented surge in available data has made the recovery of meaningful rank-one signals from noisy observations a central challenge in high-dimensional statistics. This problem often appears in Principal Component Analysis (PCA), where recovering the top eigenvalue and eigenvector allows for dimensionality reduction \cite{Jolliffe2002, Candes2011}.  In machine learning, recovering the rank-one signal from noisy observations is particularly useful for matrix denoising \cite{Bishop2006, Montanari2021, Liu2022a, Liu2022b}, or, more recently, for modelling gradient flow in teacher-student learning \cite{Coeurdoux2026}. 

Typically, the analytical framework most used to study this signal recovery problem is the spiked random matrix model \cite{Johnstone2001, Baik2005, Forrester2023}. In these models, the noise is represented via a random matrix $\bm{J}\in\mathbb{R}^{N\times N}$, and the goal is to recover a rank-one signal $\bm{x} \in \mathbb R^{N}$ from a noisy observation $\bm{A}\in \mathbb R^{N\times N}$ given by
\begin{equation} \label{eq:intro formula}
    \bm{A}=\bm{J}+\frac{\theta}{N}\bm{xx}^\top,
\end{equation}
where $\theta$ is the \emph{signal strength}. For dense, random matrices $\bm{J}$, there are many analytical results about the various spectral observables that can indicate signal recovery. Most importantly, these models exhibit a well-known BBP transition (named after the authors of \cite{Baik2005}) on the signal strength $\theta$. Namely, there exists a value $\theta_{\mathrm{crit}}$ such that for signal strengths larger than it, the top eigenvalue detaches from the right bulk-edge of the spectrum, and the top eigenvector has non-trivial overlap with the signal, making the recovery of the signal $\bm{x}$ possible via the leading principal component in the standard PCA. In the classical BBP setting, the matrix $\bm J$ belongs to a rotationally invariant ensemble of random matrices, for instance Wigner or Wishart \cite{Baik2005, Benaych-Georges2011}.

Some applications, however, require the noise background to be modelled as a \emph{sparse} matrix with finite connectivity. The spiked sparse Wigner model has many applications in the field of Principal Component Analysis \cite{Candes2011}, and was recently analytically studied, with a similar phase transition to that found in the BBP dense setting \cite{UrtePaper}. Of further interest is the case in which $\bm{J}$ is a Wishart covariance matrix of the form $\bm{X}^\top\bm{X}$. The matrix $\bm{X}\in\mathbb{R}^{M\times N}$ represents the noise contribution across $M$ measurements and $N$ variables. In this case, $\bm{J}$ encodes the pairwise correlations induced by the noise. Since in many systems each variable interacts significantly with only a limited number of others, it is natural to consider a sparse noise structure. The signal $\bm{x}$ then models a global trend affecting all variables.

This underlying sparse covariance structure with a dense global signal arises naturally in several fields. For instance in financial markets, a large number of assets in the stock market are uncorrelated \cite{Mantegna2012, Fan2016} outside of a ``market mode'' that impacts all assets globally \cite{Laloux1999, Kenett2010, Bouchaud2018}, lending itself naturally to observations of the asset correlation matrix of this form \cite{Borghesi2007, Fan2013}. Beyond finance, ecological association networks have been found to be sparse \cite{Deng2013, Kurtz2015}, while also subject to a dense signal representing global effects \cite{Tikhonov2017}. Despite these applications, the recovery threshold and spectral observables for the spiked sparse Wishart model remain analytically largely unexplored.

In this paper, we build on works such as \cite{BookParisi, Kuhn2024, Nagao2007, Kabashima2012, Susca2019, Susca2021, Budnick2025, UrtePaper}, and employ the replica formalism in order to analytically calculate spectral observables of large random symmetric matrices that consist of a sparse Wishart noise plus a dense rank-one matrix that is the outer product of the signal vector. Our aim is to characterise the recovery threshold beyond which signal inference via the top eigenvector is possible, and determine analytically the BBP-like transition line as a function of the average connectivity of the sparse noise and the ``rectangularity ratio'' $\alpha=\sqrt{M/N}$.

The paper is structured as follows. In Section \ref{sec:Lit Review}, we provide a brief overview of literature related to signal inference, spiked random matrix models, as well as applications of the replica method in Random Matrix Theory. In Section \ref{sec:The Model}, we formulate our problem and specify the matrices we are working with. In Section \ref{Main Results} we outline our main results, which we then specialise to two specific degree distributions in Section \ref{sec:special_cases}. In Section \ref{PopDyn} we present the population dynamics algorithm employed to help us solve the resulting system of self-consistent integral equations, before finally concluding and presenting our outlook for future work in Section \ref{sec:conclusions}. The full details of the replica calculations as well as technical derivations are included in the Appendices. 

\section{Literature review}\label{sec:Lit Review} 

The importance of studying detection limits in large spiked random matrices comes from their applications in a variety of fields. Community detection provides one example. Decelle et al.\ analysed detectability transitions in sparse networks generated by stochastic block models \cite{Decelle2011}, while Lelarge and Miolane established a universality connection between community detection and symmetric low-rank matrix estimation in additive Gaussian noise \cite{Lelarge2018}. In community detection as well, applications for sparse underlying graphs are found. For instance, Krzakala et al. have used spectral clustering methods in order to detect communities in sparse graphs \cite{Krzakala2013}. Abb\'e \& Sandon \cite{Abbe2015} examined recovery thresholds in general stochastic block models in both constant- and logarithmic-degree regimes. A review of developments in community detection and stochastic block models is given in \cite{Abbe2018}. 

Random-matrix models also arise naturally in neural-network connectivity. Rajan and Abbott studied random synaptic connectivity matrices in which excitatory and inhibitory columns are drawn from distributions with different means and variances \cite{Rajan2006}. Mastrogiuseppe and Ostojic subsequently considered recurrent connectivity given by the sum of a random component and a low-rank structured component, and related this structure to low-dimensional network dynamics \cite{Mastrogiuseppe2018}. Sparsity is also relevant in neural-network models. Van Vreeswijk \& Sompolinsky studied sparsely connected balanced neural networks \cite{Vreeswijk1998}, while Brunel investigated the dynamics of sparsely connected networks of excitatory and inhibitory neurons \cite{Brunel2000}. Herbert \& Ostojic subsequently studied how sparsity modifies the eigenspectrum and low-dimensional dynamics of low-rank recurrent neural networks \cite{Herbert2022}. In a similar setting, Park et al. examined the emergence of low-rank structures in neural networks, and used a linear teacher-student model to show that the dynamics of neural network learning exhibit a BBP-like transition through which isolated eigenvalues separate from the spectral bulk \cite{Park2026}.

The study of low-rank signals in sparse random backgrounds is also motivated by applications such as image reconstruction. In a related, though distinct, setting, Cand\`es et al.\ showed in their seminal work on robust PCA that a data matrix formed as the superposition of a low-rank component and a sparse corruption can, under suitable assumptions, be decomposed exactly via Principal Component Pursuit. This framework has applications in video surveillance, where the low-rank component models the stationary background and the sparse component captures moving foreground objects, as well as in face image analysis, where it can remove shadows and specularities that hinder face recognition \cite{Candes2011}. 

These applications have, in recent years, motivated the analytical study of spiked random Wigner-type models. Following the seminal BBP transition for non-null complex sample covariance matrices \cite{Baik2005}, P\'ech\'e studied the limiting eigenvalue statistics of small-rank perturbations of Gaussian Hermitian random matrices \cite{Peche2006}. F\'eral \& P\'ech\'e subsequently extended the analysis to Wigner matrices with non-Gaussian entries \cite{Peche2007}. This work was continued by Capitaine et al., who studied finite rank perturbations of the Wigner matrix, and found that as soon as certain non-zero eigenvalues of the perturbation became large enough, the largest eigenvalue of the perturbed Wigner ensemble exited the semicircle bulk \cite{Capitaine2009}. The study of the eigenvectors associated with the largest eigenvalues of spiked Wigner models was performed by Benaych-Georges \& Nadakuditi in \cite{Benaych-Georges2011} where they proposed a unified treatment of the eigenvalue and eigenvector transition. Bloemendal \& Virág then extended this work to shifted mean Gaussian orthogonal ensembles, characterising the limiting distribution of the largest eigenvalues through a stochastic operator with a boundary condition set by the perturbation \cite{Bloemendal2013}. The continuity of the BBP transition was recently studied by Bocchi et al., who found that the transition between non-recovery and recovery is discontinuous if the spectral density vanishes faster than linearly at the spectral edge \cite{Bocchi2026}. Recent work has further extended the range of spiked models that exhibit a BBP transition. Adomaityte et al. were able to generalise the BBP transition to the case of sparse Wigner noise \cite{UrtePaper}, whereas Ferreira \& Metz recently studied an inhomogeneous spiked Wigner model in a rescaled high-connectivity regime and where the variance of each noise entry is itself a random variable; for a truncated power-law distribution of these variances, they found that the BBP transition line can be non-monotonic \cite{Ferreira2026}.

The literature on dense spiked Wishart and sample-covariance models has a parallel analytical development. The BBP transition was first established by Baik et al.\ for non-null complex sample covariance matrices \cite{Baik2005}. Baik and Silverstein subsequently determined the almost-sure limits of sample eigenvalues for a broad class of spiked population models \cite{Baik2006}. Paul then studied both the limiting fluctuations of outlier sample eigenvalues and the behaviour of the associated sample eigenvectors in spiked covariance models \cite{Paul2007}. The possibility of testing for rank-one perturbations below the BBP threshold was studied by Onatski et al.\ \cite{Onatski2013}. They showed that below the threshold the joint eigenvalue distributions under the null and alternative hypotheses are mutually contiguous, so asymptotically certain discrimination is impossible, although eigenvalue-based tests can retain non-trivial asymptotic power; above the threshold, discrimination becomes asymptotically certain. For a review of exact results for rank-one perturbations in several random-matrix ensembles, we refer the reader to \cite{Forrester2023}.

To our knowledge, the finite-connectivity additive 
rank-one deformation of the diluted Wishart ensemble considered here has not previously been 
analysed at the level of its top eigenpair and recovery transition. Ref. \cite{Budnick2025} used the replica formalism to calculate the top eigenpair statistics of diluted Wishart matrices without an added signal. This formalism was originally introduced in the context of mean-field approaches to spin glasses \cite{BookParisi, Mezard2000, Nishimori2001, Kuhn2007}. It was then applied to random matrices by Edwards and Jones to retrieve the semicircle law for large symmetric matrices \cite{Edwards1976}. This methodology was then applied to sparse symmetric matrices with entries $0$ or $\pm1$ by Rodgers and Bray, who recovered the semicircle law in the dense limit, but found tails extending beyond the semicircle when the matrix is sparse \cite{BrayRodgers}. The spectral densities of sparse adjacency matrices were then studied by Rodgers and De Dominicis \cite{Rodgers1990}. The resulting integral equations, however, have proven difficult to solve, though recent work has made progress on their numerical solutions \cite{Akara2025}. Kühn \cite{Kuhn2024} proposed a different approach to compute the spectral density of sparse random matrices via replicas: making a replica-symmetric ansatz that represents the order parameters as a superposition of fluctuating Gaussians results in a set of recursive distributional equations that can subsequently be solved through a population dynamics algorithm (see also \cite{Bianconi2008}). A similar method was applied to compute the spectral density of sparse Wishart covariance matrices \cite{Nagao2007}.

\section{The Model} \label{sec:The Model}
In this Section, we define the random matrix model in Section \ref{sec:Matrix Setup}, and briefly outline the replica methodology in Section \ref{sec:Replica Methodology}.

\subsection{The sparse spiked Wishart matrix model}\label{sec:Matrix Setup}

The goal of this study is to compute the top eigenpair statistics and recovery thresholds associated with the real observation matrix defined as
\begin{equation}\label{def:model}
\bm{A} \;=\; \bm{J} \;+\; \frac{\theta}{N}\, \bm{x}\bm{x}^{\!\top} \in \mathbb R^{N \times N}\ .
\end{equation}
The vector \(\bm{x}\in\mathbb{R}^N\) is a  \emph{signal} constructed by sampling each of its components $x_i$ independently and identically from a density $\varrho_x$ defined on the real axis and with finite second moment. The scalar parameter $\theta\geq0$ regulates the strength of the rank-one signal.

The noise matrix $\bm{J}$ is an \(N\times N\) Wishart-like matrix defined as
\begin{equation}\label{eq:spike model}
\bm{J} \;=\; {\bX}^{\!\top}\,{\bX}\ ,
\end{equation}
where \(\bX \in \mathbb R^{M \times N}\) is a sparse random matrix defined as
\begin{equation} \label{X definition}
X_{ij}=c_{ij}\,W_{ij} \ ,
\end{equation}
where $i=1,\dots,M$ and $j=1,\dots,N$. 
The matrix $\bm{C} = c_{ij}\in\{0,1\}^{M\times N}$ is the biadjacency matrix of an undirected bipartite random graph with two distinct node types: $i$-nodes corresponding to its $M$ rows, and $j$-nodes corresponding to its $N$ columns. The bipartite graph is sampled from the \emph{configuration model} as follows.

\begin{figure}
    \centering
    \includegraphics[width=0.5\linewidth]{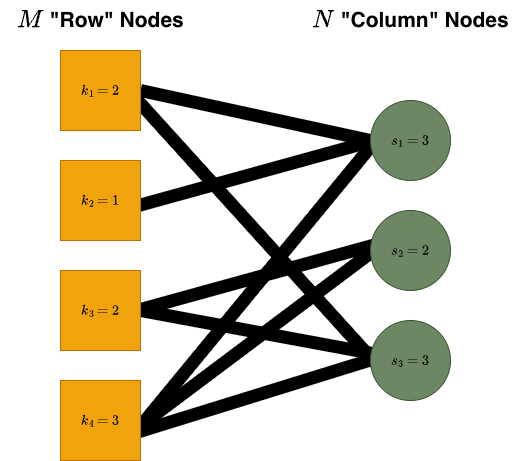}
    \caption{Schematic example of the underlying bipartite structure of the graph defining the adjacency matrix $\bm{C}$. We see the two types of nodes, row and column, as well as that the handshake constraint is satisfied since $\sum_{i=1}^4k_i=\sum_{j=1}^3s_j=8$.}
    \label{fig:bipartite example}
\end{figure}

The degrees $\bm{k} =\{k_i\}$ of the \(i\)-nodes and the degrees $\bm{s}=\{s_j\}$ of the \(j\)-nodes are jointly sampled from a distribution $p_{k,s}(\bm k,\bm s)$--symmetric under the exchange of any two $k$'s or any two $s$'s--satisfying the bipartite ``handshake'' constraint
\begin{equation}\label{eq: Handshake constraint}
\sum_{i=1}^{M}k_i=\sum_{j=1}^{N}s_j 
\end{equation}
illustrated graphically in Figure \ref{fig:bipartite example}. 
The joint distribution $p_{k,s}(\bm k,\bm s)$ is supported on $\{(k_i,s_j): 0\leq k_i\leq k_{\mathrm{max}}, 0\leq s_j\leq s_{\mathrm{max}}\}$. Imposing a maximal degree on both the columns and the rows is sufficient in order to have $\langle\lambda_1\rangle_{\bm{A}}=O(1)$, just as in \cite{Budnick2025, UrtePaper}. We define $p_k(k_1)=\sum_{\bm s,\bm k\setminus k_1} p_{k,s}(\bm k,\bm s)$ and $p_s(s_1)=\sum_{\bm k,\bm s\setminus s_1} p_{k,s}(\bm k,\bm s)$ the marginal distributions. We further assume that, in the thermodynamic limit, the empirical degree distributions self-average to these marginals, i.e.,
\begin{equation}
\frac{1}{M}\sum_{i=1}^{M}\mathbf{1}(k_i=k)
\xrightarrow[M\to\infty]{\mathbb{P}} p_k(k),
\qquad
\frac{1}{N}\sum_{j=1}^{N}\mathbf{1}(s_j=s)
\xrightarrow[N\to\infty]{\mathbb{P}} p_s(s)\ ,
\end{equation}
for every $k\in\{0,\ldots,k_{\max}\}$ and
$s\in\{0,\ldots,s_{\max}\}$.

Conditioned on specific sequences \(\bm{k}\) and \(\bm{s}\), the bipartite adjacency matrix \(\bm{C}=(c_{ij})\in\{0,1\}^{M\times N}\) is sampled uniformly from all simple bipartite graphs satisfying
\begin{equation}
\sum_{j=1}^{N}c_{ij}=k_i \ , \qquad \sum_{i=1}^{M}c_{ij}=s_j \ .
\end{equation}
The marginal distributions $p_s(s)$ and $p_k(k)$ have means $\langle s\rangle$ and $\langle k \rangle$ respectively. Combining the handshake constraint \eqref{eq: Handshake constraint}
with the assumed convergence of the empirical degree distributions gives
\begin{equation}\label{handshake constraint math}
    \langle k\rangle=\frac{N}{M}\langle s\rangle\ ,
\end{equation} 
for large $N,M$.

The joint distribution of entries for $\bm{X}$ is thus given by
\begin{equation}\label{Joint distribution Initial}
    P(\bm X,\bm C,\bm W\mid\bm k,\bm s) = P(\bm{C}\mid\bm{k},\bm{s})p(\bm{W})\prod_{ij}\delta(X_{ij}-c_{ij}W_{ij}) \ ,
\end{equation}
and the distribution of the connectivities 
\begin{align} \label{eq: C given s and k}
P(\bm C\mid \bm{k},\bm{s}) &= 
\frac{1}{\mathcal{Z}_{\mathrm{den}}}\left[\prod_{i=1}^{M}\mathbf{1}\!\left(\sum_{j=1}^{N}c_{ij}=k_i\right)\right]\left[
\prod_{j=1}^{N}\mathbf{1}\!\left(\sum_{i=1}^{M}c_{ij}=s_j\right)\right] \mathbf{1}\!\left(\sum_{i=1}^{M}k_i=\sum_{j=1}^{N}s_j\right)\ , \ 
\end{align}
where $\mathbf{1}(S)=1$ if $S$ is true and $0$ otherwise, and $\mathcal{Z}_{\mathrm{den}}$ is the normalisation constant defined by
\begin{equation}
\mathcal{Z}_{\mathrm{den}} = \sum_{\{c_{ij}\}}\left[\prod_{i=1}^{M}\mathbf{1}\!\left(\sum_{j=1}^{N}c_{ij}=k_i\right)\right]\left[
\prod_{j=1}^{N}\mathbf{1}\!\left(\sum_{i=1}^{M}c_{ij}=s_j\right)\right] \mathbf{1}\!\left(\sum_{i=1}^{M}k_i=\sum_{j=1}^{N}s_j\right)\ .\label{defZprime}
\end{equation}

We assume that $p_{k,s}(\bm{k},\bm{s})$ has support only on degree sequences for which the constraint set defining $\mathcal{Z}_{\rm den}(\bm{k},\bm{s})$ is non-empty, so that
$$
\mathcal{Z}_{\rm den}(\bm{k},\bm{s})>0\ .
$$
In other words, the prescribed row and column degrees are assumed to admit at least one simple bipartite adjacency matrix.

The bond weights are sampled independently from the density $W_{ij} \sim \varrho_W$ with compact support and strictly positive mean. Equivalently, the induced distribution of \(\bm X\) conditioned on $\bm s$ and $\bm k$ is
\begin{align} \label{eq: X given degree sequences}
P(\bm X\mid \bm k, \bm s) &= 
\sum_{\bm C\in\{0,1\}^{M\times N}} P(\bm C\mid \bm k,\bm s)
\int \mathrm{d}\bm W\,p(\bm W) \prod_{i=1}^{M}\prod_{j=1}^{N}
\delta(X_{ij}-c_{ij}W_{ij}) \ ,
\end{align}
with $p(\bm W) = \prod_{i=1}^{M}\prod_{j=1}^{N}\varrho_W(W_{ij})$. Another crucial parameter is the ``rectangularity ratio''
\begin{equation}
\alpha=\sqrt{\frac{M}{N}} \ ,
\end{equation}
which stays fixed as the dimensionality of the matrix grows large $M,N\to\infty$. We thus say that the corresponding matrix $\bm{X}$ is sparse in the sense that the average number $\langle k \rangle$  of non-zero elements per row and $\langle s \rangle$ per column do not scale with either $M$ or $N$.

\subsection{Replica Methodology} \label{sec:Replica Methodology}
Here, we sketch out how the replica method can be used to calculate a variety of observables related to our matrix $\bm{A}$ in the thermodynamic limit. 

We can evaluate the largest eigenvalue $\lambda_1$ of our matrix $\bm{A}$ by framing it in terms of a Courant-Fischer maximisation problem. Namely, we have that

\begin{equation}
    \lambda_1 = \frac{1}{N}\max_{\bm{v}\in \mathbb{R}^N,\ |\bm{v}|^2=N} \langle\bm{v}, \bm{A} \bm{v}\rangle\ ,
    \label{eq:Lambda1 Opt}
\end{equation}

\noindent where here $\langle\cdot,\cdot\rangle$ stands for the standard dot product among vectors in $\mathbb{R}^N$. We can now define the following partition function
\begin{equation}
     \mathcal{Z}_{\beta}=\int \mathrm{d}\bm{v}~ \exp\left(\frac{\beta}{2}\langle\bm{v},\bm{A}\bm{v}\rangle\right)\delta\left(|\bm{v}|^2-N\right)\ .\label{canonicalpart}
\end{equation}
Then, through a saddle-point approach in the zero-temperature limit $\beta\to\infty$, the integral's mass concentrates around the eigenvector associated with the largest eigenvalue. Thus we have that, applying \eqref{eq:Lambda1 Opt}
\begin{equation}
    \mathcal{Z}_{\beta}\approx \exp\left(\frac{\beta}{2}\max_{\bm{v}\in \mathbb{R}^N,\ |\bm{v}|^2=N} \left\langle\bm{v},\bm{A} \bm{v}\right\rangle\right)=\exp\left(\frac{\beta}{2}N\lambda_1\right)\ .
\end{equation}

\noindent Therefore, after taking the average over the matrix $\bm A$ (which entails averaging over the randomness from the signal $\bm x$ as well as the matrix $\bm X$ at fixed $\bm s$ and $\bm k$ given by \eqref{eq: X given degree sequences}), we can write the following relation for the average largest eigenvalue of $\bm A$:

\begin{equation}
    \Big\langle\lambda_1\Big\rangle_{\bm{A}} = \lim_{\beta\rightarrow\infty} \frac{2}{\beta N} \Big\langle \log \mathcal{Z}_{\beta}\Big\rangle_{\bm{A}}\ . 
    \label{eq:Lambda1 StatMech}
\end{equation}

\noindent In order to tackle the average on the r.h.s of Eq. \eqref{eq:Lambda1 StatMech} we invoke the replica trick \cite{BookParisi}

\begin{equation}
    \Big\langle\lambda_1\Big\rangle_{\bm{A}} = \lim_{\beta\rightarrow\infty}\lim_{N\to \infty} \frac{2}{\beta N}  \lim_{n\rightarrow 0} \frac{1}{n}\log\Big\langle \mathcal{Z}^n_{\beta} \Big\rangle_{\bm A}\ ,
    \label{eq:Lambda_Replica}
\end{equation}
where $n$ is initially promoted to an integer, and then analytically continued to the vicinity of zero. The problem thus amounts to computing the average of the \emph{replicated} partition function $\mathcal{Z}^n_{\beta}$ (an $n$-fold integral). We will also assume that the $n$ and $N$ limits commute harmlessly, so the $N$ limit can be taken first.

In order to study further observables related to $\bm{A}$, let \(F:\mathbb{R}\times\mathbb{R}^N\times\mathbb{R}^N\to\mathbb{R}\) be a function of the scalar $u$, the signal vector \(\bm{x}\) and the top eigenvector \(\bm{v}_{\mathrm{top}}\) of \(\bm{A}\). We want to calculate
\begin{equation}
\langle F(u,\bm{v}_{\mathrm{top}},\bm{x})\rangle_{\bm{A}}\ .
\end{equation}
To do so, we define the auxiliary partition function \cite{Susca2019,Susca2021}
\begin{equation}
\mathcal{Z}_{\beta}(u;\bm{A};t;F)=\int \mathrm{d}\bm{v}\;\exp\!\Bigg[\frac{\beta}{2}\langle \bm{v},\bm{A}\bm{v}\rangle+\beta t N F(u,\bm{v},\bm{x})\Bigg]\;\delta\big(|\bm{v}|^{2}-N\big)\ .\label{tauxiliarypartfun}
\end{equation}
We note that we can retrieve the partition function used for the calculation of the top eigenvalue in \eqref{canonicalpart} by setting \(t=0\) (or $F(u,\bm{v},\bm{x})=0$). Using the chain rule, we can take the partial derivative w.r.t. $t$ of the logarithm of the auxiliary partition function to find that
\begin{equation}
\frac{\partial}{\partial t}\log \mathcal{Z}_{\beta}
=\int \mathrm{d}\bm{v}\;\beta  N F(u,\bm{v},\bm{x})\frac{
\exp\!\Big[\frac{\beta}{2}\langle \bm{v},\bm{A}\bm{v}\rangle+\beta t N F(u,\bm{v},\bm{x})\Big]\;
\delta(|\bm{v}|^{2}-N)}
{\displaystyle\int \mathrm{d}\bm{v}'\;
\exp\!\Big[\frac{\beta}{2}\langle \bm{v}',\bm{A}\bm{v}'\rangle+\beta t N F(u,\bm{v}',\bm{x})\Big]\;
\delta(|\bm{v}'|^{2}-N)}\ .    
\end{equation}
Evaluating this partial derivative at $t=0$ gives us the following:
\begin{equation}
\left.\frac{\partial}{\partial t}\log\mathcal{Z}_{\beta}\right|_{t=0}
=
\int \mathrm{d}\bm{v}\; \beta N\, F(u,\bm{v},\bm{x})\;
\frac{\exp\!\big[\tfrac{\beta}{2}\langle \bm{v},\bm{A}\bm{v}\rangle\big]}
{\int\mathrm{d}\bm{v}'\,\exp\!\big[\tfrac{\beta}{2}\langle \bm{v}',\bm{A}\bm{v}'\rangle\big]\delta(|\bm{v}'|^{2}-N)}
\;\delta\big(|\bm{v}|^2-N\big)\ ,
\end{equation}
where we recognise the Gibbs--Boltzmann distribution of vectors $\bm v$ on the sphere in the form
\begin{equation}
P_{\beta,\bm{A}}(\bm{v})
=
\frac{\exp\!\big[\tfrac{\beta}{2}\langle \bm{v},\bm{A}\bm{v}\rangle\big]}
{\int \mathrm{d}\bm{v}'\,\exp\!\big[\tfrac{\beta}{2}\langle \bm{v}',\bm{A}\bm{v}'\rangle\big]\;
\delta\big(|\bm{v}'|^2-N\big)}
\;\delta\big(|\bm{v}|^2-N\big)\ .
\end{equation}
In the limit $\beta\to\infty$ the measure again concentrates around the top eigenvector of the matrix $\bm A$, and thus we have that the average of our function $F(u,\bm{v}_{\mathrm{top}},\bm{x})$ can be written as
\begin{equation}
\langle F(u,\bm{v}_{\mathrm{top}},\bm{x})\rangle_{\bm{A}}
= \lim_{\beta\to\infty}\lim_{N\to \infty}\frac{1}{N\beta}
\left.\frac{\partial}{\partial t}\langle \log(\mathcal{Z}_{\beta})\rangle_{\bm{A}}\right|_{t=0}\ .
\end{equation}
Appealing to the replica trick \cite{BookParisi} once again, we have that the average of our function of interest can be written as
\begin{equation}\label{eq:Function average replics}
\langle F(u,\bm{v}_{\mathrm{top}},\bm{x})\rangle_{\bm{A}}
= \left.\lim_{\beta\to\infty}\lim_{n\to 0}\lim_{N\to \infty}
\frac{1}{nN\beta}\frac{\partial}{\partial t}
\log\big\langle \mathcal{Z}_{\beta}^n(u;\bm A;t;F)\big\rangle_{\bm{A}}\right|_{t=0} 
\end{equation}
in terms of the auxiliary partition function \eqref{tauxiliarypartfun}, where again the limit $N\rightarrow\infty$ is taken first.

We are thus able to calculate a host of different observables related to the top eigenvector $\bm{v}_{\mathrm{top}}$ and the signal vector $\bm{x}$. For example, setting
\begin{equation}\label{eq:Top eigenvector F choice}
F(u,\bm{v},\bm{x})\rightarrow F_{\mathrm{top}}(u,\bm{v}) := \frac{1}{N}\sum_{i=1}^N\delta(u-v_i) 
\end{equation}
\noindent allows us to compute the density of the top eigenvector components
\begin{equation}\label{eq:Average top eigenvector initial}
    \rho_{\mathrm{top}}(u)=\left\langle\frac{1}{N}\sum_{i=1}^N\delta(u-v_{\mathrm{top},i})\right\rangle_{\bm{A}}\ .
\end{equation}
Similarly, if we are interested in computing the average overlap between the top eigenvector $\bm{v}_{\mathrm{top}}$ and the signal $\bm{x}$, then by choosing
\begin{equation}\label{eq:F choice overlap}
    F(u,\bm{v}, \bm{x})\rightarrow F_{\mathrm{ov}}(u,\bm{v},\bm{x}):=\frac{1}{N}\sum_{i=1}^N\delta(u-x_iv_i)\ ,
\end{equation}
we can obtain the overlap density
\begin{equation} \label{def:overlap density}
    \rho_{\mathrm{ov}}(u)=\left\langle\frac{1}{N}\sum_{i=1}^N\delta(u-x_iv_{\mathrm{top},i})\right\rangle_{\bm{A}}\ ,
\end{equation}
which by integrating over $u$ gives us the following expression for the average overlap between the top eigenvector and the signal
\begin{equation}\label{eq:overlap density initial}
    \lim_{N\rightarrow\infty}\left\langle\frac{\langle\bm{x},\bm{v}_{\mathrm{top}}\rangle}{N}\right\rangle_{\bm{A}}=\lim_{N\rightarrow\infty}\int \mathrm{d}u \ \rho_\mathrm{ov}(u)u\ .
\end{equation}
To compute the squared overlap between the top eigenvector $\bm{v}_{\mathrm{top}}$ and the signal $\bm x$, we use the definition of $\bm{A}$ from \eqref{def:model} and insert it into \eqref{canonicalpart} giving
\begin{equation} \label{partition function t is 0}
    \mathcal{Z}_{\beta}=\int \mathrm{d}\bm{v}\;\exp\!\Bigg[\frac{\beta}{2}\langle \bm{v},\bm{J}\bm{v}\rangle+\frac{\beta\theta}{2N}\langle\bm x,\bm v\rangle^2\Bigg]\;\delta\big(|\bm{v}|^{2}-N\big)\ .
\end{equation}
Furthermore, recalling \eqref{eq:Lambda1 StatMech}, we can take the derivative of the logarithm of \eqref{partition function t is 0} to obtain the following relation for the squared overlap
\begin{equation}
    \frac{\mathrm{d}\langle\lambda_1\rangle_{\bm{A}}}{\mathrm{d}\theta}=\lim_{\beta\to\infty}\frac{2}{\beta}\frac{1}{N}\frac{\mathrm{d}}{\mathrm{d}\theta}\left\langle \log Z_{\beta} \right\rangle_{\bm A}=\left\langle \frac{\langle\bm x,\bm v_{\mathrm{top}}\rangle^2}{N^2} \right\rangle_{\bm A} \ .\label{eq:HF}
\end{equation}
The full replica calculation for all such observables is presented in \ref{replica calculations}. In the next section, we present a summary of our main results.

\section{Main Results} \label{Main Results}
In this section, we will provide the main results of our replica calculations that led to the calculation of the observables discussed in Section \ref{sec:Replica Methodology}. The calculations are presented in full in \ref{replica calculations}. 

We also introduce the following shorthand notation $\{\mathrm{d}\pi\}_s:=\prod_{l=1}^s \pi(\omega_l,h_l)\mathrm{d}\omega_l\mathrm{d}h_l$ for a given joint density $\pi$, and $\{h\}_s=\sum_{l=1}^sh_l$ for an indexed collection of scalar variables. Additionally, $\langle ...\rangle_x$ denotes an average taken over the random variable $x$, whereas $\langle...\rangle_{\{x\}_s}$ denotes averaging over $s$ instances of the random variable, and $\langle ...\rangle_{x,\{y\}_s}$ denotes averaging over one instance of $x$ and $s$ instances of $y$.
\begin{result}[Self-consistency equations and the top eigenvalue]
    Let the signal $\bm{x}$ have i.i.d. entries $x\sim\varrho_x$. Let the matrix $\bm X$ be as defined in \eqref{X definition} with column degree sequence $\bm s$ and row degree sequence $\bm k$ jointly drawn from $p_{k,s}(\bm k,\bm s)$ with marginals $p_k(k)$ and $p_s(s)$ and bond weights $W\sim\varrho_W$. Let $\langle k\rangle=\sum_k k p_k(k)$ and $\langle s\rangle=\sum_s s p_s(s)$, with $\langle k\rangle = \frac{1}{\alpha^2}\langle s\rangle$. Let $\pi(\omega,h)$ and $\rho(\sigma,\mu)$ be the joint probability densities of the pairs $(\omega,h)$ and $(\sigma,\mu)$ respectively, with both $\omega>0$ and $\sigma>0$, that solve the following coupled Recursive Distributional Equations (RDE)\footnote{These are coupled integral equations for joint probability densities.}
    \begin{equation}
    \pi(\omega, h) = \sum_{s=1}^{s_{\mathrm{max}}} \frac{s p_s(s)}{\langle s\rangle} \int \{ \mathrm{d}\rho \}_{s-1} \Bigg\langle \delta\left( \omega - \left(\lambda_\theta - \left\{ \frac{W^2}{\sigma}\right\}_{s-1}\right) \right)
    \delta\left( h - \left(\theta x q + \left\{ \frac{\mu W}{\sigma} \right\}_{s-1}\right) \right) \Bigg\rangle_{x,\{W\}_{s-1}} \ ,\label{eq:pi recursive}
    \end{equation} 
    \begin{equation}\label{eq:rho recursive}
        \rho(\sigma, \mu) = \sum_{k=1}^{k_{\mathrm{max}}} \frac{k p_k(k)}{\langle k\rangle } \int \{ \mathrm{d}\pi \}_{k-1} \Bigg\langle\delta\left( \sigma - \left(1 - \left\{ \frac{W^2}{\omega} \right\}_{k-1}\right) \right)
        \delta\left( \mu - \left\{ \frac{hW}{\omega} \right\}_{k-1} \right)\Bigg\rangle_{\{W\}_{k-1}} \ ,
    \end{equation}
    supplemented by the following two constraints that fix the parameters $\lambda_\theta$ and $q$
    \begin{equation}\label{eq:lambda condition}
        1 = \sum_{s=0}^{s_{\mathrm{max}}} p_s(s) \int \left\{ \mathrm{d}\rho \right\}_s \left\langle \left( \frac{\theta qx + \left\{ \frac{\mu W}{\sigma} \right\}_s}{\lambda_\theta - \left\{ \frac{W^2}{\sigma} \right\}_s} \right)^2 \right\rangle_{x,\{W\}_{s}}\ ,
    \end{equation}
    \begin{equation}\label{eq:q condition}
        q = \sum_{s=0}^{s_{\mathrm{max}}} p_s(s) \int \{ \mathrm{d}\rho \}_s \left\langle \frac{\theta  qx^2 + x \left\{ \frac{\mu W}{\sigma} \right\}_s}{\lambda_\theta - \left\{ \frac{W^2}{\sigma} \right\}_s} \right\rangle_{x,\{W\}_{s}}\ . 
    \end{equation}
Then the average largest eigenvalue $\langle\lambda_1\rangle_{\bm{A}}$
of the matrix $\bm{A}$ defined in \eqref{def:model} is equal to the value
$\lambda_\theta$ of the spectral parameter at the extremising
saddle-point solution of
\crefrange{eq:pi recursive}{eq:q condition}. Namely,
    \begin{equation}
        \langle\lambda_1\rangle_{\bm{A}}=\lambda_\theta \ .
    \end{equation}
\end{result}
We can numerically solve \crefrange{eq:pi recursive}{eq:q condition} efficiently using a population dynamics algorithm outlined in Section \ref{PopDyn}. Having found the joint probability densities $\pi$ and $\rho$ that satisfy these equations, we can then compute other spectral properties of the matrix $\bm{A}$.

\begin{result}[Top eigenvector component density]
    The average top eigenvector component density defined in \eqref{eq:Average top eigenvector initial} is, for large $N,M$, given by
    \begin{equation}\label{eq:eigenvector density}
        \rho_{\mathrm{top}}(u)  = \sum_{s=0}^{s_{\mathrm{max}}} p_s(s) \int \{ \mathrm{d}\rho \}_s \left\langle \delta \left( u - \frac{\theta  qx + \left\{ \frac{\mu W}{\sigma} \right\}_s}{\lambda_\theta - \left\{ \frac{W^2}{\sigma} \right\}_s} \right) \right\rangle_{x,\{W\}_s}\ .
    \end{equation}
\end{result} 
\begin{result}[Eigenvector component overlap density]
    The average overlap density given by \eqref{def:overlap density} is given by
    \begin{equation} \label{eq:overlap formula}
        \rho_{\mathrm{ov}}(u) = \sum_{s=0}^{s_{\mathrm{max}}} p_s(s) \int \{ \mathrm{d}\rho \}_s \left\langle \delta \left( u - \frac{\theta  qx^2 + x \left\{ \frac{\mu W}{\sigma} \right\}_s}{\lambda_\theta - \left\{ \frac{W^2}{\sigma} \right\}_s} \right) \right\rangle_{x,\{W\}_s}\ .
    \end{equation}
\end{result}
From the overlap density \eqref{eq:overlap formula}, we obtain the average value of the overlap between $\bm{v}_{\mathrm{top}}$ and $\bm{x}$.
\begin{result}[Average overlap]
     The average overlap between the top eigenvector and the signal vector is given by
     \begin{equation} \label{eq:overlap}
         \lim_{N \to \infty} \left\langle \frac{\langle \bm{x}, \bm{v}_{\mathrm{top}}\rangle}{N} \right\rangle_{\bm{A}} = \sum_{s=0}^{s_{\mathrm{max}}} p_s(s) \int \{ \mathrm{d}\rho \}_s \left\langle \frac{\theta q x^2 + x \left\{ \frac{\mu W}{\sigma} \right\}_s}{\lambda_\theta - \left\{ \frac{W^2}{\sigma} \right\}_s} \right\rangle_{x,\{W\}_s}.
     \end{equation}
\end{result}
\begin{remark}[Interpretation of $q$]
    We note the equivalence between \eqref{eq:overlap} and \eqref{eq:q condition}. This implies that similarly to the parameter $\lambda_\theta$ being interpreted as the average largest eigenvalue, the parameter $q$ can be interpreted as the average overlap between the top eigenvector and the signal vector.
\end{remark}

\begin{remark}[Sign of $q$]
We note that the overlap \eqref{eq:overlap density initial} between the signal and the top eigenvector (and thus $q$) is a signed quantity by definition. Since the spike enters the model through $\bm x\bm x^{\!\top}$, the signal is identifiable only up to a global sign, while $\bm v_{\mathrm{top}}$ itself is also defined only up to sign. We therefore fix the orientation of $\bm v_{\mathrm{top}}$ relative to $\bm x$ such that\footnote{Equivalently, one may introduce a symmetry-breaking protocol, introducing an infinitesimal aligning field and removing it after taking the thermodynamic limit.}
\[
\lim_{N\to\infty}\left\langle \frac{\langle \bm x,\bm v_{\mathrm{top}}\rangle}{N}\right\rangle_{\bm A}
= q \ge 0\ .
\]
\end{remark}

\begin{figure}[h!]
    \centering
    \begin{subfigure}{0.32\textwidth}
        \centering
        \includegraphics[width=\linewidth]{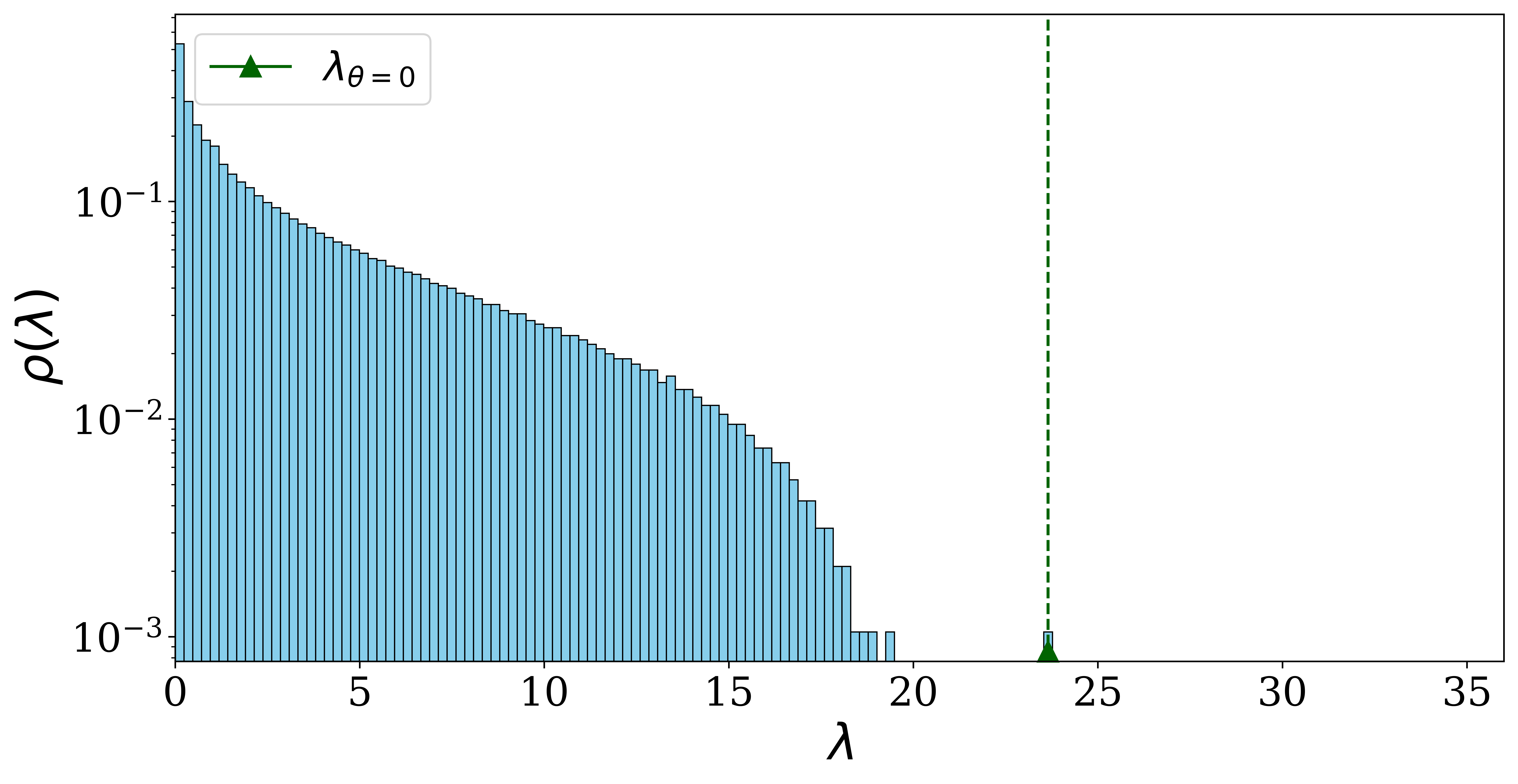}
        \label{fig:spectral0}
    \end{subfigure}
    \hfill
    \begin{subfigure}{0.32\textwidth}
        \centering
        \includegraphics[width=\linewidth]{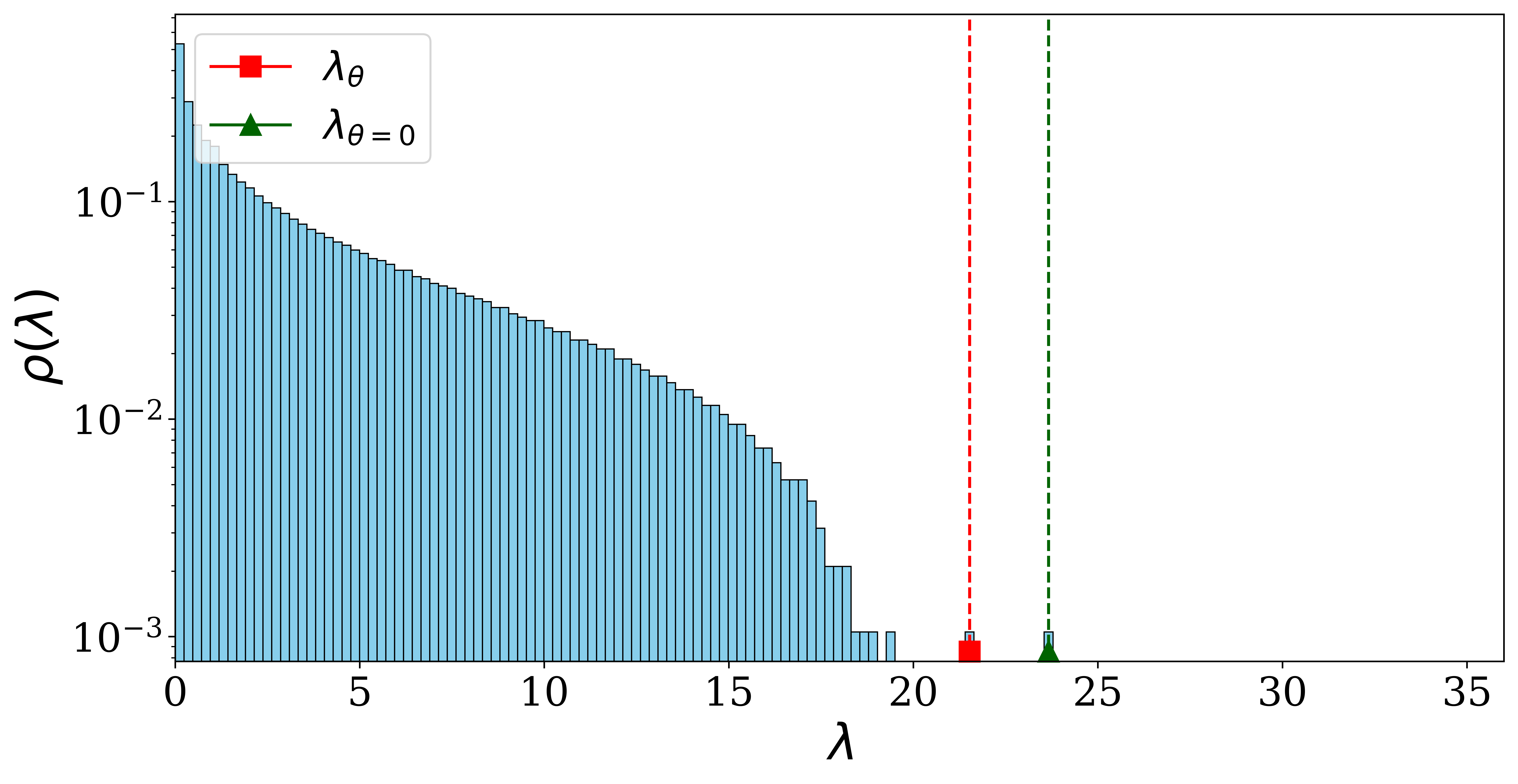}
        \label{fig:spectral16}
    \end{subfigure}
    \hfill
    \begin{subfigure}{0.32\textwidth}
        \centering
        \includegraphics[width=\linewidth]{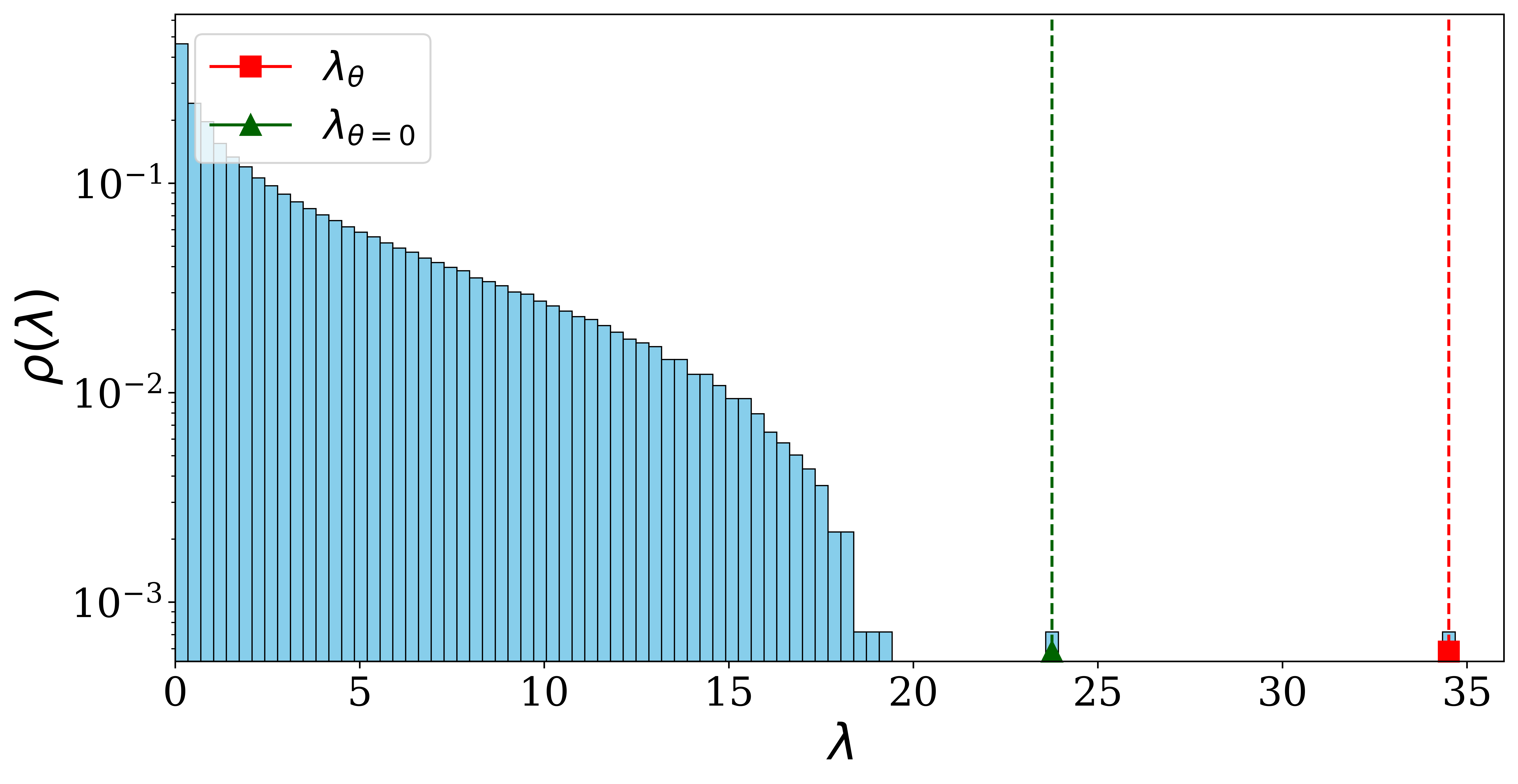}
        \label{fig:spectral30}
    \end{subfigure}
    \caption{Eigenvalue density of $1000$ matrices $\bm{A}$ of size $N\times N=4000\times4000$ as defined in \eqref{def:model}, with $\bm{X}\in\mathbb{R}^{M\times N}$ a pure $5000\times 4000$ adjacency matrix of a Poissonian random graph with mean column degree $\langle s\rangle=4$, mean row degree $\langle k\rangle= 3.2$ (following $\langle k\rangle =\frac{1}{\alpha^2}\langle s\rangle$ as per \eqref{handshake constraint math}), and weight distribution $\varrho_W(W)=\delta_{W,1}$. The signal $\bm{x}$ has i.i.d. entries $x_i\sim\mathcal{N}(0,1)$. (\textit{Left}) Density for $\theta=0$ corresponding to the matrix defined in \cite{Budnick2025}, where there is one sole outlier from the bulk located at $\lambda_{\theta=0}\simeq 23.65$. The sole outlier is due to the bond weights having a non-zero mean. (\textit{Centre}) Density for $0<\theta=16<\theta_{\mathrm{crit}}$. Here the signal-associated eigenvalue has separated from the bulk, yet
remains smaller than $\lambda_{\theta=0}$.}(\textit{Right}) Density for the recovery phase, when $\theta_{\mathrm{crit}}<\theta=30$, resulting in the eigenvalue associated with the signal overtaking the noise eigenvalue, and becoming the top eigenvalue. The adjacency matrices are generated using an unbiased configuration algorithm that can be found in Ref.~\cite{Annibale2017}. 
    \label{fig:spectral_density}
\end{figure}

\begin{remark}[Reduction to the pure noise matrix]
    We note that setting $\theta=0$, or $\varrho_x=\delta_0$, removes the signal in \eqref{def:model} and we are left with the pure noise matrix. In this case, \crefrange{eq:pi recursive}{eq:lambda condition} and Eq. \eqref{eq:eigenvector density} recover the formulas for $\langle\lambda_1\rangle_{\bm{A}}$ and $\rho_{\mathrm{top}}(u)$ from \cite{Budnick2025}, as expected.
\end{remark}
Finally, we are also interested in the BBP-like phase transition of the top eigenvalue and associated eigenvector statistics as a function of $\theta$. Namely, we can identify the critical threshold $\theta_{\mathrm{crit}}$ depending on the average connectivity $\langle k\rangle$ (or $\langle s\rangle$) and the rectangularity ratio $\alpha$ such that $\forall \ \theta\leq\theta_{\mathrm{crit}}$ we have $\langle\lambda_1\rangle_{\bm{A}}=\lambda_{\theta=0}$ and $\lim_{N \to \infty} \left\langle \langle \bm{x}, \bm{v}_\mathrm{top} \rangle/N \right\rangle_{\bm{A}} = 0$, meaning that the signal is not recoverable from the top eigenvector of $\bm A$. Past this threshold however  $\forall \ \theta>\theta_{\mathrm{crit}}$ we have $\langle\lambda_1\rangle_{\bm{A}}=\lambda_{\theta}$ and the average overlap $\lim_{N \to \infty} \left\langle \langle \bm{x}, \bm{v}_\mathrm{top} \rangle/N \right\rangle_{\bm{A}} \neq0$, indicating that (partial) signal recovery is possible through the top eigenvector. This phase transition is illustrated graphically in Figure \ref{fig:spectral_density}.

\begin{result}[Recovery threshold]
    Under the assumption that the signal distribution $\varrho_x$ is such that $\langle x\rangle_x=0$, and $\langle x^2\rangle_x=\sigma_x^2$, the recovery threshold $\theta_\mathrm{crit}$ is given by
    \begin{equation}\label{eq:recovery threshold}
    \theta_{\mathrm{crit}} = \frac{1}{\sigma_x^2 Q(\lambda_{\theta=0})}\ ,
    \end{equation}
    where $\lambda_{\theta=0}$ is the top eigenvalue of the noise matrix $\bm{J}$ as obtained in \cite{Budnick2025}, and the function $Q(\lambda)$ is given by
    \begin{equation}\label{eq:Big Q definition}
        Q(\lambda) = \sum_{s=0}^{s_{\mathrm{max}}} p_s(s) \int \{ \mathrm{d}\rho \}_s \left\langle\frac{1}{\lambda - \left\{ \frac{W^2}{\sigma} \right\}_s}\right\rangle_{\{W\}_s}\ .
    \end{equation}
\end{result}

\begin{remark}[Overlap condition]
    We note that \eqref{eq:q condition}, written in terms of $Q(\lambda)$, can be expressed (for a signal with zero-centered mean) as
    \begin{equation}\label{eq:transition equation set to 0}
        q\bigl(1-\theta\sigma_x^2Q(\lambda)\bigr)=0\ .
    \end{equation}
This equation admits a solution with $q=0$, corresponding to the
non-recovery phase of the leading eigenvector, and a non-trivial branch
with $q\neq 0$, for which
\[
    \theta\sigma_x^2Q(\lambda)=1\ .
\]
The recovery threshold for the leading eigenvector is reached when this
non-trivial branch meets the top eigenvalue $\lambda_{\theta=0}$ of the
noise matrix. Evaluating it at $\lambda=\lambda_{\theta=0}$ therefore
gives \eqref{eq:recovery threshold}.
\end{remark}

\begin{figure}
    \centering
    \includegraphics[width=0.7\linewidth]{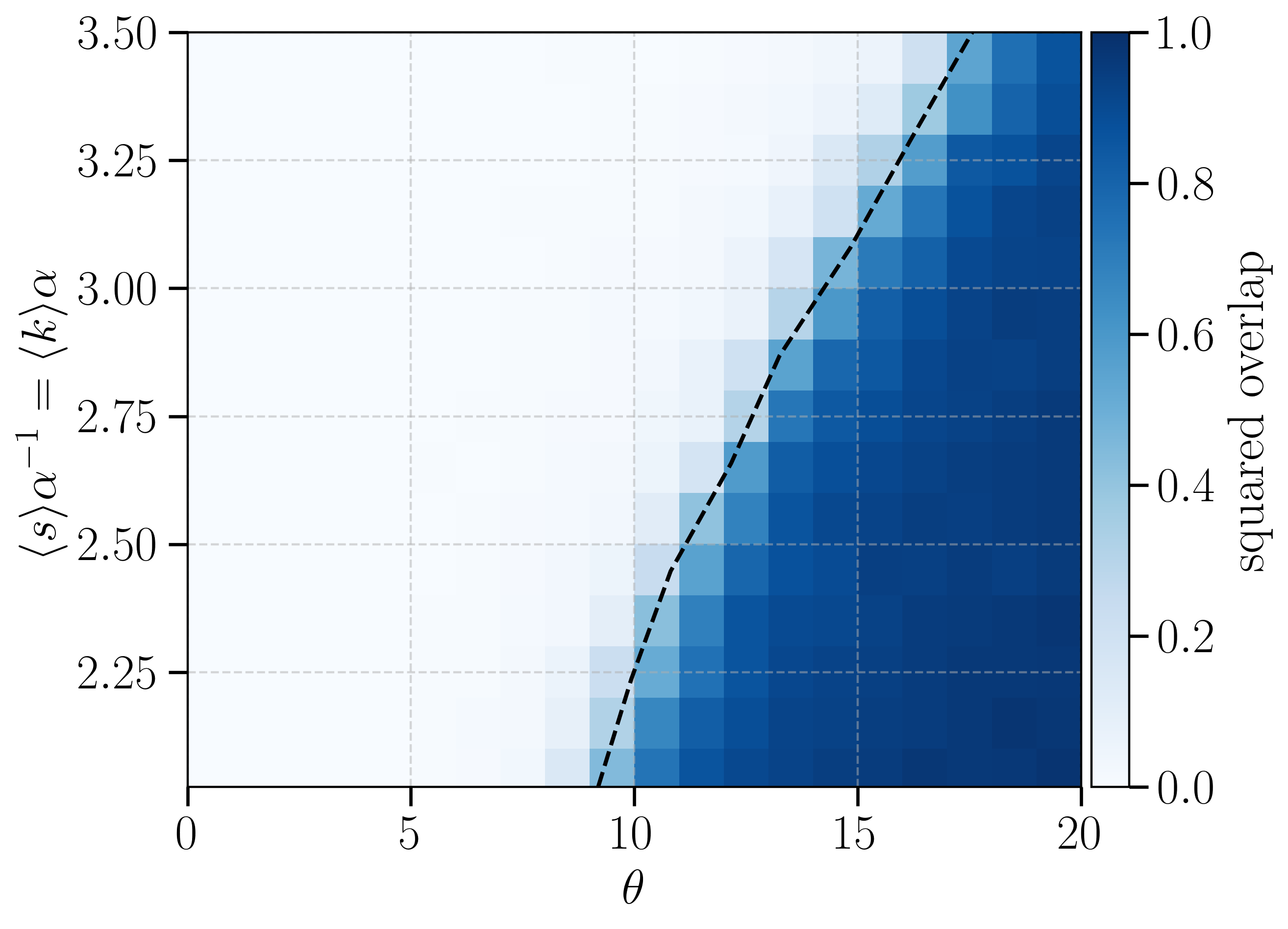}
    \caption{Heat map of the average squared overlap between the top eigenvector $\bm{v}_\mathrm{top}$ and the signal vector $\bm{x}$ obtained for the truncated Poisson setup with $k_{\rm max} = s_{\rm max} = 20$ for various values of $\langle s\rangle \alpha^{-1} = \langle k \rangle \alpha$ with $\alpha=\sqrt{5/4}$ and signal strength $\theta$. The top eigenvector is obtained via numerical diagonalisation of $50$ matrices $\bm A$ of size $N=1000$ (implying $M=\alpha^2 N=1250$) as defined in \eqref{def:model}. The spike components are standard Gaussian, $x_i \sim \mathcal N(0,1)$, and we work with a pure adjacency matrix, $W_{ij}=1$. The dashed line represents the critical threshold value $\theta_\mathrm{crit}$ in \eqref{eq:Poisson_Thetacrit} as a function of the rescaled average connectivities, obtained via the numerical solution of the more explicit form for $\tilde Q^{(P)}(\lambda)$ in \eqref{eq:QW} with $P=15$.}
    \label{fig:Poisson_phase_diagram}
\end{figure}

\begin{remark}[Dense Limit]
    In the large connectivity limit $\langle s\rangle\rightarrow\infty$, after a suitable rescaling of the matrix entries, the critical value $\theta_\mathrm{crit}$ in \eqref{eq:recovery threshold} converges to $\theta_{\mathrm{crit}} = \frac{1+\alpha}{\sigma_x^2\alpha^2} $, and the top eigenvalue to $\lambda_\theta = \theta \sigma_x^2 \left( 1 + \frac{\alpha^2 }{\alpha^2\theta \sigma_x^2 - 1} \right)$, retrieving the ``classical'' BBP result for additive deformations of dense Wishart matrices \cite{Baik2005, Forrester2023,Benaych-Georges2011}. This is shown in \ref{sec:dense limit}.
\end{remark}
\begin{remark}[Recovery and non-recovery phases]
    Figure \ref{fig:Poisson_phase_diagram} shows that increasing $\langle s\rangle$ increases the necessary signal strength in order to reach the recovery phase. Intuitively, this phenomenon can be understood since higher values of $\langle s\rangle$ lead to denser noise matrices, obscuring the signal and making signal recovery more difficult.
\end{remark}

\section{Special Cases} \label{sec:special_cases}
We notice that the RDE solutions in \eqref{eq:pi recursive} and \eqref{eq:rho recursive} only depend on the marginal single-degree distributions $p_k(k)$ and $p_s(s)$ and not on the original joint distribution $p_{k,s}(\bm k,\bm s)$. Therefore, irrespective of what this original distribution $p_{k,s}(\bm k,\bm s)$ is (and whether it can be uniquely reconstructed from the marginals), we now specialise the results of Section \ref{Main Results} by working directly with the marginal degree distributions $p_k(k)$ and $p_s(s)$. We consider in particular the truncated Poisson and random biregular graph type, with fixed noise weights. We are working under the assumption that the signal distribution $\varrho_x$ is zero-centered $\langle x\rangle_x=0$, with $\langle x^2\rangle_x=\sigma_x^2>0$. A detailed derivation can be found in \ref{app:sec:special_cases}.

\begin{figure}
    \centering
    \includegraphics[width=0.8\linewidth]{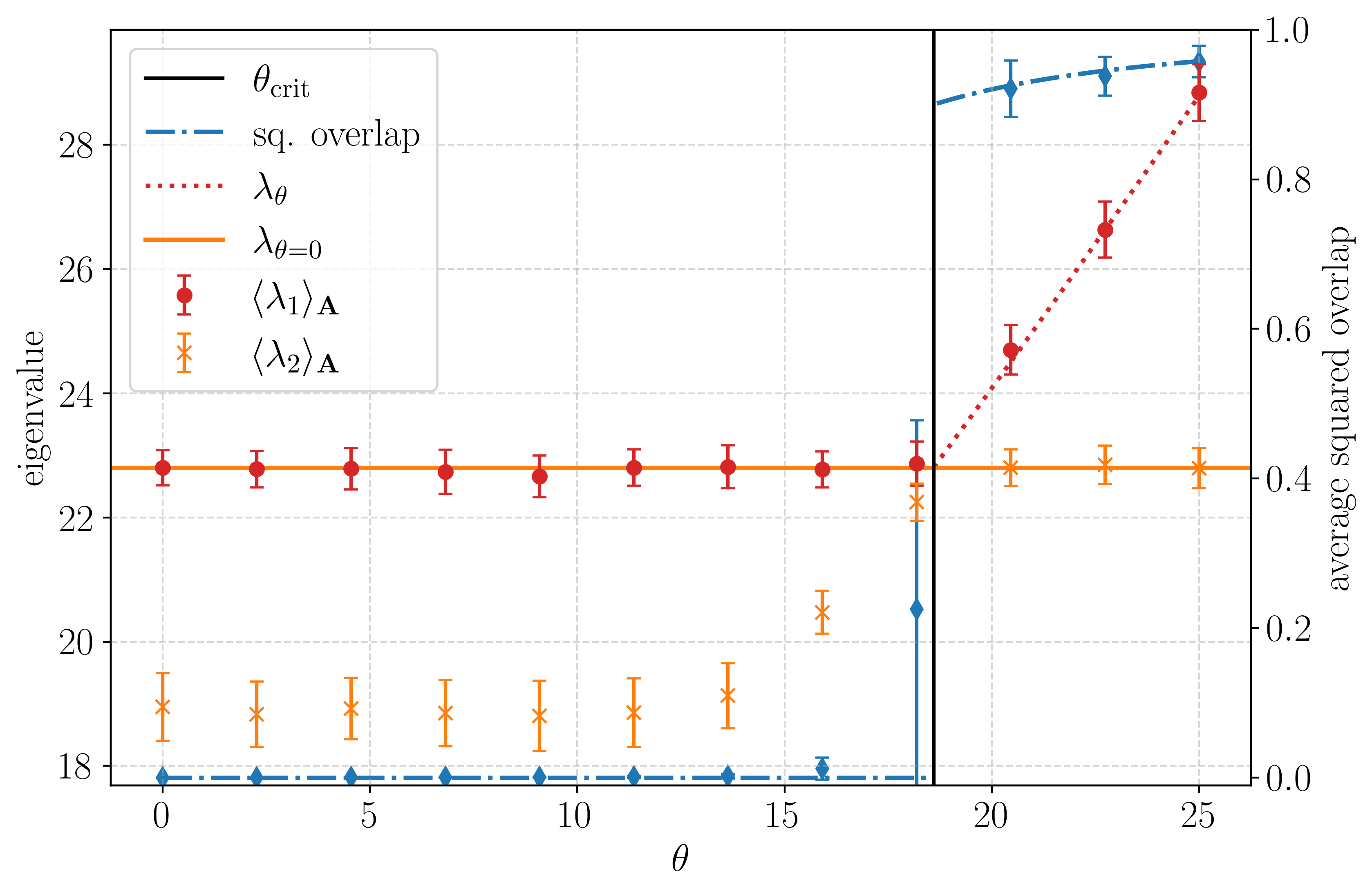}
    \caption{Truncated Poisson setup with $ \langle s \rangle = 3, \langle k \rangle = 4$, $s_{\mathrm{max}}=k_{\mathrm{max}}=15$ and $\alpha = \sqrt{3/4}$. The lines correspond to theoretical predictions presented in Section~\ref{sec5:truncated_poisson_special} and the dots are results from 50 numerical simulations via direct diagonalisation of the matrix $\bm A$ of size $N \times N = 5000 \times 5000$, with standard deviation error bars. The spike components are $x_i \sim \mathcal N(0,1)$, and the bond weights are $W_{ij}=1$. The metrics are presented as a function of the signal strength $\theta$. The solid black vertical line represents the value $\theta_\mathrm{crit}$ as predicted by \eqref{eq:Poisson_Thetacrit}, and we observe the phase transition where the squared overlap of the top eigenvector and the signal vector \eqref{eq:Poisson_overlap} jumps discontinuously from a zero to a non-zero value (blue dash-dot line), and the eigenvalue associated with the signal \eqref{eq:Poisson_eig} (red dotted line) overtakes the structural eigenvalue $\lambda_{\theta=0}$ (orange solid line) and becomes the top eigenvalue that further grows with $\theta$. The metrics are computed with a truncation at $P=7$.
    }
    \label{fig:Poiss_transition}
\end{figure}

\subsection{Truncated Poisson}\label{sec5:truncated_poisson_special}
In this section, we consider the column degree distribution
\begin{align}
 p_s^{(\rm Poi)}(s) = \frac{\bar c^s}{\Gamma_s s!} \ , \quad s \in \{ 0,\ldots,s_{\max} \} \ ,
\end{align}
where the normalisation factor is $\Gamma_s = \sum_{s=0}^{s_{\max}}\frac{\bar c^s}{s!}$, the average column degree is defined as $\langle s\rangle = \sum_{s=0}^{s_{\max}} s p_s^{(\rm Poi)}(s)$, which gives the relation between \(\langle s\rangle\) and \(\bar c\) as
\[ \langle s\rangle = \frac{
\sum_{s=0}^{s_{\max}} s\,\dfrac{\bar c^s}{s!} }{ \sum_{s=0}^{s_{\max}} \dfrac{\bar c^s}{s!}
}\ . \]
As $s_{\max}\to+\infty$, the expected average degree satisfies $\langle s\rangle \to \bar c$.
The associated degree-corrected distribution that appears in the RDEs in \eqref{eq:pi recursive} and \eqref{eq:rho recursive} reads
\begin{align}
\frac{s p_s^{(\rm Poi)}(s)}{\langle s \rangle} &= \frac{\bar c^{s}}{\langle s \rangle\Gamma_s (s-1)!} \ , \quad s \in \{ 1,\ldots,s_{\max} \} \ .
\end{align}
The row degree distribution $p_{k}^{(\rm Poi)}(k) = \frac{\bar d^k}{\Gamma_k k!} , \ k \in \{ 0,\ldots,k_{\max} \}$ is defined in an analogous way, with $\Gamma_k = \sum_{k=0}^{k_{\max}}\frac{\bar d^k}{k!}$ and the average degree $\langle k \rangle = \sum_{k=0}^{k_{\max}} k p_k^{(\rm Poi)}(k)=\frac{1}{\alpha^2}\langle s\rangle$ through the handshake constraint \eqref{handshake constraint math}. 

\begin{result}[Truncated Poisson]
    
Let $m_{\sigma,\tau}^{(P)}(\lambda_\theta)$ and $m_{\omega,\tau}^{(P)}(\lambda_\theta)$ be the solutions of the following coupled self-consistent equations for $\tau = 1,\dots,P$,
    \begin{align}\label{eq:hierarchy_sigma}
    m_{\sigma,\tau}^{(P)}(\lambda_{\theta})
    = \sum_{k=1}^{k_{\rm max}} \frac{k p_k(k)}{\langle k \rangle}
    \sum_{p=0}^{P} \binom{\tau+p-1}{p}
    \sum_{\substack{r_1+\dots+r_{k-1}=p\\ r_\ell \ge 0}}
    \frac{p!}{r_1!\cdots r_{k-1}!}
    \prod_{\ell=1}^{k-1}\left[\langle W^{2r_\ell}\rangle_W\,
    m_{\omega,r_\ell}^{(P)}\right] 
    \end{align}
    and
    \begin{align}\label{eq:hierarchy_omega}
    m_{\omega,\tau}^{(P)}(\lambda_{\theta})
    = \sum_{s=1}^{s_{\rm max}} \frac{s p_s(s)}{\langle s \rangle}
    \sum_{p=0}^{P} \binom{\tau+p-1}{p}
    \frac{1}{\lambda_{\theta}^{\,\tau+p}}
    \sum_{\substack{r_1+\dots+r_{s-1}=p\\ r_\ell \ge 0}}
    \frac{p!}{r_1!\cdots r_{s-1}!}
    \prod_{\ell=1}^{s-1}\left[\langle W^{2r_\ell}\rangle_W\,
    m_{\sigma,r_\ell}^{(P)}\right] \ .
    \end{align}
    It is a system of $2P$ equations with $2P$ unknowns and is solved numerically. Then, the function $Q(\lambda_\theta)$ given in the general form in \eqref{eq:Big Q definition} that encodes the observables of interest admits the approximate form
    \begin{align}\label{eq:QW}
    \tilde Q(\lambda_{\theta}) \simeq \frac{\langle s \rangle}{\bar c}m_{\omega,1}^{(P)}(\lambda_{\theta})+ R^{(P)}(\lambda_{\theta},s_{\rm max}) =: \tilde Q^{(P)}(\lambda_{\theta}) \ ,
    \end{align}
    with
    \begin{align}
    R^{(P)}(\lambda_{\theta},s_{\rm max}) =  \frac{\bar c^{s_{\max}}}{\Gamma_s s_{\max}!}
    \sum_{p=0}^{P} \frac{1}{\lambda_{\theta}^{\,1+p}}
    \sum_{\substack{r_1+\dots+r_{s_{\max}}=p\\ r_\ell \ge 0}}
    \frac{p!}{r_1!\cdots r_{s_{\max}}!}
    \prod_{\ell=1}^{s_{\max}}\left[\langle W^{2r_\ell}\rangle_W\, m_{\sigma,r_\ell}^{(P)}\right] \ .
    \end{align}
    The critical value of the signal strength is
    \begin{equation}\label{eq:Poisson_Thetacrit}
    \theta_{\rm crit} = \frac{1}{\sigma_x^2 \tilde Q(\lambda_{\theta=0})} \simeq \frac{1}{\sigma_x^2 \tilde Q^{(P)}(\lambda_{\theta=0})}  \ ,
    \end{equation}
    where $\lambda_{\theta=0}$ is the outlier eigenvalue associated with the noise, i.e., the top eigenvalue of the sparse noise matrix $\bm{J}$ as obtained in \cite{Budnick2025}.
    In the recovery phase for a given $\theta>\theta_{\rm crit}$, the top eigenvalue is
    \begin{align}\label{eq:Poisson_eig}
        \langle\lambda_1\rangle_{\bm{A}} &= \lambda_\theta \simeq (\tilde Q^{(P)})^{-1}\left(\frac{1}{\theta \sigma^2_x}\right) \ ,
    \end{align}
   where we have denoted $\left(Q^{(P)}\right)^{-1}$ for the inverse function of $Q^{(P)}$. The squared overlap between the signal and the top eigenvector is given by
    \begin{align}\label{eq:Poisson_overlap}
        \lim_{N \to \infty}\left\langle \frac{\langle \bm{x}, \bm{v}_{\mathrm{top}} \rangle^2}{N^2} \right\rangle_{\bm{A}} &\simeq -\frac{1}{\theta^2\sigma_x^2 \tilde Q{^{(P)}}' (\lambda_\theta)} \ .
    \end{align}
    The quality of the approximation increases with $P$.
\end{result}

In Fig.~\ref{fig:Poiss_transition}, we plot the above theoretical results as functions of the signal strength $\theta$ for $P=7$. We find good agreement with results from direct diagonalisation of 50 large matrices $\bm A$ of size $N = 5000$. The value of the transition $\theta_{\rm crit}$ in \eqref{eq:Poisson_Thetacrit} is indicated as a black solid vertical line. For $\theta\leq\theta_{\rm crit}$, the top eigenvalue indicated as red dots sits at the value $\lambda_{\theta=0}$ (orange line) which is the top eigenvalue of the noise matrix $\bm{J}$ (as obtained by \cite{Budnick2025} for the pure noise case $\theta=0$).
For $\theta>\theta_{\rm crit}$, the top eigenvalue is correctly predicted by \eqref{eq:Poisson_eig} and the squared overlap jumps discontinuously to a positive value correctly predicted by \eqref{eq:Poisson_overlap}, indicating the onset of the recovery regime via the top eigenvector. This scenario is illustrated in Fig.~\ref{fig:spectral_density} (right).

\subsection{Random Biregular}\label{sec:random regular main body}
In this special case, we set the marginal degree distributions to $p_{s}^{(R)}(s) = \delta_{s,\langle s \rangle}$ and $p_{k}^{(R)}(k) = \delta_{k,\langle k \rangle}$, with $\langle k \rangle =\frac{1}{\alpha^2}\langle s\rangle$. We also use unit weights for the noise matrix, $W_{ij}=1$.
\begin{result}[Random biregular setup]
    The critical value of the signal strength from \eqref{eq:recovery threshold} can be computed in closed form as
    \begin{equation}\label{eq:RR_crit}
        \theta_{\mathrm{crit}}=\frac{\langle s \rangle (\langle s \rangle \langle k \rangle-\langle s \rangle-\langle k \rangle)}{\sigma_x^2(\langle s \rangle-1)}\ .
    \end{equation}
    For $\theta>\theta_\mathrm{crit}$, the average largest eigenvalue is given by
    \begin{equation}\label{eq:RR_eig}
         \langle\lambda_1\rangle_{\bm{A}}= \lambda_\theta=\frac{1}{2} \left(\langle s \rangle \langle k \rangle + (2-\langle s \rangle)\,\theta \sigma_x^2+\langle s \rangle \sqrt{(\langle k \rangle-\theta\sigma_x^2)^2 + 4\theta\sigma_x^2} \right) \ ,
    \end{equation}
    and the average squared overlap between the signal and the top eigenvector is equal to
    \begin{equation}\label{eq:RR_ov}
        \left\langle \frac{\langle \bm{v}_{\rm top},\bm{x} \rangle^2}{N^2} \right\rangle_{\bm{A}} = \frac{\sigma_x^2}{2}\left[2-\langle s \rangle + \langle s \rangle \ \frac{\theta \sigma_x^2 + 2 - \langle k \rangle} {\sqrt{(\langle k \rangle - \theta\sigma_x^2)^2 + 4\theta\sigma_x^2}} \right]\ .
    \end{equation}
    Moreover, setting $\theta=0$ in \eqref{eq:RR_eig} gives the noise outlier $\lambda_{\theta=0}=\langle s \rangle \langle k \rangle$, and the right bulk edge $\lambda_{\rm b}$, first computed by Godsil and Mohar~\cite{GODSIL1988191}, is used to obtain the transition value at which the signal-associated eigenvalue first exits the bulk 
    \begin{equation}\label{eq:RR_bulk}
    \theta_{\mathrm{b}} =
    \frac{\langle k \rangle-2+ \sqrt{(\langle s \rangle-1)(\langle k \rangle-1)}-\sqrt{\frac{\langle k \rangle-1}{\langle s \rangle-1}}}{\sigma_x^2} \ .
    \end{equation}
    The details of the derivation and the value of $\lambda_{\rm b}$ can be found in \ref{app:sec:random_regular}.
\end{result}

\begin{remark}[Bulk statistics]
In the random biregular case, the population distributions are point masses, so
$Q^{RR}(\lambda)$, the analogue of \eqref{eq:Big Q definition}, is available in
closed form and the relevant observables can be computed exactly. For
non-degenerate degree distributions, such as the truncated Poisson case, an
exact determination of the bulk transition would instead require a second
largest eigenpair analysis, as in Ref.~\cite{Susca2020b}. We nevertheless
expect the same three-phase scenario, with an intermediate regime
$\theta_{\rm b}<\theta\leq\theta_{\rm crit}$ in which the signal-associated
eigenvalue is an outlier below $\lambda_{\theta=0}$. In this regime the signal
is not recoverable from the top eigenvector, but partial recovery through the
second eigenvector is expected. Determining $\lambda_{\rm b}$ and
$\theta_{\rm b}$ for general degree distributions is left to future work.
\end{remark}

\begin{figure}
    \centering
    \includegraphics[width=0.8\linewidth]{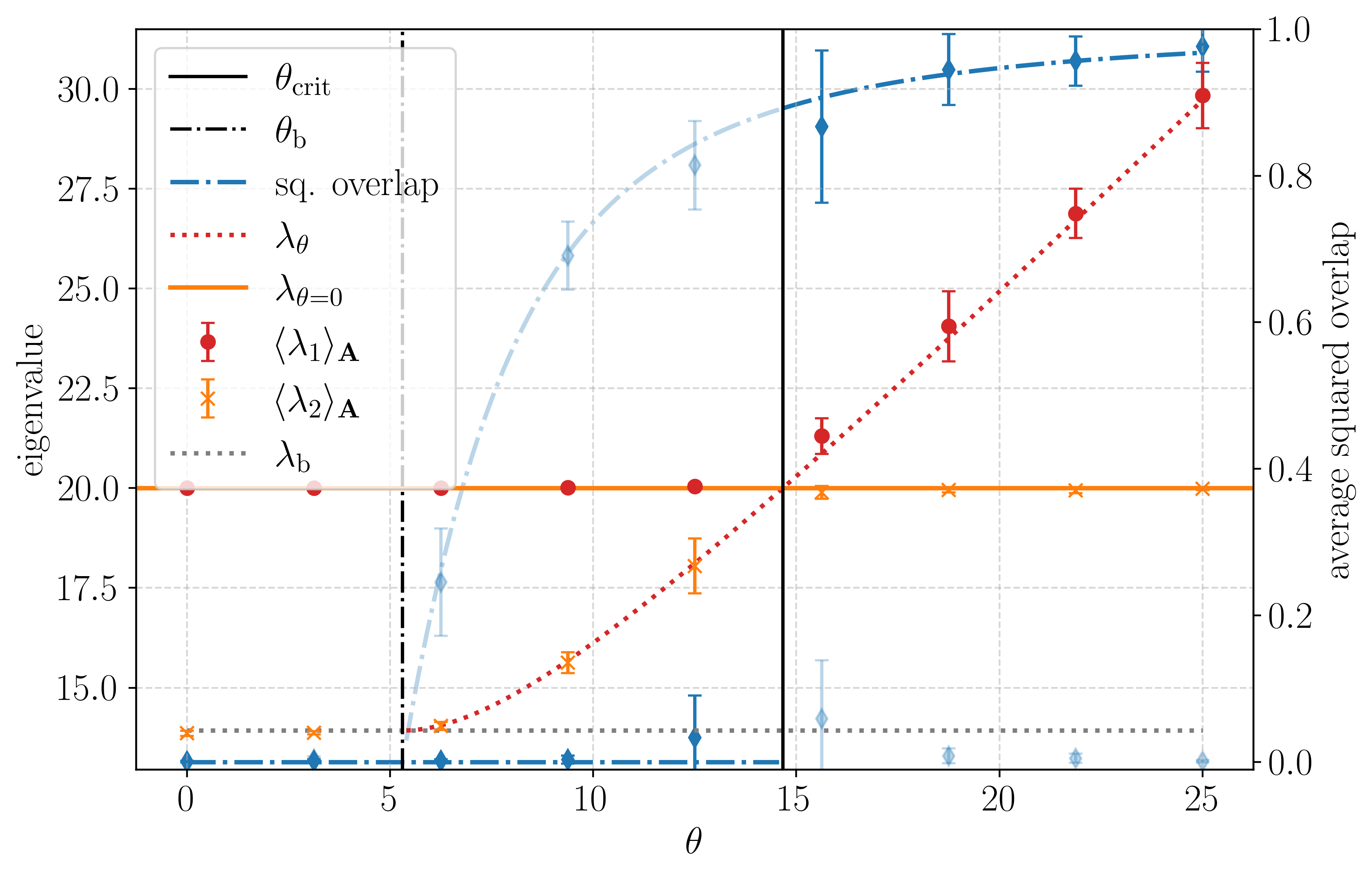}
    \caption{Random Biregular setup with $\alpha = \sqrt{4/5}$ and $ \langle s \rangle = 4, \langle k \rangle = 5$. The lines correspond to theoretical predictions outlined in Section~\ref{sec:random regular main body} and the dots are results from 20 numerical simulations via direct diagonalisation of matrices $\bm A$ of size $N \times N = 1200 \times 1200$, with standard deviation error bars. The spike components are $x_i \sim \mathcal N(0,1)$, and the bond weight distribution of the noise matrix $\bm X$ is $\varrho_W(W)=\delta_{W,1}$. The metrics are presented as a function of the signal strength $\theta$. Red dots report the top eigenvalue, which conforms to the theoretical predictions of the noise eigenvalue (orange line) for $\theta\leq\theta_{\rm crit}$ and the $\lambda_\theta$ \eqref{eq:RR_eig} (red dotted line) for $\theta>\theta_{\rm crit}$. For $\theta_{\rm b}<\theta\leq\theta_{\rm crit}$, the signal has a non-zero squared overlap with the eigenvector associated with the \textit{second} largest eigenvalue (light blue diamonds), while above $\theta_{\rm crit}$, the squared overlap between the top eigenvector and the signal is non-zero. Both values of the overlap are correctly predicted by the formula in \eqref{eq:RR_ov}.}
    \label{fig:RR_transition}
\end{figure}

Figure~\ref{fig:RR_transition} shows this three-phase behaviour in the random
regular case. The closed-form predictions agree well with direct
diagonalisation of large matrices $\bm A$, averaged over 20 realisations. For
$\theta\leq\theta_{\rm b}$, the top eigenvalue remains the structural noise
eigenvalue $\lambda_{\theta=0}$ (red dots and solid orange line). At
$\theta=\theta_{\rm b}$, given by \eqref{eq:RR_bulk}, the analytical
continuation of the signal-associated eigenvalue in \eqref{eq:RR_eig}
exits the right bulk edge (gray dotted line) and becomes the second largest
eigenvalue (orange crosses). It remains below $\lambda_{\theta=0}$ for
$\theta_{\rm b}<\theta<\theta_{\rm crit}$ and crosses it at
$\theta=\theta_{\rm crit}$, as in the centre panel of
Fig.~\ref{fig:spectral_density}.

\begin{remark}[Bulk transition]
The closed-form expression \eqref{eq:RR_ov} can be analytically continued to
$\theta<\theta_{\rm crit}$, where it predicts the squared overlap between the
signal and the eigenvector associated with the second largest eigenvalue, in
agreement with the simulations (light blue diamonds). Its zero coincides with
$\theta_{\rm b}$. Whether an analogous continuation holds for the Poissonian
case remains an open question.
\end{remark}

\section{Population Dynamics}\label{PopDyn}

In order to solve \crefrange{eq:pi recursive}{eq:q condition}, we appeal to a population dynamics (PD) algorithm \cite{Kuhn2007, Mezard2000, Zdeborov2016} that has been used in various similar settings to solve a number of problems in the field of random matrix theory \cite{Susca2019, Susca2021, Kuhn2017, Kuhn2024, Bartolucci2024}. In this section we will briefly present the algorithm we used in order to solve \crefrange{eq:pi recursive}{eq:q condition}, in the recovery phase $\theta>\theta_\mathrm{crit}$  (otherwise we are in the noise-only scenario already considered in \cite{Budnick2025}). For the purposes of our algorithm, we assume that the spike distribution is zero-centered, i.e. $\langle x\rangle_x$=0.

The joint probability densities $\pi(\omega,h)$ and $\rho(\sigma,\mu)$ are represented as histograms of two ``populations'' of variables, each comprising a set of $N_p$ pairs $\mathcal{P}_1 = \{(\omega_i,h_i)\}_{i=1}^{N_p}$ and $\mathcal{P}_2 = \{(\sigma_i,\mu_i)\}_{i=1}^{N_p}$ respectively. They are initialised by drawing from a random uniform distribution for each variable.  Given a set of inputs $ \theta, \alpha, \ p_s(s), \ p_k(k) \ \text{ (with $p_k(k)$ and $p_s(s)$ satisfying the handshake constraint \eqref{handshake constraint math})}, \ \varrho_x, $ and $ \varrho_W(W)$, the algorithm will output $\langle\lambda_1\rangle_{\bm{A}}$ and $q = \left\langle\frac{\langle \bm{v}_{\mathrm{top}},\bm{x}\rangle}{N}\right\rangle$. We then implement the following steps, suggested by the recursive form of \eqref{eq:pi recursive} and \eqref{eq:rho recursive}:

\begin{enumerate}

    \item Randomly initialise real parameters $\lambda$ and $q$.
    
    \item Generate a random $s\sim\frac{sp_s\left(s\right)}{\langle s\rangle}$.

    \item Draw $s-1$ i.i.d. random variables $W_\ell$ from $\varrho_W(W)$.

    \item Select $s-1$ random pairs $\{(\sigma_\ell,\mu_\ell)\}_{\ell=1}^{s-1}$ from the population $\rho(\sigma,\mu)$, and then compute

    \begin{gather}
        \omega^{(\mathrm{new})} = \lambda_{\theta} -\sum_{\ell=1}^{s-1} \frac{W_\ell^2}{\sigma_\ell}\ \label{popdyn omega new} ,\\
        h^{(\mathrm{new})} = \theta  qx +\sum_{\ell=1}^{s-1} \frac{W_\ell \mu_\ell}{\sigma_\ell}\ \label{popdyn h new},
    \end{gather}

    with $x$ sampled from $\varrho_x$, and replace a randomly selected pair $(\omega_r,h_r)$ with $(\omega^{(\mathrm{new})},h^{(\mathrm{new})})$.

    \item Generate a random $k\sim \frac{kp_k(k)}{\langle k\rangle}$.

    \item Draw $k-1$ i.i.d. random variables $W_\ell$ from $\varrho_W(W)$.
    
    \item Select $k-1$ random pairs $\{(\omega_\ell,h_\ell)\}_{\ell=1}^{k-1}$ from the population $\pi(\omega, h)$, and then compute
    
    \begin{gather}
        \sigma^{(\mathrm{new})} = 1 -\sum_{\ell=1}^{k-1} \frac{W_\ell^2}{\omega_\ell}\ \label{popdyn sigma new},\\
        \mu^{(\mathrm{new})} = \sum_{\ell=1}^{k-1} \frac{W_\ell h_\ell}{\omega_\ell}\ \label{popdyn mu new},
    \end{gather}

    and replace a randomly selected pair $(\sigma_r,\mu_r)$ with $(\sigma^{(\mathrm{new})},\mu^{(\mathrm{new})})$.
    \item After every sweep (an entire replacement of the population members), we monitor the populations' first four moments. If they have plateaued, then the algorithm ends, otherwise return to (ii).
\end{enumerate}
\color{black}
This algorithm replicates and solves the RDEs \eqref{eq:pi recursive} and \eqref{eq:rho recursive}, since the delta constraints in the RDEs precisely impose the recursive relations between population members that the algorithm imposes in \eqref{popdyn omega new}, \eqref{popdyn h new}, \eqref{popdyn sigma new}, and \eqref{popdyn mu new}. Thus, as we gradually impose these delta constraints through the population replacements, the population distributions are expected to converge (for the right set of input parameters) to the distributions defined by \eqref{eq:pi recursive} and \eqref{eq:rho recursive}. Once the two populations' moments have plateaued we have reached equilibrium. We then compute the following two integral constraints using a Monte Carlo sampling method over the stabilised populations:
\begin{equation}\label{eq:PopDyn:lambda condition}
    \alpha_1 = \sum_{s=0}^{s_{\mathrm{max}}} p_s(s) \int \left\{ \mathrm{d}\rho \right\}_s \left\langle \left( \frac{\theta  qx + \left\{ \frac{\mu W}{\sigma} \right\}_s}{\lambda_\theta - \left\{ \frac{W^2}{\sigma} \right\}_s} \right)^2 \right\rangle_{x,\{W\}_s} , 
\end{equation}
and
\begin{equation}\label{eq:PopDyn:q condition}
    \alpha_2 =\sum_{s=0}^{s_{\mathrm{max}}} p_s(s) \int \{ \mathrm{d}\rho \}_s \left\langle \frac{\theta x^2}{\lambda_\theta - \left\{ \frac{W^2}{\sigma} \right\}_s} \right\rangle_{x,\{W\}_s}, 
\end{equation}
which correspond to the two conditions on $\lambda_\theta$ and $q$ given by \eqref{eq:lambda condition} and \eqref{eq:q condition} (provided we are in the recovery phase with $q \neq 0$, and where we have divided \eqref{eq:q condition} by $q$). Given some pre-determined tolerance, if $\alpha_1$ and $\alpha_2$ both equal $1$, then we have obtained the populations $\pi(\omega,h)$ and $\rho(\sigma,\mu)$, and parameters $\lambda_\theta$ and $q$ that solve \crefrange{eq:pi recursive}{eq:q condition}.
If $\alpha_1\neq1$ \emph{or} $\alpha_2\neq1$, then we follow \cite{UrtePaper}, and instead of searching over the full grid $(q,\lambda_\theta)$ we rescale the input parameters as
\begin{align}
    q_{\mathrm{new}}&\rightarrow \frac{1}{\sqrt{\alpha_1}}q_{\mathrm{old}},
    \\
    \lambda_{\theta,\mathrm{new}} &\rightarrow \alpha_2\lambda_{\theta,\mathrm{old}}.
\end{align}
These rescalings are informed by $\alpha_1$ and $\alpha_2$, and are motivated by the RDE forms. To see why $\lambda_{\theta,\mathrm{new}} \rightarrow \alpha_2\lambda_{\theta,\mathrm{old}}$ is an appropriate rescaling, note that through \eqref{popdyn omega new} the $\omega$ terms will be increased by a factor of around $\alpha_2$.  In turn, the denominator of \eqref{eq:PopDyn:lambda condition} will be scaled by a factor similar to $\alpha_2$. This will thus drive the condition towards $1$. Similar motivations are behind the rescaling of $q_{\mathrm{new}}$. With these rescalings, we thus repeat the PD algorithm outlined in steps (i)-(viii) until we have obtained $\alpha_1=1$ and $\alpha_2=1$ within a suitable tolerance threshold. 

\begin{figure}
     \centering
     \includegraphics[width=1\linewidth]{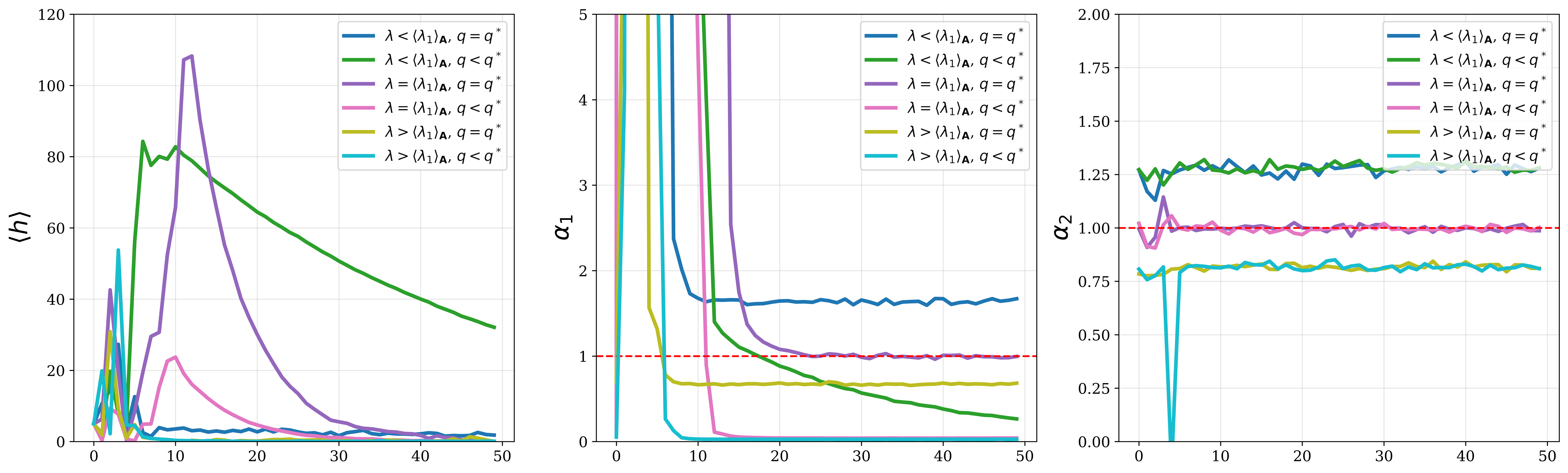}
     \caption{Graphical representation of the PD algorithm outlined in Section \ref{PopDyn}, with parameters $N_p=10^5$, $\alpha=\sqrt{5/4}, \ \varrho_W(W)=\delta_{W,1}$ and $\varrho_x\sim\mathcal{N}(0,1)$, with $p_s(s)$ a Poissonian distribution with mean $\langle s\rangle =4$ and $p_k(k)$ a Poissonian with mean $\langle k \rangle=\alpha^{-2}\langle s\rangle$ as per \eqref{handshake constraint math}, and with respective maximal degrees $s_{\mathrm{max}}=k_{\mathrm{max}}=30$. Population statistics for various different combinations of parameters $\lambda$ and $q$ are plotted. For ease of notation, in the legend we have used the shorthand $q^*=\left\langle\frac{\langle\bm{v}_{\mathrm{top}},\bm{x}\rangle}{N}\right\rangle_{\bm{A}}$ for the numerical value of the average overlap between top eigenvector and signal. (\textit{Left}) The first moment of $h$ as the number of population sweeps ($x$-axis) increases. We see (regardless of the chosen values of parameters $\lambda$ or $q$) that the population averages shrink to $0$. (\textit{Centre}) The value of $\alpha_1$ as given by \eqref{eq:PopDyn:lambda condition} as the number of sweeps increases. We see that it is only when $\lambda = \langle\lambda_1\rangle_{\bm{A}}$ and $q = \left\langle\frac{\langle\bm{v}_{\mathrm{top}},\bm{x}\rangle}{N}\right\rangle_{\bm{A}}$ that $\alpha_1\rightarrow1$ as the population is updated. (\textit{Right}) The value of $\alpha_2$ as given by \eqref{eq:PopDyn:q condition} as the number of sweeps increases. We see that it is only when $\lambda = \langle\lambda_1\rangle_{\bm{A}}$ that $\alpha_2\rightarrow1$ as the population is updated. The added constraint \eqref{eq:PopDyn:lambda condition} selects the correct population.}
     \label{fig:PopDyn Graphics}
 \end{figure}

 It is important to note that in contrast to the population dynamics in \cite{Budnick2025, Susca2019}, the populations' stability is not determined by the parameter $\lambda$. In contrast to previous PD implementations (where the $h$ populations exploded if $\lambda<\langle\lambda_1\rangle_{\bm{A}}$, converged if $\lambda=\langle\lambda_1\rangle_{\bm{A}}$, and shrunk to $0$ if $\lambda>\langle\lambda_1\rangle_{\bm{A}}$), in the recovery phase $\theta>\theta_{\mathrm{crit}}$ we find that for all values of $\lambda>\lambda_{\theta=0}$, the first moment of the $h$ populations shrinks to zero. However, it is \emph{only} when the parameters are set at $\lambda=\langle\lambda_1\rangle_{\bm{A}}$ and $q=\left\langle \frac{\langle \bm{x}, \bm{v}_\mathrm{top} \rangle}{N} \right\rangle_{\bm{A}}$ that the constants given by \eqref{eq:PopDyn:lambda condition} and \eqref{eq:PopDyn:q condition} are equal to $1$. This phenomenon is shown in Figure \ref{fig:PopDyn Graphics}.

\begin{figure}[h!]
    \centering
    \begin{subfigure}{0.49\textwidth}
        \centering
        \includegraphics[width=\linewidth]{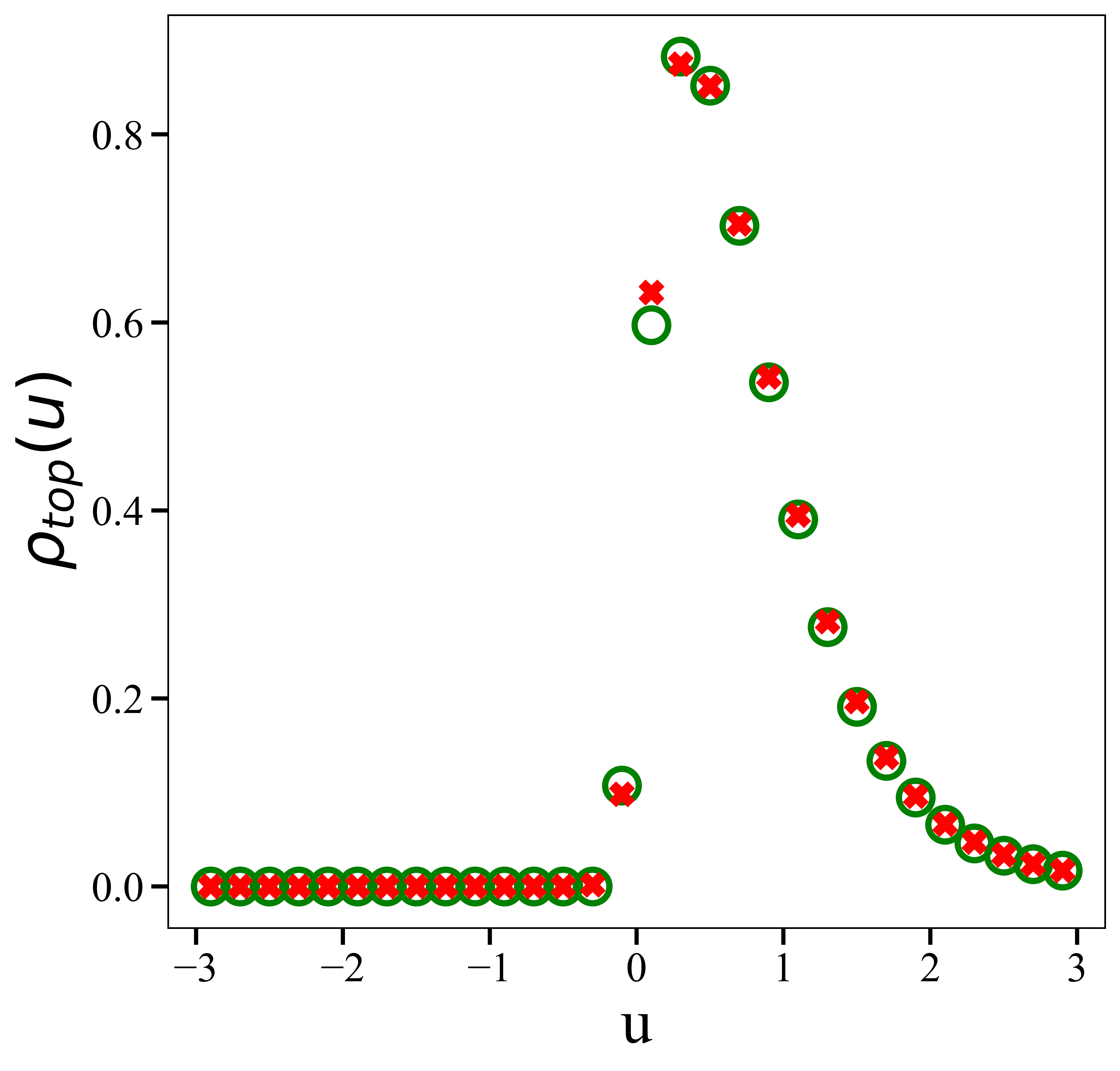}
        \label{fig:eigvec_poisson_16}
    \end{subfigure}
    \hfill
    \begin{subfigure}{0.49\textwidth}
        \centering
        \includegraphics[width=\linewidth]{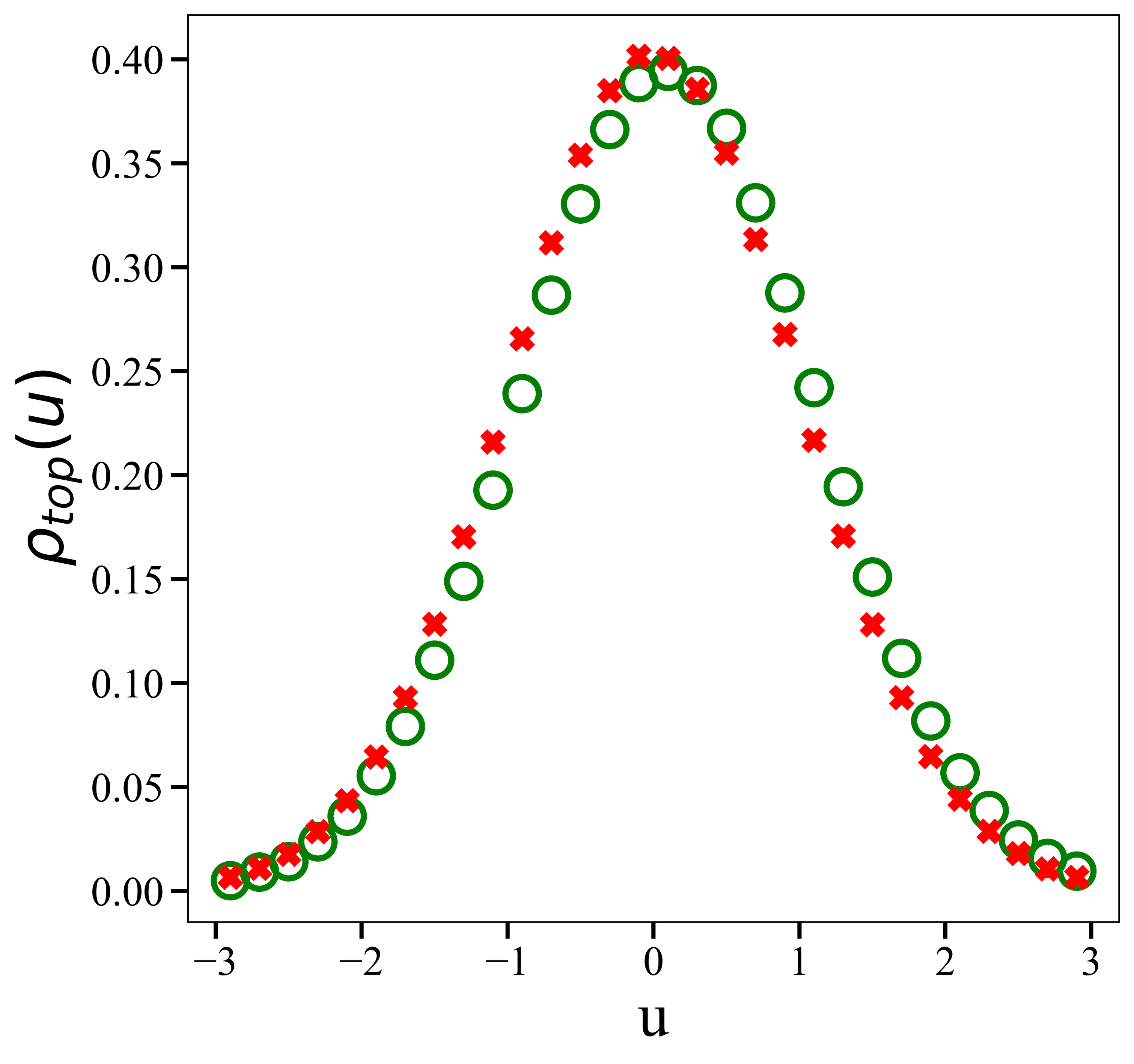}
        \label{fig:eigvec_poisson_30}
    \end{subfigure}
    \caption{Top eigenvector component density of matrices of the form \eqref{def:model} with parameters as in Figure \ref{fig:spectral_density}, comparing numerical diagonalisation (green circles) and the result of the population dynamics using \eqref{eq:eigenvector density} (red crosses). The target error tolerance was set to $0.2$. (\textit{Left}) Non-recovery phase $\theta=16<\theta_\mathrm{crit}$. (\textit{Right}) Recovery phase $\theta=20>\theta_\mathrm{crit}$ showing that the top eigenvector component density closely matches the distribution of the spike components.}
    \label{fig:eigenvector density_Poisson}
\end{figure}

\begin{figure}[h!]
    \centering
    \begin{subfigure}{0.49\textwidth}
        \centering
        \includegraphics[width=\linewidth]{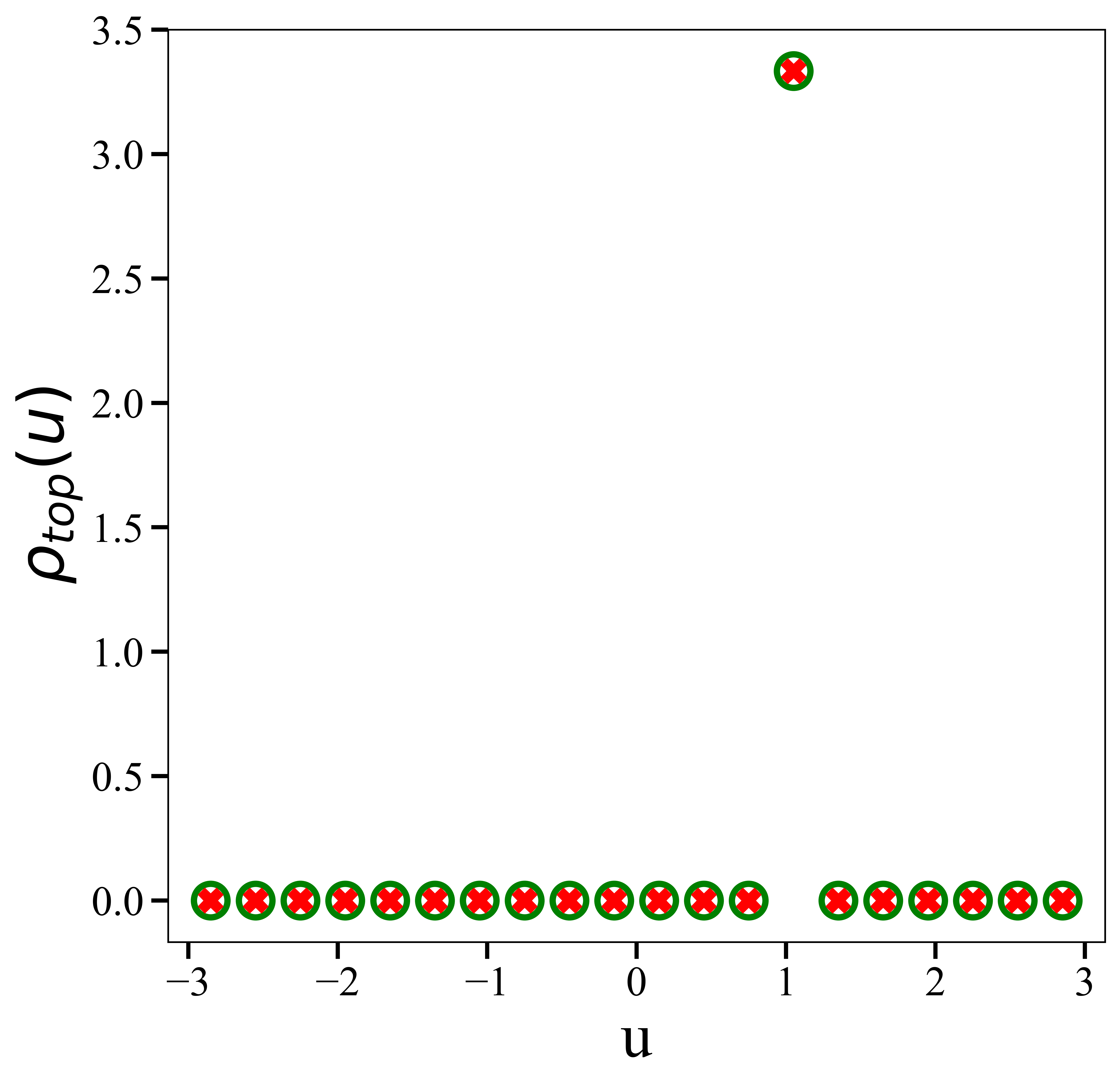}
        \label{fig:eigvec_regular_0}
    \end{subfigure}
    \hfill
    \begin{subfigure}{0.49\textwidth}
        \centering
        \includegraphics[width=\linewidth]{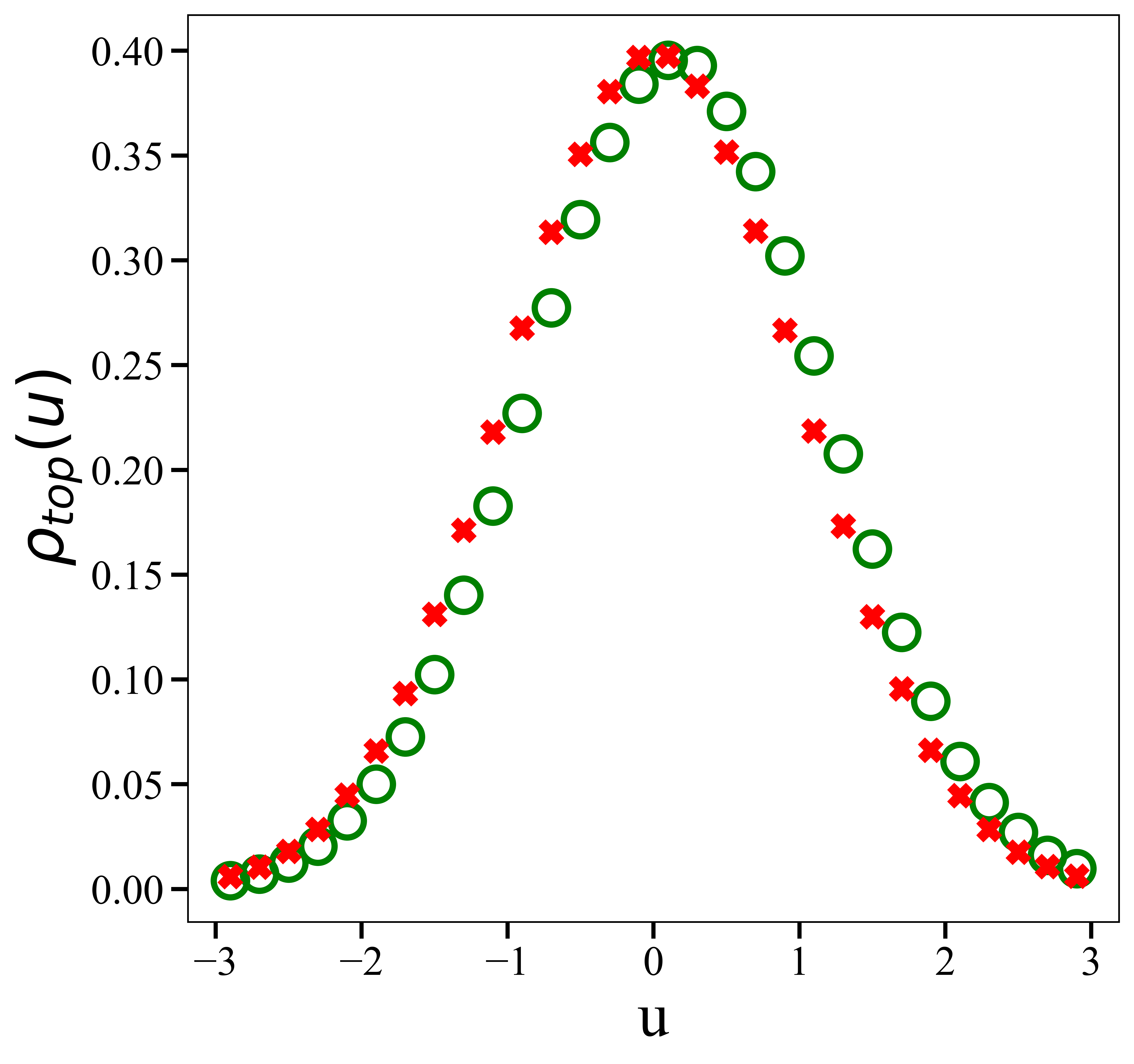}
        \label{fig:eigvec_regular_10}
    \end{subfigure}
    \caption{Equivalent figure to Figure \ref{fig:eigenvector density_Poisson}, when $\bm{X}$ is a $4000\times 2000$ pure adjacency matrix of a random biregular graph with column degree $\langle s\rangle=4$, and row degree $\langle k\rangle=2$. The other parameters are the same as in Figure \ref{fig:spectral_density}. (\textit{Left}) Non-recovery phase $\theta=1<\theta_\mathrm{crit}$. (\textit{Right}) Recovery phase $\theta=10>\theta_\mathrm{crit}$.}
    \label{fig:eigenvector density_Regular}
\end{figure}

With the correct populations $\pi(\omega,h)$ and $\rho(\sigma,\mu)$ generated through this algorithm, we are now able to numerically evaluate the top eigenvector component density from \eqref{eq:eigenvector density}. To do so, we initially choose a resolution for our density, denoted by $\Delta u$, and then slice up the numerical support of $\rho_{\mathrm{top}}(u)$, $[-a,a]$, into bins of size $\Delta u $. We then randomly sample members of the population in order to evaluate the quantity
\begin{equation}
    \frac{\theta  qx + \left\{ \frac{\mu W}{\sigma} \right\}_s}{\lambda - \left\{ \frac{W^2}{\sigma} \right\}_s},
\end{equation}
along with randomly sampled $x$ and $W$ from $\varrho_x$ and $\varrho_W(W)$ respectively. Each time this value fell within a  given bin, a count of one was added. After performing this many times, the bins were normalised which produced our numerical density $\rho_{\mathrm{top}}(u)$. The results of this algorithm are shown in Figures \ref{fig:eigenvector density_Poisson} and \ref{fig:eigenvector density_Regular}, which additionally show the phase transition between $\theta<\theta_\mathrm{crit}$ and $\theta>\theta_\mathrm{crit}$, with the top eigenvector overlapping with the spike distribution $\mathcal{N}(0,1)$ in the right panels. 

\begin{remark}[Perron--Frobenius]
For the non-negative bond weights considered in Figures~\ref{fig:eigenvector density_Poisson} and~\ref{fig:eigenvector density_Regular}, the noise matrix $\bm J=\bm X^\top\bm X$ is entrywise non-negative. The Perron--Frobenius theorem therefore guarantees a non-negative leading eigenvector (strictly positive when $\bm J$ is irreducible). By contrast, the rank-one term $\bm x\bm x^\top$ generally contains entries of both signs for the zero-centered signals considered here, so $\bm A$ need not be entrywise non-negative. Perron--Frobenius then no longer constrains the sign of its leading eigenvector, whose components may consequently take both positive and negative values, as seen in the right panels of Figures~\ref{fig:eigenvector density_Poisson} and~\ref{fig:eigenvector density_Regular}.
\end{remark}

\section{Summary and conclusions} \label{sec:conclusions}
We have used the replica method in order to study the problem of recovering a rank-one signal $\frac{\theta}{N}\bm{x}\bm{x}^\top\in\mathbb{R}^{N\times N}$ from a noisy sparse Wishart-like background $\bm{J}=\bm{X}^\top\bm{X}\in\mathbb{R}^{N\times N}$ where the matrix $\bm{X}\in\mathbb{R}^{M\times N}$ is sparse, and the nonzero entries of $\bm{X}$ have a bond weight drawn from the weight distribution $\varrho_W(W)$. The replica methodology allowed us to study the average largest eigenvalue of the matrix $\bm{A}=\bm{J}+\frac{\theta}{N}\bm{xx}^\top$, as well as the average top eigenvector component density and the average overlap between the signal $\bm{x}$ and the top eigenvector.

We were able to link all of these spectral observables to a system of self-consistent recursive distributional equations given by \crefrange{eq:pi recursive}{eq:q condition}. Notably, the two order parameters $\lambda$ and $q$ introduced during the replica calculations correspond at equilibrium exactly to the average largest eigenvalue, and the average overlap between the top eigenvector and the signal. We were then able to solve the system of self-consistent equations using a population dynamics algorithm. Numerical simulations confirmed excellent agreement between the order parameters that solve the equations, and the observables themselves calculated from direct numerical diagonalisation. 

The values of our spectral observables were found to depend on the signal strength $\theta$, on the average ``sparsity'' of the noise, encoded in $\langle k\rangle$ or $\langle s\rangle$, and on the rectangularity ratio $\alpha$. We find that there is an analogue to the BBP-transition for rank-one signal recovery with sparse Wishart noise. The transition happens around a critical value of the signal strength $\theta_{\mathrm{crit}}$. When we have $\theta<\theta_{\mathrm{crit}}$, then the average largest eigenvalue is that of the noise matrix $\bm{J}$, and the average overlap between the top eigenvector of $\bm{A}$ and the signal $\bm{x}$ is zero. However when the signal strength is larger than this critical value ($\theta>\theta_{\mathrm{crit}}$), then the top eigenvalue changes, and becomes the eigenvalue associated with the signal, and we have non-zero overlap between the top eigenvector and the signal. We are thus in the recovery phase, and we can recover our signal from the noisy background via standard PCA.

We then provided results for the case when the matrix $\bm{X}$ is constructed as the biadjacency matrix of a $(\langle k \rangle, \langle s \rangle)$-biregular bipartite graph with $M$ row nodes with degree $
\langle k \rangle$, and $N$ column nodes with degree $\langle s \rangle$, as well as the adjacency matrix of a truncated Poissonian bipartite graph. We provide an analytical formula for the critical signal strength, as well as for the average largest eigenvalue and the average overlap. These formulae showed excellent agreement with direct numerical diagonalisation. In the case of the random biregular graph, we were also able to analytically extend the formula for the overlap in order to identify the transition from bulk eigenvalue to outlier.

Future work could explore constructing the model with sparse signal plus a sparse noise, as opposed to the dense signal used here. While this may lead to identifiability issues concerning two sparse components \cite{Candes2011}, recent progress has been made concerning finding a BBP analogue for the doubly sparse model of a sparse Wigner matrix plus a sparse signal \cite{Dumitriu2026}. Another natural extension is a signal with non-zero mean, relevant for a global ``market mode'', where the signal overlaps with the Perron eigenvector of $\bm J$ and the threshold analysis above no longer applies. Additionally, the study of the second eigenpair statistics as in \cite{Susca2020b} could be motivated by our exploration of the bulk to outlier transition in the random biregular case in order to formalise the bulk transition formulae for $\lambda_b$ and $\theta_b$ in the non-degenerate case.

\section*{Acknowledgments}

P.V. acknowledges support from UKRI FLF Scheme (No. MR/X023028/1).

\color{black}

\newpage

\appendix

\section{Replica Calculation}\label{replica calculations}
The replica calculation used to derive the results of Section \ref{Main Results} can be split up into six main steps: (1) Dealing with the disorder by averaging the replicated partition function $\mathcal{Z}^n_{\beta}$ (see Eq. \eqref{tauxiliarypartfun}) over the entries of $\bm{X}$ and $\bm{x}$; (2) Introducing functional order parameters in order to reformulate the replicated partition function in terms of a functional integral amenable to a saddle point analysis; (3) Making use of a replica symmetric ansatz that defines the order parameters as a superposition of an uncountably infinite set of Gaussians, allowing for a more natural saddle point evaluation; (4) Evaluating the normalisation constant from the degree distributions of the configuration model; (5) Deriving the saddle point equations in terms of the replica-symmetric Gaussian ansatz; (6) Evaluating the replicated partition function at the saddle point in the limit \(\beta\rightarrow\infty\). 

\subsection{Averaging the replicated partition function over the disorder} \label{sub sec averaging}
Our goal is now to find the average replicated partition function $\mathcal{Z}_{\beta}(u;\bm{A};t;F)$ in Eq. \eqref{tauxiliarypartfun} given by
\begin{equation}\label{eq:Average Partition Function Original}
\big\langle \mathcal{Z}^n_{\beta}\big\rangle_{\bm{A}}
=
\int\Big(\prod_{a=1}^n \mathrm{d}\bm{v}_a\Big)
\Big\langle
\exp\!\Big[ \frac{\beta}{2}\sum_{a=1}^n\langle\bm{v}_a,\bm{A}\bm{v}_a\rangle
+ \beta t N \sum_{a=1}^n F(u,\bm{v}_a,\bm{x})\Big]
\Big\rangle_{\bm{A}}\prod_{a=1}^n \delta\big(|\bm{v}_a|^2-N\big)
\ ,
\end{equation}
and to write it as the integral of an exponential function. 

Recalling the definition of both $\bm{A}$ and $\bm{J}$, we write: 
\begin{align}
\nonumber\frac{\beta}{2}\sum_{a=1}^n\sum_{\substack{i,j=1}}^N v_{ia}A_{ij}v_{ja}&=\frac{\beta}{2}\sum_{a=1}^n\sum_{i,j=1}^N v_{ia}J_{ij}v_{ja}
+\frac{\beta\theta}{2N}\sum_{a=1}^n\sum_{i,j=1}^N v_{ia}(\bm{xx}^\top)_{ij}v_{ja}
\\
\nonumber &=\frac{\beta}{2}\sum_{a=1}^n\sum_{i,j=1}^N\sum_{l=1}^M v_{ia}\,X_{l i}\,X_{l j}\,v_{ja}
+ \frac{\beta\theta}{2N}\sum_{a=1}^n \langle \bm{v}_a,\, \bm{x} \bm{x}^\top \bm{v}_a\rangle
\\
&=\frac{\beta}{2}\sum_{a=1}^n\sum_{l=1}^M\left(\sum_{i=1}^N v_{i a} X_{l i}\right)^2
+ \frac{\beta\theta}{2N}\sum_{a=1}^n \langle \bm{v}_a,\bm{x}\rangle^2\ , 
\end{align}
where we have denoted by $\langle\cdot,\cdot\rangle$ the standard dot product on $\mathbb{R}^N$. We can now use a Hubbard--Stratonovich transformation of the form
\begin{equation}
\int_{-\infty}^{\infty} \mathrm{d}x\; \mathrm{e}^{-a x^{2} + b x}
= \sqrt{\frac{\pi}{a}}\; \mathrm{e}^{\frac{b^{2}}{4a}}
\end{equation}
on the squared $v_{ia}X_{li}$ sum to write

\begin{align}
\exp\!\Big[ \frac{\beta}{2}\sum_{a=1}^n
\langle\bm{v}_a,\bm{A}\bm{v}_a\rangle\Big]
\propto&
\int \prod_{a=1}^n\mathrm{d}\bm{u}_a\,
\exp\!\Bigg[
-\frac{\beta}{2}\sum_{a=1}^n\sum_{l=1}^M u_{la}^2
+\beta\sum_{a=1}^n\sum_{l=1}^M\sum_{i=1}^N
u_{la}v_{ia}X_{li}
\nonumber\\
&\hspace{4cm}
+\frac{\beta\theta}{2N}\sum_{a=1}^n
\langle\bm{v}_a,\bm{x}\rangle^2
\Bigg]\ ,
\end{align}
where we have ignored pre-factors whose logarithm vanishes in the limit.

We can now take the average over $\bm{A}$ of our simplified term
\begin{equation}\label{eq:average intermediary 2}
    \left\langle
\exp\Big[\sum_{a=1}^n\sum_{l=1}^M\sum_{i=1}^N \beta\, u_{la}\, v_{i a}\, X_{l i}\Big]
\;
\exp\Big[\frac{\beta\theta}{2N}\sum_{a=1}^n \langle \bm{v}_a, \bm{x}\rangle^2
+\sum_{a=1}^n \beta t N F(u,\bm{v}_a,\bm{x})\Big]\right\rangle_{\bm{A}}.
\end{equation}
Using the law of total expectation, \(\langle{...\rangle_{\bm{A}}}=\langle\langle...\rangle_{\bm{J}|\bm{x}}\rangle_{\bm{x}}\) \cite{Kuhn2024},  we note that last two terms of \eqref{eq:average intermediary 2} are independent of the noise matrix \(\bm{J}\). Hence the $\bm{J}|\bm{x}$  average simplifies to
\begin{equation}\label{eq:average intermediate 1}
\Big\langle
\exp\Big[\sum_{a=1}^n\sum_{l=1}^M\sum_{i=1}^N \beta\, u_{l a}\, v_{i a}\, X_{l i}\Big]
\Big\rangle_{\bm{J}}
\;
\exp\Big[\frac{\beta\theta}{2N}\sum_{a=1}^n \langle \bm{v}_a, \bm{x}\rangle^2
+\sum_{a=1}^n \beta t N F(u,\bm{v}_a,\bm{x})\Big]\ .
\end{equation}
We recall that our matrix ensemble probability given degree sequences $\bm{ k}$ and $\bm{s}$ is given by \eqref{eq: C given s and k} and \eqref{eq: X given degree sequences}. We can equivalently rewrite \eqref{eq: C given s and k} as:
\begin{align} \label{eq: C given s and k with b}
P(\bm C\mid \bm{k},\bm{s}) =& 
\frac{1}{\mathcal{Z}_{\mathrm{den}}}\left[\prod_{i=1}^{M}\mathbf{1}\!\left(\sum_{j=1}^{N}c_{ij}=k_i\right)\right]\left[
\prod_{j=1}^{N}\mathbf{1}\!\left(\sum_{i=1}^{M}c_{ij}=s_j\right)\right] \mathbf{1}\!\left(\sum_{i=1}^{M}k_i=\sum_{j=1}^{N}s_j\right) \nonumber
\\
&\times \left[\prod_{i=1}^M\prod_{j=1}^N\left(\frac{b}{\sqrt{NM}}\delta_{c_{ij},1}+\left(1-\frac{b}{\sqrt{NM}}\right)\delta_{c_{ij},0}\right)\right].
\end{align}
The insertion of the ``Erd\H{o}s-R\'enyi'' (ER) $b$-dependent term does not change the ensemble we are averaging over, since, as in \cite{UrtePaper}, that term is constant for all matrices in the original ensemble. Nevertheless, averaging over the ER disorder is a convenient choice that leads to a simpler and more compact expression. Obviously, the final result must not depend on the spurious connectivity factor $b$, which is indeed the case as we show below. In \eqref{eq: C given s and k with b}, to avoid more cumbersome notation, we have kept the same symbol $\mathcal{Z}_{\mathrm{den}}$ to now indicate the corresponding $b$-dependent normalisation.

We note that the $\bm{J}$ randomness from the $\langle..\rangle_{\bm{J}}$ average in \eqref{eq:average intermediate 1} arises simply from $\bm{X}$, and thus we can perform this average using \eqref{eq: X given degree sequences} and the version of $P(\bm{C}|\bm k, \bm s)$ given in \eqref{eq: C given s and k with b}: 
\begin{align}
    &\int\mathrm{d}\bm{W}p(\bm{W})\int \mathrm{d} \underline{\hat{k}} \, \mathrm{d} \underline{\hat{s}} \, \mathrm{d} \xi \, \frac{1}{\mathcal{Z}_{\mathrm{den}}} \sum_{\{c_{li}\}}\prod_{l=1}^M\prod_{i=1}^N\left(\frac{b}{\sqrt{NM}}\delta_{c_{li},1}+\left(1-\frac{b}{\sqrt{NM}}\right)\delta_{c_{li},0}\right) \nonumber
    \\
    &\times\exp \left[ \mathrm{i} \sum_{l=1}^M \hat{k}_l \left( \sum_{i=1}^N c_{li} - k_l \right) \right] \exp \left[ \mathrm{i} \sum_{i=1}^N \hat{s}_i \left( \sum_{l=1}^M c_{li} - s_i \right) \right] \nonumber \\
    & \times \exp \left[ \mathrm{i} \, \xi \left( \sum_{l=1}^M k_l - \sum_{i=1}^N s_i \right) \right] \exp \left[ \beta  \sum_{a=1}^n \sum_{l=1}^M \sum_{i=1}^N u_{la} v_{ia} c_{li}W_{li} \right], \label{eq:X average intermediate}
\end{align}
where we have opened up the indicator functions in \eqref{eq: C given s and k with b} using their Fourier representation, and dropped constants whose logarithm vanishes in the limit. We note that the Lagrange multiplier $\xi$ enforces the handshake constraint \eqref{eq: Handshake constraint}, and whose saddle point equations (looking ahead) will reproduce our consistency constraint \eqref{handshake constraint math}.

The average over the ER connectivity \eqref{eq: C given s and k with b} and $\bm W$ can be performed in the standard way (see Appendix D in \cite{Susca2021}) to give

\begin{align}
    &\frac{1}{\mathcal{Z}_{\mathrm{den}}} \int \mathrm{d} \underline{\hat{k}} \, \mathrm{d} \underline{\hat{s}} \, \mathrm{d} \xi \exp \left[ -\mathrm{i} \sum_{l=1}^M (\hat{k}_l - \xi) k_l\right] \exp \left[ -\mathrm{i} \sum_{i=1}^N (\hat{s}_i + \xi) s_i \right] \nonumber \\
    & \times \exp \left[ \frac{b}{\sqrt{NM}} \sum_{l=1}^M \sum_{i=1}^N \left( \left\langle \exp \left[ \beta W \sum_{a=1}^n u_{la} v_{ia}\right]\right\rangle_{{W}} \exp\left[ \mathrm{i} (\hat{k}_l + \hat{s}_i) \right]  - 1 \right) \right]
    \ ,
\end{align}
where $\langle\cdot\rangle_W$ denotes averaging with respect to a single variable $W$ distributed as $\varrho_W$.

Furthermore, we can then use a similar Fourier representation for the normalisation delta functions (with each replicated representation having Lagrange multiplier $\lambda_a$) to write the average replicated partition function as:
\begin{equation}\label{Partition function as ratio of num and den}
    \left\langle \mathcal{Z}^n_{\beta} \right\rangle_{\bm A} \propto \frac{\mathcal{Z}_{\mathrm{num}}}{\mathcal{Z}_{\mathrm{den}}}\ ,
\end{equation}
with
\begin{align} \label{average replicated partition function 1}
    \mathcal{Z}_{\mathrm{num}} &=\int \left( \prod_{a=1}^n \mathrm{d} \bm{v}_a \, \mathrm{d} \bm{u}_a \, \mathrm{d} \lambda_a \right) \exp \left[ -\frac{\beta}{2} \sum_{a=1}^n \sum_{l=1}^M u_{la}^2 + \mathrm{i} \frac{\beta}{2} N \sum_{a=1}^n \lambda_a - \mathrm{i} \frac{\beta}{2} \sum_{a=1}^n\sum_{i=1}^N v_{ia}^2 \lambda_a \right] \nonumber \\
    &\quad \times \left\langle \exp \left[ \frac{\theta}{N} \frac{\beta}{2} \sum_{a=1}^n \langle \bm{v}_a, \bm{x} \rangle^2 + \sum_{a=1}^n \beta t N F(u, \bm{v}_a, \bm{x}) \right] \right\rangle_{\bm{x}} \nonumber \\
    &\quad \times \int \mathrm{d} \underline{\hat{k}} \, \mathrm{d} \underline{\hat{s}} \, \mathrm{d} \xi~\exp \left[ -\mathrm{i} \sum_{l=1}^M (\hat{k}_l - \xi) k_l\right] \exp \left[ -\mathrm{i} \sum_{i=1}^N (\hat{s}_i + \xi) s_i \right] \nonumber \\
    &\quad \times \exp \left[ \frac{b}{\sqrt{NM}} \sum_{l=1}^M \sum_{i=1}^N  \left( \left\langle \exp \left[ \beta W \sum_{a=1}^n u_{la} v_{ia}\right]\right\rangle_{{W}} \exp\left[ \mathrm{i} (\hat{k}_l + \hat{s}_i) \right]  - 1 \right) \right]\ .
\end{align}

\subsection{$\mathcal{Z}_{\mathrm{num}}$: Functional integral representation} \label{subsec functional integral}
Next, we will express $\mathcal{Z}_{\mathrm{num}}$ through a functional integral over the following three order parameters:
\begin{equation} \label{order param 1 unintegrated}
    \phi(\vec{v}, \hat{s}) = \frac{1}{N} \sum_{i=1}^N \delta(\hat{s} - \hat{s}_i) \prod_{a=1}^n \delta(v_a - v_{ia})
\end{equation}
\begin{equation} \label{order param 2 unintegrated}
    \psi(\vec{u}, \hat{k}) = \frac{1}{M} \sum_{l=1}^M \delta(\hat{k} - \hat{k}_l) \prod_{a=1}^n \delta(u_a - u_{la})
\end{equation}
\begin{equation} \label{order param 3}
\varphi(\vec v)=\frac{1}{N}\sum_{i=1}^N\prod_{a=1}^n \delta(v_a - x_iv_{ia})\ ,
\end{equation}
where $\vec v,\vec u\in\mathbb{R}^n$ are $n$-dimensional vectors in replica-space. We will make use of the integrated versions of \eqref{order param 1 unintegrated} and \eqref{order param 2 unintegrated}, which are given by 
\begin{align}
    \phi(\vec{v}) = \int \mathrm{d} \hat{s} \, \mathrm{e}^{\mathrm{i} \hat{s}} \phi(\vec{v}, \hat{s}) &= \frac{1}{N} \int \mathrm{d} \hat{s} \, \mathrm{e}^{\mathrm{i} \hat{s}} \sum_{i=1}^N \delta(\hat{s} - \hat{s}_i) \prod_{a=1}^n \delta(v_a - v_{ia}) \nonumber \\
    &= \frac{1}{N} \sum_{i=1}^N \mathrm{e}^{\mathrm{i} \hat{s}_i} \prod_{a=1}^n \delta(v_a - v_{ia})\ , \label{order param 1}
\end{align}
and similarly for \eqref{order param 2 unintegrated}. We introduce the order parameters into the replicated partition function \eqref{average replicated partition function 1}, by multiplying it by functional integral representations of the identity
\begin{equation}
    1 = \int \mathcal{D}\varphi \mathcal{D}\hat{\varphi} \ N\exp \left[ -\mathrm{i} \int \mathrm{d}\vec{v} \, \hat{\varphi}(\vec{v}) \left( N \varphi(\vec{v}) - \sum_{i=1}^N \prod_{a=1}^n \delta(v_a - x_i v_{ia}) \right) \right]
\end{equation}

\begin{equation}
    1 = \int \mathcal{D}\phi \mathcal{D}\hat{\phi} \ N \exp \left[ -\mathrm{i} \int \mathrm{d}\vec{v} \, \hat{\phi}(\vec{v}) \left( N \phi(\vec{v}) - \sum_{i=1}^N \mathrm{e}^{\mathrm{i} \hat{s}_i} \prod_{a=1}^n \delta(v_a - v_{ia}) \right) \right]
\end{equation}

\begin{equation}
    1 = \int \mathcal{D}\psi \mathcal{D}\hat{\psi} \ M \exp \left[ -\mathrm{i} \int \mathrm{d}\vec{u} \, \hat{\psi}(\vec{u}) \left( M \psi(\vec{u}) - \sum_{l=1}^M \mathrm{e}^{\mathrm{i} \hat{k}_l} \prod_{a=1}^n \delta(u_a - u_{la}) \right) \right]\ .
\end{equation}
Hence, our replicated partition function becomes:
\begin{align}
   \mathcal{Z}_{\mathrm{num}} &= \int \mathcal{D}\phi \mathcal{D}\hat{\phi} \mathcal{D}\psi \mathcal{D}\hat{\psi} \mathcal{D}\varphi \mathcal{D}\hat{\varphi} \, \mathrm{d}\vec{\lambda} \exp \left[ -\mathrm{i} N \int \mathrm{d}\vec{v} \left( \hat{\phi}(\vec{v}) \phi(\vec{v}) + \hat{\varphi}(\vec{v}) \varphi(\vec{v}) \right) - \mathrm{i} M \int \mathrm{d}\vec{u} \, \hat{\psi}(\vec{u}) \psi(\vec{u}) \right] \nonumber \\
    &\quad \times \exp \left[ \mathrm{i} \frac{\beta}{2} N \sum_{a=1}^n \lambda_a \right] \nonumber \\
    &\quad \times \int \mathrm{d}\underline{\hat{k}} \, \mathrm{d}\underline{\hat{s}} \, \mathrm{d}\xi \exp \left[ -\mathrm{i} \sum_{l=1}^M (k_l \hat{k}_l - k_l \xi) - \mathrm{i} \sum_{i=1}^N (s_i \hat{s}_i + s_i \xi) \right] \nonumber \\
    &\quad \times \int \left( \prod_{a=1}^n \mathrm{d}\bm{v}_a \right) \exp \left[ -\mathrm{i} \frac{\beta}{2} \sum_{i=1}^N \sum_{a=1}^n v_{ia}^2 \lambda_a + \mathrm{i} \int \mathrm{d}\vec{v} \, \hat{\phi}(\vec{v}) \sum_{i=1}^N \mathrm{e}^{\mathrm{i} \hat{s}_i} \prod_{a=1}^n \delta(v_a - v_{ia}) \right] \nonumber \\
    &\quad \times \int \left( \prod_{a=1}^n \mathrm{d}\bm{u}_a \right) \exp \left[ -\frac{\beta}{2} \sum_{a=1}^n \sum_{l=1}^M u_{la}^2 + \mathrm{i} \int \mathrm{d}\vec{u} \, \hat{\psi}(\vec{u}) \sum_{l=1}^M \mathrm{e}^{\mathrm{i} \hat{k}_l} \prod_{a=1}^n \delta(u_a - u_{la}) \right] \nonumber \\
    &\quad \times \left\langle \exp \left[ \frac{\theta}{N} \frac{\beta}{2} \sum_{a=1}^n \langle \bm{v}_a, \bm{x} \rangle^2 + \sum_{a=1}^n \beta t N F(u, \bm{v}_a, \bm{x}) + \mathrm{i} \int \mathrm{d}\vec{v} \, \hat{\varphi}(\vec{v}) \sum_{i=1}^N \prod_{a=1}^n \delta(v_a - x_i v_{ia}) \right] \right\rangle_{\bm{x}} \nonumber \\
    &\quad \times \exp \left[ \frac{b}{\sqrt{NM}} \sum_{l=1}^M \sum_{i=1}^N \left( \left\langle \exp \left[ \beta W \sum_{a=1}^n u_{la} v_{ia} \right] \right\rangle_W \exp[\mathrm{i} \hat{k}_l] \exp[\mathrm{i} \hat{s}_i] - 1 \right) \right]\ .
\end{align}
We note that we can use the definition of the integrated order parameters to write:
\begin{align}
    &\frac{b}{\sqrt{NM}}\sum_{l=1}^M \sum_{i=1}^N \left( \left\langle \exp \left[ \beta W \sum_{a=1}^n u_{la} v_{ia} \right] \right\rangle_W \exp[\mathrm{i} \hat{k}_l] \exp[\mathrm{i} \hat{s}_i] - 1 \right) \\
    &= b\sqrt{NM} \left( \left(\int \mathrm{d}\vec{v} \mathrm{d}\vec{u} \, \phi(\vec{v}) \psi(\vec{u}) \left\langle \exp \left[ \beta W \sum_{a=1}^n u_a v_a \right] \right\rangle_W\right) - 1 \right)\ .
\end{align}
Similarly, we can also use them to rewrite the inner product between $\bm{v}_a$ and the signal $\bm{x}$ as follows
\begin{align}
\sum_{a=1}^n \langle \bm{v}_a,\bm{x}\rangle^2= N^2\int\mathrm{d}\vec v\,\mathrm{d}\vec w\ \Big(\sum_{a=1}^n v_a w_a\Big)\,\varphi(\vec v)\,\varphi(\vec w)\ .
\end{align}

Furthermore, we can kill the $\mathrm{d}\vec u$ integrals with the corresponding delta functions, and then recognise that the remaining $\mathrm{d} \bm{u}_a$ and $\mathrm{d}\underline{\hat{k}}$ integrals factorise over the $M$ sites as
\begin{equation}
    \prod_{l=1}^M \int \mathrm{d}\hat{k} \, \mathrm{d}\vec{u} \exp \left[ -\frac{\beta}{2} \sum_{a=1}^n u_{a}^2 + \mathrm{i} \hat{\psi}(\vec{u}) \exp[\mathrm{i} \hat{k}] - \mathrm{i} k_l \hat{k} \right]\ .
\end{equation}
We rewrite this in exponential form
\begin{align}
    \exp \left[ \sum_{l=1}^M \log \left( \int \mathrm{d}\vec{u} \exp \left[ -\frac{\beta}{2} \sum_{a=1}^n u_{a}^2 \right] \int \mathrm{d}\hat{k} \exp \left[ \mathrm{i} \hat{\psi}(\vec{u}) \exp[\mathrm{i} \hat{k}] - \mathrm{i} k_l \hat{k} \right] \right) \right]\ .\label{ExpformSingleSiteIntegral}
\end{align}
Now consider
\begin{equation}
    I[k_l, \vec{u}] = \int \mathrm{d}\hat{k} \exp \left[ \mathrm{i} \hat{\psi}(\vec{u}) \exp[\mathrm{i} \hat{k}] - \mathrm{i} k_l \hat{k} \right]\ .
\end{equation}
Using a Taylor expansion, we can express this as:
\begin{align}
    I[k_l, \vec{u}] = \sum_{r=0}^\infty \frac{(\mathrm{i} \hat{\psi}(\vec{u}))^r}{r!} \int \mathrm{d}\hat{k} \exp \left[ \mathrm{i} \hat{k} (r - k_l) \right]
    =\frac{(\mathrm{i} \hat{\psi}(\vec{u}))^{k_l}}{k_l!}\ ,
\end{align}
where in the last line we identified the Fourier representation of the Kronecker delta. Hence, our single site integral \eqref{ExpformSingleSiteIntegral} becomes:
\begin{equation}
 \exp \left[ \sum_{l=1}^M \log \left( \int \mathrm{d}\vec{u} \exp \left[ -\frac{\beta}{2} \sum_{a=1}^n u_{a}^2 \right] \frac{(\mathrm{i} \hat{\psi}(\vec{u}))^{k_l}}{k_l!} \right) \right]\ .
\end{equation}
By the assumed self-averaging of the empirical row-degree distribution, in the thermodynamic limit $M\to\infty$ we have
\begin{equation}
    \frac{1}{M} \sum_{l=1}^M \log(f(k_l)) \simeq \sum_{k=0}^{k_{\mathrm{max}}} p_k(k) \log (f(k))\ ,
\end{equation}
to rewrite our single site integral \eqref{ExpformSingleSiteIntegral} as
\begin{equation}
   \exp \left\{ M \sum_{k=0}^{k_{\mathrm{max}}} p_k(k) \left[ \log \left( \int \mathrm{d}\vec{u} \exp \left[ -\frac{\beta}{2} \sum_{a=1}^n u_a^2 \right] (\mathrm{i} \hat{\psi}(\vec{u}))^k \right) - \log(k!) \right] \right\}\ .
\end{equation}
An identical calculation gives us the following expression for the remaining $\mathrm{d}\bm v_a$ and $\mathrm{d}{\underline{\hat s}}$ integrals assuming for $F$ the form $F(u,\bm v_a,\bm x)=\frac{1}{N}\sum_{i=1}^N f(u, v_{ia},x_i)$

\begin{align}
 \nonumber &\exp \left\{ N \sum_{s=0}^{s_{\mathrm{max}}} p_s(s) \left[ \log \left(  \int \mathrm{d}\vec{v}\left\langle \exp \left[ -\mathrm{i} \frac{\beta}{2} \sum_{a=1}^n v_a^2 \lambda_a + \sum_{a=1}^n \beta t  f(u, v_a, x) + \mathrm{i} \hat{\varphi}(x\vec{v}) \right] \right\rangle_x\times\right.\right.\right.\\
 &\times\left.\left.\left.(\mathrm{i} \hat{\phi}(\vec{v}))^s \right) - \log(s!) \right] \right\}\ ,
\end{align}
where $\langle...\rangle_x$ stands for averaging over $\varrho_x$.
Similarly, by self-averaging of the empirical degree distributions,
\begin{equation}
    \sum_{i=1}^N s_i = N \sum_{s=0}^{s_{\mathrm{max}}} p_s(s) s \quad , \quad \sum_{l=1}^M k_l = M \sum_{k=0}^{k_{\mathrm{max}}} p_k(k) k \quad.
\end{equation}
Eventually we can write our average replicated partition function in a form amenable to a saddle point evaluation in the $N,M\rightarrow\infty$ limit
\begin{equation}
    \mathcal{Z}_{\mathrm{num}} \simeq \int \mathcal{D}\phi \mathcal{D}\hat{\phi} \mathcal{D}\varphi \mathcal{D}\hat{\varphi} \mathcal{D}\psi \mathcal{D}\hat{\psi} \, \mathrm{d}\vec{\lambda} \, \mathrm{d}\xi \exp \left[ \sqrt{NM} \, S\left[ \phi, \hat{\phi}, \varphi, \hat{\varphi}, \psi, \hat{\psi}, \vec{\lambda}, \xi \right] \right]\ , \label{eq:pre ansatz action VERY general}
\end{equation}
where
\begin{align}
    S\left[ \phi, \hat{\phi}, \varphi, \hat{\varphi}, \psi, \hat{\psi}, \vec{\lambda}, \xi \right] =&  S_1[\phi, \hat{\phi}] + S_2[\varphi, \hat{\varphi}] + S_3[\psi, \hat{\psi}] + S_4[\vec{\lambda}] + S_5[\phi, \psi] \nonumber
    \\
    &+ S_6[\varphi] + S_7[\hat{\psi}] + S_8[\hat{\varphi}, \hat{\phi}, \vec{\lambda}] + S_9[\xi]\label{oldactionSassumofnine}
\end{align}
with:
\begin{align}
    S_1[\phi, \hat{\phi}] &= -\mathrm{i} \frac{1}{\alpha} \int \mathrm{d}\vec{v} \, \hat{\phi}(\vec{v}) \phi(\vec{v}) \label{S1 field}\\
    S_2[\varphi, \hat{\varphi}] &= -\mathrm{i} \frac{1}{\alpha} \int \mathrm{d}\vec{v} \, \hat{\varphi}(\vec{v}) \varphi(\vec{v}) \\
    S_3[\psi, \hat{\psi}] &= -\mathrm{i} \alpha \int \mathrm{d}\vec{u} \, \hat{\psi}(\vec{u}) \psi(\vec{u}) \\
    S_4[\vec{\lambda}] &= \mathrm{i} \frac{\beta}{2\alpha} \sum_{a=1}^n \lambda_a \\
    S_5[\phi, \psi] &= b  \left(\left(\int\mathrm{d}\vec{v} \, \mathrm{d}\vec{u} \, \phi(\vec{v}) \psi(\vec{u}) \left\langle \exp \left[ \beta W \sum_{a=1}^n u_a v_a \right] \right\rangle_W\right) - 1 \right)\label{S5 field} \\
    S_6[\varphi] &= \frac{\theta\beta}{2\alpha} \int \mathrm{d}\vec{v} \, \mathrm{d}\vec{w} \, \varphi(\vec{v}) \varphi(\vec{w}) \sum_{a=1}^n v_a w_a \\
    S_7[\hat{\psi}] &= \alpha \sum_{k=0}^{k_{\mathrm{max}}} p_k(k) \left[ \log \left( \int \mathrm{d}\vec{u} \exp \left[ -\frac{\beta}{2} \sum_{a=1}^n u_a^2 \right] (\mathrm{i} \hat{\psi}(\vec{u}))^k \right) - \log(k!) \right] \\
    S_8[\hat{\varphi}, \hat{\phi}, \vec{\lambda}] &= \frac{1}{\alpha} \sum_{s=0}^{s_{\mathrm{max}}} p_s(s) \Biggl[ \log\biggl(  \int \mathrm{d}\vec{v} ~\Bigl\langle\exp \Bigl[ -\mathrm{i} \frac{\beta}{2} \sum_{a=1}^n v_a^2 \lambda_a + \sum_{a=1}^n \beta t f(u, v_a, x) \nonumber \\
    & \qquad \qquad \qquad + \mathrm{i} \hat{\varphi}(x\vec{v}) \Bigr] \Bigr\rangle_x (\mathrm{i} \hat{\phi}(\vec{v}))^s \biggr) - \log(s!) \Biggr] \label{S8 field}
    \\
    S_9[\xi] &= \mathrm{i} \xi \left( \alpha \sum_{k=0}^{k_{\mathrm{max}}} p_k(k) k - \frac{1}{\alpha} \sum_{s=0}^{s_{\mathrm{max}}} p_s(s) s \right)\ . \label{S9 Field}
\end{align}
We first write the saddle point equations for the fields $\varphi$ and $\hat\varphi$, and multipliers $\vec\lambda$ and $\xi$. Before deriving the saddle point equations for the remaining fields, we will perform first a functional change of variables in \ref{sec:replica sym ansatz}, and then derive their saddle point equations in terms of the new functional variables.  

By taking the partial derivative of the action w.r.t. $\varphi, \hat\varphi, \vec\lambda, \mathrm{ and } \ \xi$ and setting it to zero, we obtain, respectively, the following saddle point equations:
\begin{align}
  & 1 = \sum_{s=0}^{s_{\mathrm{max}}} p_s(s) \frac{\int \mathrm{d}\vec{v} \left\langle \exp\left[ -\mathrm{i}\frac{\beta}{2} \sum_{a=1}^n v_a^2 \lambda_a + \sum_{a=1}^n \beta t f(u, v_a, x) + \mathrm{i}\hat{\varphi}^*(x\vec{v}) \right] \right\rangle_x v_a^2 \, (\mathrm{i}\hat{\phi}(\vec{v}))^s}{\int \mathrm{d}\vec{v}' \left\langle \exp\left[ -\mathrm{i}\frac{\beta}{2} \sum_{a=1}^n {v'_a}^2 \lambda_a + \sum_{a=1}^n \beta t f(u, v'_a, x)  + \mathrm{i}\hat{\varphi}^*(x\vec{v}') \right] \right\rangle_x (\mathrm{i}\hat{\phi}(\vec{v}'))^s}\ , \label{lambda saddle point 1}
   \\
   &\mathrm{i} \hat{\varphi}^*(\vec{v}) = \theta \beta \int \mathrm{d}\vec{w} \, \langle \vec{v}, \vec{w} \rangle \varphi^*(\vec{w}) \label{varphi saddle point}\ ,
    \\
    & \varphi^*(\vec{v}) = \sum_{s=0}^{s_{\mathrm{max}}} p_s(s) \frac{\left\langle |x|^{-n} \exp\left[ -\mathrm{i}\frac{\beta}{2} \sum_{a=1}^n \frac{v_a^2}{x^2} \lambda_a + \sum_{a=1}^n \beta t  f\!\left(u, \frac{v_a}{x}, x\right) + \mathrm{i}\hat{\varphi}^*(\vec{v}) \right] \big(\mathrm{i}\hat{\phi}(\vec{v}/x)\big)^{s} \right\rangle_x}{\int \mathrm{d}\vec{v}' \left\langle \exp\left[ -\mathrm{i}\frac{\beta}{2} \sum_{a=1}^n {v'_a}^2 \lambda_a + \sum_{a=1}^n \beta t f(u, v'_a, x) + \mathrm{i}\hat{\varphi}^*(x\vec{v}') \right] \right\rangle_x (\mathrm{i}\hat{\phi}(\vec{v}'))^s}\ , \label{q saddle point initial}
    \\
    &\langle k \rangle = \frac{1}{\alpha^2} \langle s \rangle \label{xi saddle point}\ ,
\end{align}
where $^*$ denotes the field at the saddle point. We note that the saddle point equation obtained by differentiation w.r.t. $\xi$ correctly reproduces \eqref{handshake constraint math}. Using the linearity of the inner product, it follows from \eqref{varphi saddle point} that $\mathrm{i}\,\hat\varphi^*(\vec v)$ is a linear function of its arguments. Hence it can be written in terms of a $t$ dependent $n$-dimensional vector $\vec q(t)=(q_a(t))$ for $a=1,...,n$

\begin{equation}
\mathrm{i}\,\hat\varphi^*(\vec v)=\theta\beta\sum_{a=1}^nq_a(t)v_a, \qquad \vec q(t)=\int \mathrm{d}\vec w\;\vec w\;\varphi^*(\vec w)\ . \label{q definition initial}
\end{equation}

\subsection{$\mathcal{Z}_{\mathrm{num}}$: Replica-symmetric functional change of variables}\label{sec:replica sym ansatz}

Now, we first adopt a replica symmetric (RS) functional change of variables in the integral \eqref{eq:pre ansatz action VERY general} for the remaining fields $\phi,\hat\phi,\psi$ and $\hat\psi$, before taking a functional derivative of the action $S$ w.r.t. the new fields.

We follow the methodology first introduced in \cite{Kuhn2007} and later employed in \cite{Susca2019, Budnick2025, UrtePaper, Susca2021}: this functional change of variables assumes that the dependence of the fields on the vector components $\vec v$ and $\vec u$ is only through a permutation-symmetric function of the vector components--namely the fields are represented as superposition of uncountably infinite Gaussians with random non-zero means and variances. For instance, we would make the change of variable $\phi\to\pi$
\begin{equation}
\phi(\vec{v}) = r \int \mathrm{d}\pi \prod_{a=1}^n \frac{1}{\mathcal{Z}(\omega, h)} \exp\left[ -\frac{\beta}{2}\omega v_a^2 + \beta h v_a \right],\label{phipichangevar}
\end{equation}
where we have introduced $\pi(\omega,h)$ as the joint probability density of the fluctuating variance and mean ($\int\mathrm{d}\pi=1$), and its associated measure $\mathrm{d}\pi=\mathrm{d}\omega\,\mathrm{d}h\,\pi(\omega,h)$. Furthermore we have also introduced
\begin{equation}\label{Z definition}
    \mathcal{Z}(x,y) = \sqrt{\frac{2\pi}{\beta x}}\exp\left[\frac{\beta y^2}{2x}\right]=\int\mathrm{d}v\exp\left[-\frac{\beta}{2}xv^2+\beta y v\right]
\end{equation}
as the Gaussian normalising factor. The prefactor $r$ is introduced to account for the fact that the field $\phi$ need not be normalised. The representation \eqref{phipichangevar} treats the $\vec{v}$-Gaussians as functional basis vectors on which the field $\phi$ can be developed, and it has the advantage that every $v$-integral in the action where the field $\phi$ appears effectively turns into an easily solvable Gaussian integral. The resulting saddle point equations in terms of the density $\pi$ (and the corresponding densities for the other fields) further acquire a particularly appealing RDE form, as detailed below.

We implement our RS functional change of variables for our remaining fields by multiplying \eqref{eq:pre ansatz action VERY general} by the following representation of the identity
\begin{align}
1=&\int  \mathcal{D}\pi \;J_\phi[\pi] \delta\left(\phi(\vec{v}) - r \int \mathrm{d}\pi \prod_{a=1}^n \frac{1}{\mathcal{Z}(\omega, h)} \exp\left[ -\frac{\beta}{2}\omega v_a^2 + \beta h v_a \right]\right) \delta\left(\int \mathrm{d}\pi - 1\right)\ , \label{app:eqfirst_A} 
\end{align}
where $J_\phi[\pi]=|\det\frac{\delta\phi}{\delta\pi}|$ is the $N$-independent Jacobian that connects the two fields.

Similar changes of variables are adopted for the other fields:
\begin{align}
1=&\int  \mathcal{D}\hat{\pi} \;J_{\hat\phi}[\hat\pi] \delta\left(\mathrm{i}\hat{\phi}(\vec{v}) - \hat{r} \int \mathrm{d}\hat{\pi} \prod_{a=1}^n \exp\left[ \frac{\beta}{2}\hat{\omega} v_a^2 + \beta \hat{h} v_a \right]\right) \delta\left(\int \mathrm{d}\hat{\pi} - 1\right)\ , \\
1=&\int \mathcal{D}\rho \;J_{\psi}[\rho] \delta\left(\psi(\vec{u}) - \tau \int \mathrm{d}\rho \prod_{a=1}^n \frac{1}{\mathcal{Z}(\sigma, \mu)} \exp\left[ -\frac{\beta}{2}\sigma u_a^2 + \beta \mu u_a \right]\right) \delta\left(\int \mathrm{d}\rho - 1\right)\ , \\
1=&\int  \mathcal{D}\hat{\rho} \; J_{\hat\psi}[\hat\rho] \delta\left(\mathrm{i}\hat{\psi}(\vec{u}) - \hat{\tau} \int \mathrm{d}\hat{\rho} \prod_{a=1}^n \exp\left[ \frac{\beta}{2}\hat{\sigma} u_a^2 + \beta \hat{\mu} u_a \right]\right) \delta\left(\int \mathrm{d}\hat{\rho} - 1\right)\ . \label{app:eqlast_A}
\end{align}

We have further introduced normalised densities $\hat{\pi}(\hat{\omega},\hat{h})$, $\rho(\sigma,\mu)$, $\hat{\rho}(\hat{\sigma},\hat{\mu})$ and their respective measures. We can then open up the delta functions using their Fourier (and functional) integral representations in order to bring them into the exponential and deform the action subject to these additional constraints. 
Taking the field $\phi$ as representative example to show how the functional integral \eqref{eq:pre ansatz action VERY general} gets modified, we have
\begin{align}
\mathcal{Z}_{\mathrm{num}}=\int &\mathcal{D}\phi \mathcal{D}{\pi} \mathcal{D}\Lambda_\phi \mathrm{d}\gamma_\phi J_\phi[\pi]\dots \nonumber \\
&\exp\Biggl( \sqrt{N M}\biggl( S\left[\phi, \dots\right] - \int \mathrm{d}\vec{v} \; \Lambda_\phi(\vec{v}) \biggl[ \phi(\vec{v}) - r \int \mathrm{d}\pi \prod_{a=1}^n \frac{1}{\mathcal{Z}(\omega, h)} \exp\left[ -\frac{\beta}{2}\omega v_a^2 + \beta h v_a \right] \biggr] \nonumber \\
&\qquad \ \ - n\gamma_\phi \left[ \int \mathrm{d}\pi - 1 \right] - \dots \biggr) \Biggr) \nonumber \\
= \int &\mathcal{D}\phi \mathcal{D}{\pi} \mathcal{D}\Lambda_\phi \mathrm{d}\gamma_\phi J_\phi[\pi]\dots \exp\left[ \sqrt{N M} \; \mathcal{A}\left[\phi, \Lambda_\phi, \gamma_\phi, \pi, \dots\right] \right]\ ,
\end{align}
where the new action $\mathcal{A}$ is an additive modification of the old action $S$ in \eqref{oldactionSassumofnine} where two multipliers per field ($\Lambda_\phi$ and $\gamma_\phi$) are added to enforce the functional change of variables, and normalisation of the joint probability density function $\pi$. 

Finally, we also make the following Replica-Symmetric Ansatz concerning the multipliers $\lambda_a$ and $\vec{q}(t)$:
\begin{align}
    \mathrm{i} \lambda_a &= \lambda(t) \quad , \quad \forall a=1,\dots,n \ , \label{lambda ansatz}\\
    \vec{q}(t) &= \mathbf{1}_n \, q(t)\ . \label{q ansatz}
\end{align}
We note that this is not a replica-symmetric functional change of variables, but is instead a pure ansatz that the Lagrange multipliers be replica-symmetric. 

We will now write the stationarity conditions of the new action $\mathcal{A}$ with respect to the six new replica-symmetric fields/multipliers, still using $\phi$ as representative example (the treatment for all other fields is completely analogous).

First, taking $\frac{\delta \mathcal{A}}{\delta \Lambda_\phi}$ and setting it to zero correctly reproduces the functional change of variables
\begin{equation}
\phi(\vec{v}) = r \int \mathrm{d}\pi \prod_{a=1}^n \frac{1}{\mathcal{Z}(\omega,h)} \exp\left[-\frac{\beta}{2}\omega v_a^2 + \beta h v_a\right]\ .\label{phi RS Ansatz saddle point}
\end{equation}
The stationarity constraint $\frac{\partial \mathcal{A}}{\partial \gamma_\phi}=0$ reproduces the $\pi$ density normalisation
\begin{equation}
\int \mathrm{d}\pi = 1\ . \label{pi normalisation saddle point}
\end{equation}
Next, we obtain
\begin{equation}
\frac{\delta \mathcal{A}}{\delta \phi} = \frac{\delta S}{\delta \phi} - \Lambda_\phi=0\Rightarrow \Lambda_\phi = \frac{\delta S}{\delta \phi}\ .\label{Big Lambda stationarity constraint}
\end{equation}
Finally the stationarity constraint $\frac{\delta \mathcal{A}}{\delta \pi}=0$ yields:
\begin{align}
\nonumber n\gamma_\phi =& r\int \mathrm{d}\vec{v} \; \Lambda_\phi(\vec{v}) \prod_{a=1}^n \frac{1}{\mathcal{Z}(\omega,h)} \exp\left[-\frac{\beta}{2}\omega v_a^2 + \beta h v_a\right] \label{gamma_phi saddle point}\\
\nonumber =& r\int \mathrm{d}\vec{v} \left( \mathrm{i}\hat{\phi}(\vec{v}) - b\alpha \int \mathrm{d}\vec{u} \; \psi(\vec{u}) \left\langle \exp\left[\beta W\sum_{a=1}^n v_a u_a\right] \right\rangle_W \right)\times\\
&\times\prod_{a=1}^n \frac{1}{\mathcal{Z}(\omega,h)} \exp\left[-\frac{\beta}{2}\omega v_a^2 + \beta h v_a\right],
\end{align}
where in the last line we used \eqref{Big Lambda stationarity constraint} and performed the partial derivatives of \eqref{S1 field} and \eqref{S5 field} w.r.t. $\phi(\vec v)$. Analogous calculations related to the other three RS fields ($\hat\phi, \psi$ and $\hat\psi$) will give the saddle point equations in the exact same form as \eqref{phi RS Ansatz saddle point}, \eqref{pi normalisation saddle point}, \eqref{Big Lambda stationarity constraint}, and \eqref{gamma_phi saddle point} but with their respective densities and Lagrange parameters.

We can now use \eqref{phi RS Ansatz saddle point}, and the respective other versions for $\hat\phi, \psi$ and $\hat\psi$, as well as \eqref{lambda ansatz} and \eqref{q ansatz} to write the action $S$ (and thus our replicated partition function at the saddle point) in terms of the densities $\pi(\omega,h)$, $\hat\pi(\hat\omega,\hat h)$, $\rho(\sigma,\mu)$, and $\hat\rho(\hat\sigma,\hat\mu)$, as well as the parameters $\lambda$ and $q$. For instance, the contribution $S_1$ to the action in \eqref{S1 field} can be written as follows
\begin{align}
-\frac{1}{\alpha}\int \mathrm{d}\vec{v}\;{\phi}(\vec{v})(\mathrm{i}\hat{\phi}(\vec{v}))=&
-\frac{r\hat{r}}{\alpha}\int \mathrm{d}\pi\,\mathrm{d}\hat{\pi}\int \mathrm{d}\vec{v}\;\prod_{a=1}^{n}\frac{1}{\mathcal{Z}(\omega,h)}
\exp\!\Big[-\frac{\beta}{2}(\omega-\hat{\omega})\,v_{a}^{2}+\beta\,(h+\hat{h})\,v_{a}\Big]
\nonumber \\=&
-\frac{r\hat{r}}{\alpha}\int \mathrm{d}\pi\,\mathrm{d}\hat{\pi}\;
\left[\frac{\mathcal{Z}(\omega-\hat{\omega},\,h+\hat{h})}{\mathcal{Z}(\omega,h)}\right]^{n} 
\nonumber\\
=& -\frac{r\hat{r}}{\alpha} - n\frac{r\hat{r}}{\alpha}
\int \mathrm{d}\pi\,\mathrm{d}\hat{\pi}\;\log\!\left[\frac{\mathcal{Z}(\omega-\hat{\omega},\,h+\hat{h})}{\mathcal{Z}(\omega,h)}\right] + o(n)\ ,
\end{align}
where in the last step we used that for \(n\ll1\), \(x^{n}\approx 1 + n\log x\). The same calculation allows us to express $S_3[\psi,\hat{\psi}]$ in terms of $\rho$ and $\hat{\rho}$ as
\begin{equation}
    S_3[\rho,\hat{\rho}] = -\alpha\hat{\tau}\tau\,- n\alpha\hat{\tau}\tau\, \int \mathrm{d}\rho \mathrm{d}\hat{\rho} \log\left[ \frac{\mathcal{Z}(\sigma-\hat{\sigma}, \mu+\hat{\mu})}{\mathcal{Z}(\sigma,\mu)} \right]\ .\label{S3 with ansatz}
\end{equation}
Similar manipulations of $S_5, S_7$ and $S_8$ yield the following expressions in terms of the replica symmetric fields, where we keep terms up to $o(n)$
\begin{align}
    S_{5}[{\pi},\rho]=& b(r\tau-1)+nr\tau b\int \mathrm{d}{\pi}\,\mathrm{d}\rho\;\left\langle\log \frac{\mathcal{Z}\left(\omega-\frac{W^2}{\sigma},\,h+\frac{\mu W}{\sigma}\right)} {\mathcal{Z}(\omega,h)} \right\rangle_{W}\ ,
    \\
    S_7[\hat{\rho}] =& \alpha \log(\hat{\tau}) \langle k\rangle - \alpha \sum_{k=0}^{k_{\mathrm{max}}} p_k(k) \log(k!) + n \alpha \sum_{k=0}^{k_{\mathrm{max}}} p_k(k) \int \{\mathrm{d}\hat{\rho}\}_{k}\log\left(\mathcal{Z}\left(1-\{\hat{\sigma}\}_k,\{\hat{\mu}\}_k\right)\right)\ ,
    \\
    S_8[\hat{\pi}, \lambda, q] =& \frac{\log(\hat{r})}{\alpha} \langle s\rangle - \frac{1}{\alpha} \sum_{s=0}^{s_{\mathrm{max}}} p_s(s) \log(s!) +  n\mathcal{I} \left( \hat{\pi}, \lambda(t), q(t), t \right)\ ,
\end{align}
with
\begin{align}
\mathcal{I} \left( \hat{\pi}, \lambda(t), q(t), t \right) &= \frac{1}{\alpha} \sum_{s=0}^{s_{\mathrm{max}}} p_s(s)\int \{\mathrm{d}\hat{\pi}\}_s \log \Biggl \langle\biggl( \int \mathrm{d}v \exp \biggl[ -\frac{\beta}{2} v^2 (\lambda(t) - \{\hat{\omega}\}_s) \nonumber \\
&\qquad + \beta v (\theta q(t)x + \{\hat{h}\}_s) + \beta t f(u, v, x) \biggr] \biggr) \Biggr\rangle_x\ , \label{eq:I_s definition}
\end{align}
and recalling the shorthands $\{h\}_s=\sum_{l=1}^sh_l$ and equivalent versions for other variables.

Furthermore, recalling the definition of $q(t)$ from \eqref{q definition initial} allows us to write $S_2$ and $S_6$ as
\begin{equation}
    S_2[q(t)] = -\frac{\theta\beta}{\alpha} n q(t)^2\ , 
\end{equation}
\begin{equation}
    S_6[q(t)]=\frac{\theta\beta}{2\alpha}nq(t)^2\ .
\end{equation}
Finally, the remaining two action terms are given by:
\begin{align}
    S_{4}[\lambda(t)]=&\frac{\beta}{2\alpha}\;n\,\lambda(t)\ ,
    \\
    S_9[\xi] =& \mathrm{i} \xi \left( \alpha \langle k\rangle - \frac{1}{\alpha} \langle s\rangle \right)\ .
\end{align}
The action $S_9[\xi]$ vanishes at the saddle point as a result of \eqref{xi saddle point}.
 Therefore, we conclude that $\mathcal{Z}_{\mathrm{num}}$ has the following asymptotics for large $N,M$
\begin{equation}
  \mathcal{Z}_{\mathrm{num}}\approx \mathrm{e}^{\sqrt{NM} (\mathcal{A}_0+n \mathcal{A}_1(\beta)+o(n))} \label{Z num action written out}
\end{equation}
with
\begin{align}
  \mathcal{A}_0 =&- \frac{r\hat r}{\alpha}-\alpha\tau\hat \tau +b(r\tau-1) +\alpha\log(\hat \tau)\langle k\rangle- \alpha \sum_{k=0}^{k_{\mathrm{max}}} p_k(k) \log(k!) \nonumber \\
  & +\frac{\log(\hat{r})}{\alpha} \langle s\rangle - \frac{1}{\alpha} \sum_{s=0}^{s_{\mathrm{max}}} p_s(s) \log(s!) \label{S num A0}
  \\
  \mathcal{A}_1(\beta) =& - \frac{r\hat{r}}{\alpha}
\int \mathrm{d}\pi\,\mathrm{d}\hat{\pi}\;\log\!\left[\frac{\mathcal{Z}(\omega-\hat{\omega},\,h+\hat{h})}{\mathcal{Z}(\omega,h)}\right] - \alpha\hat{\tau}\tau\, \int \mathrm{d}\rho \mathrm{d}\hat{\rho} \log\left[ \frac{\mathcal{Z}(\sigma-\hat{\sigma}, \mu+\hat{\mu})}{\mathcal{Z}(\sigma,\mu)} \right] \nonumber\\
& +r\tau b\int \mathrm{d}{\pi}\,\mathrm{d}\rho\;\left\langle\log \frac{\mathcal{Z}\left(\omega-\frac{W^2}{\sigma},\,h+\frac{\mu W}{\sigma}\right)} {\mathcal{Z}(\omega,h)} \right\rangle_{W} + \alpha \sum_{k=0}^{k_{\mathrm{max}}} p_k(k) \int \{\mathrm{d}\hat{\rho}\}_{k}\log\left(\mathcal{Z}\left(1-\{\hat{\sigma}\}_k,\{\hat{\mu}\}_k\right)\right) \nonumber\\
&+\mathcal{I} \left( \hat{\pi}, \lambda(t), q(t), t \right)-\frac{\theta\beta}{2\alpha} q(t)^2+\frac{\beta}{2\alpha}\;\,\lambda(t)\ .\label{eq:A1betadef}
\end{align}
The parameters $r,\hat{r},\tau,\hat\tau,\ldots$ need to satisfy simultaneous equations arising from the saddle point conditions on the action $\mathcal{A}$. These conditions are derived below (see Section \ref{Densities at the saddle point}, equations \eqref{c hat relation to b alpha v2}, \eqref{t hat and c relation to alpha b v2}, \eqref{t anf langle rangle k}, and \eqref{c hat c to langle rangle s}).

\subsection{$\mathcal{Z}_{\mathrm{den}}$: asymptotic behaviour}\label{sec:normalisation constraint}

We now calculate the large $N,M$ asymptotics of the normalisation constant $\mathcal{Z}_{\mathrm{den}}$ originally defined by:
\begin{equation} \label{defZprimeapp}
    {\mathcal{Z}_{\mathrm{den}} = \sum_{\{c_{ij}\}}}\left[\prod_{i=1}^{M}\mathbf{1}\!\left(\sum_{j=1}^{N}c_{ij}=k_i\right)\right]\left[
\prod_{j=1}^{N}\mathbf{1}\!\left(\sum_{i=1}^{M}c_{ij}=s_j\right)\right] \mathbf{1}\!\left(\sum_{i=1}^{M}k_i=\sum_{j=1}^{N}s_j\right)\ ,
\end{equation}
and later re-defined as
\begin{align}
\nonumber    \mathcal{Z}_{\mathrm{den}} =& \sum_{\{c_{ij}\}}\left[\prod_{i=1}^M\prod_{j=1}^N\left(\frac{b}{\sqrt{NM}}\delta_{c_{ij},1}+\left(1-\frac{b}{\sqrt{NM}}\right)\delta_{c_{ij},0}\right)\right]\left[\prod_{i=1}^M\mathbf{1}\left(\sum_{j=1}^Nc_{ij}=k_i\right)\right]\\ &\times\left[\prod_{j=1}^N\mathbf{1}\left(\sum_{i=1}^Mc_{ij}=s_j\right)\right]\mathbf{1}\left(\sum_{i=1}^Mk_i=\sum_{j=1}^Ns_j\right)\ ,
\end{align}
after the insertion of the $b$-dependent ER factor in Eq. \eqref{eq: C given s and k with b}. Specifically, we want to calculate the limit $\lim_{N\to\infty}\log \mathcal{Z}_{\mathrm{den}}/N$ for a fixed instance of $\bm k$ and $\bm s$ drawn from $p_{k,s}(\bm k,\bm s)$ with marginals $p_k(k)$ and $p_s(s)$ respectively. 

Introducing the order parameters
\begin{align}
    \eta &= \frac{1}{N} \sum_{i=1}^N \exp[\mathrm{i} \hat{s}_i]\ , \\
    \chi &= \frac{1}{M} \sum_{l=1}^M \exp[\mathrm{i} \hat{k}_l]\ ,
\end{align}
we can incorporate their definition into $\mathcal{Z}_{\mathrm{den}}$, by using the following representations of the unity
\begin{equation}
    1 = \int N \, \mathrm{d}\eta \, \mathrm{d}\hat{\eta} \exp \left[ -\mathrm{i} \hat{\eta} \left( N \eta - \sum_{i=1}^N \exp[\mathrm{i} \hat{s}_i] \right) \right]\ ,
\end{equation}
\begin{equation}
    1 = \int M \, \mathrm{d}\chi \, \mathrm{d}\hat{\chi} \exp \left[ -\mathrm{i} \hat{\chi} \left( M \chi - \sum_{l=1}^M \exp[\mathrm{i} \hat{k}_l] \right) \right]\ .
\end{equation}
Multiplying our expression for $\mathcal{Z}_{\mathrm{den}}$ by these identities, and using Fourier expressions for the indicator functions, allows us to rewrite $\mathcal{Z}_{\mathrm{den}}$ in the following manner:
\begin{align}
    \mathcal{Z}_{\mathrm{den}} &\propto \int \mathrm{d}\eta \, \mathrm{d}\hat{\eta} \, \mathrm{d}\chi \, \mathrm{d}\hat{\chi} \, \mathrm{d}\underline{\hat{k}} \, \mathrm{d}\underline{\hat{s}} \, \mathrm{d}\xi \exp \left[ \mathrm{i} \xi \left( \sum_{l=1}^M k_l - \sum_{i=1}^N s_i \right) + \frac{b}{\sqrt{NM}} \sum_{l=1}^M \sum_{i=1}^N \left( \exp[\mathrm{i}(\hat{k}_l + \hat{s}_i)] - 1 \right) \right] \nonumber \\
    &\quad \times \exp \left[ -\mathrm{i} \sum_{l=1}^M \hat{k}_l k_l - \mathrm{i} \sum_{i=1}^N \hat{s}_i s_i - \mathrm{i} \hat{\eta} \eta N + \mathrm{i} \hat{\eta} \sum_{i=1}^N \exp[\mathrm{i} \hat{s}_i] - \mathrm{i} \hat{\chi} \chi M + \mathrm{i} \hat{\chi} \sum_{l=1}^M \exp[\mathrm{i} \hat{k}_l] \right].
\end{align}
We note clearly that we can use the definitions of $\eta$ and $\chi$ to write
\begin{equation}
    \sum_{l=1}^M \sum_{i=1}^N \left( \exp \left[ \mathrm{i} (\hat{k}_l + \hat{s}_i) \right] - 1 \right) = NM (\eta\chi - 1)\ .
\end{equation}
Now considering the remaining $\hat{k}_l$ integrals we see that we can write them in exponential form as follows:
\begin{equation}
    \int \mathrm{d}\underline{\hat{k}} \exp \left[ -\mathrm{i} \sum_{l=1}^M \hat{k}_lk_l + \mathrm{i} \hat{\chi} \sum_{l=1}^M \exp[\mathrm{i} \hat{k}_l] \right]
    = \exp \left[ \sum_{l=1}^M \log \left( \int \mathrm{d}\hat{k} \exp \left[ -\mathrm{i} \hat{k} k_l + \mathrm{i} \hat{\chi} \exp[\mathrm{i} \hat{k}] \right] \right) \right] \label{remaining k hat integrals}\ .
\end{equation}
We can use a Taylor series expansion of $\exp \left[ \mathrm{i} \hat{\chi} \exp[\mathrm{i} \hat{k}] \right]$ to write the $\hat{k}$ integral as follows:
\begin{equation}
    \int \mathrm{d}\hat{k} \exp \left[ -\mathrm{i} \hat{k} k_l + \mathrm{i} \hat{\chi} \exp[\mathrm{i} \hat{k}] \right] = \sum_{r=0}^\infty \frac{(\mathrm{i} \hat{\chi})^r}{r!} \int \mathrm{d}\hat{k} \exp \left[ \mathrm{i} \hat{k} (r - k_l) \right]= \frac{(\mathrm{i} \hat{\chi})^{k_l}}{k_l!}\ ,
\end{equation}
where we have identified the Fourier representation of a delta function. Hence inserting this expression into \eqref{remaining k hat integrals}, we find that the $\underline{\hat{k}}$ integral is equal to
\begin{equation}
     \exp \left[ \sum_{l=1}^M \log(\mathrm{i} \hat{\chi})^{k_l} - \sum_{l=1}^M \log(k_l!) \right],
\end{equation}
which, using the assumed self-averaging of the empirical row-degree distribution, becomes for large $M$
\begin{equation}
    \exp \left[ M \log(\mathrm{i} \hat{\chi})\sum_{k=0}^{k_{\mathrm{max}}} k p_k(k)  - M \sum_{k=0}^{k_{\mathrm{max}}} p_k(k) \log(k!) \right]\ .
\end{equation}
An identical procedure also gives us the following expression for the $\underline{\hat s}$ integrals:
\begin{equation}
    \exp \left[ N \log(\mathrm{i} \hat{\eta}) \sum_{s=0}^{s_{\mathrm{max}}} s p_s(s) - N \sum_{s=0}^{s_{\mathrm{max}}} p_s(s) \log(s!) \right]\ .
\end{equation}
Thus we can write the normalisation constant in the following form which is amenable to a saddle point evaluation
\begin{equation}
    \mathcal{Z}_{\mathrm{den}} \simeq \int \mathrm{d}\eta \, \mathrm{d}\hat{\eta} \, \mathrm{d}\chi \, \mathrm{d}\hat{\chi} \, \mathrm{d}\xi \exp \left[ \sqrt{NM} \, S_{\mathcal{Z}_{\mathrm{den}}}(\eta, \hat{\eta}, \chi, \hat{\chi}, \xi) \right],
\end{equation}
where
\begin{align}
    S_{\mathcal{Z}_{\mathrm{den}}}(\eta, \hat{\eta}, \chi, \hat{\chi}, \xi) &= \mathrm{i} \xi \left( \alpha \langle k\rangle - \frac{1}{\alpha} \langle s\rangle \right) - \mathrm{i} \frac{\eta \hat{\eta}}{\alpha} - \mathrm{i} \alpha \chi \hat{\chi} \nonumber \\
    &\quad + \alpha \, \log(\mathrm{i} \hat{\chi}) \langle k\rangle - \alpha \sum_{k=0}^{k_{\mathrm{max}}} p_k(k) \, \log(k!) \nonumber \\
    &\quad + \frac{\log(\mathrm{i} \hat{\eta})}{\alpha} \langle s\rangle - \frac{1}{\alpha} \sum_{s=0}^{s_{\mathrm{max}}} \log(s!) \, p_s(s) + b(\eta\chi - 1)\ ,
\end{align}
where we have also used the self-averaging of the empirical degree distributions in the $\xi$ terms.

The system of saddle-point equations for the normalisation term, found by taking $\frac{\partial}{\partial \xi} S_{\mathcal{Z}_{\mathrm{den}}}$, $\frac{\partial}{\partial \eta} S_{\mathcal{Z}_{\mathrm{den}}}$, $\frac{\partial}{\partial \hat{\eta}} S_{\mathcal{Z}_{\mathrm{den}}}$, $\frac{\partial}{\partial \chi} S_{\mathcal{Z}_{\mathrm{den}}}$, and $\frac{\partial}{\partial \hat{\chi}} S_{\mathcal{Z}_{\mathrm{den}}}$ is given by:
\begin{align}
    \langle k \rangle &= \frac{1}{\alpha^2} \langle s \rangle\ , \label{xi saddle point v2}\\
    \mathrm{i} \frac{\hat{\eta}}{\chi} &= \alpha b\ , \label{c hat relation to t alpha b}\\
    \mathrm{i} \eta \hat{\eta} &= \langle s \rangle\ , \label{eta to langle rangle s}\\
    \mathrm{i} \frac{\hat{\chi}}{\eta} &= \frac{b}{\alpha}\ , \label{cb relation to t hat}\\
    \mathrm{i} \chi \hat{\chi} &= \langle k \rangle\ . \label{chi to langle rangle k}
\end{align}

Therefore, we conclude that $\mathcal{Z}_{\mathrm{den}}$ has the following asymptotics for large $N,M$
\begin{equation}
\mathcal{Z}_{\mathrm{den}}\approx\mathrm{e}^{\sqrt{NM}\tilde{\mathcal{A}}_0}\ , \label{Z den written out}
\end{equation}
with
\begin{align}
    \tilde{\mathcal{A}}_0 &=  - \mathrm{i}\frac{\eta\hat\eta}{\alpha} - \mathrm{i}\alpha \chi\hat\chi + \alpha \, \log(\mathrm{i} \hat{\chi}) \langle k\rangle - \alpha \sum_{k=0}^{k_{\mathrm{max}}} p_k(k) \, \log(k!)\nonumber
    \\
    &\quad + \frac{\log(\mathrm{i} \hat{\eta})}{\alpha} \langle s\rangle - \frac{1}{\alpha} \sum_{s=0}^{s_{\mathrm{max}}} \log(s!) \, p_s(s) + b\left(\eta\chi - 1\right)\ , \label{Z dem A0}
\end{align}
where the parameters satisfy the saddle point equations above, and the $\xi$ dependent term has vanished at the saddle point as a result of \eqref{xi saddle point v2}.

\subsection{Full Asymptotics of $\langle \mathcal{Z}^n_{\beta}\rangle_{\bm A}$}\label{sec:fullasymptoticsZbetan}

We can now combine our asymptotic expressions for $\mathcal{Z}_{\mathrm{num}}$, given by \eqref{Z num action written out}, and $\mathcal{Z}_{\mathrm{den}}$, given by \eqref{Z den written out}, to write the full asymptotics of the average replicated partition function $\langle \mathcal{Z}^n_\beta \rangle_{\bm{A}}$ using \eqref{Partition function as ratio of num and den} as
\begin{equation}
    \left\langle \mathcal{Z}^n_{\beta} \right\rangle_{\bm A} \propto \frac{\mathcal{Z}_{\mathrm{num}}}{\mathcal{Z}_{\mathrm{den}}}\approx \exp\left[\sqrt{NM}\left( \mathcal{A}_0- \tilde{\mathcal{A}}_0+n \mathcal{A}_1(\beta)+o(n)\right)\right]\ .\label{eq:partfunasymptwithzeroorder}
\end{equation}
Moreover, we will shortly prove that $\mathcal{A}_0$ and $\tilde{\mathcal{A}_0}$ are equal, leaving behind only the $\mathcal{O}(n)$ term $\mathcal{A}_1(\beta)$. We first proceed to derive the RDEs satisfied by the joint probability densities at the saddle point.

\color{black}
\subsection{Densities at the saddle point} \label{Densities at the saddle point}

We are now able to use \eqref{gamma_phi saddle point}, as well as the equivalent forms for the other fields, to ascertain the structure of the normalised densities ${\pi}(\omega,h)$, $\hat{\pi}(\hat\omega,\hat{h})$, $\rho(\sigma,\mu)$ and $\hat{\rho}(\hat{\sigma},\hat{\mu})$ at the saddle point. We will begin by setting $t=0$ (for which we denote $\lambda(0)=\lambda_\theta$ and $q(0)=q$), and study the corresponding saddle point structure of the densities introduced in \ref{sec:replica sym ansatz} to find the average largest eigenvalue $\langle\lambda_1\rangle_{\bm{A}}$ from the formula \eqref{eq:Lambda_Replica}.

Inserting the $\hat\phi$ and $\psi$ versions of \eqref{phi RS Ansatz saddle point} into \eqref{gamma_phi saddle point} gives
\begin{align}
n\gamma_\phi=&\int \mathrm{d}\vec{v} \; \Biggl(\hat{r} \int \mathrm{d}\hat{\pi} \prod_{a=1}^n \exp\left[-\frac{\beta}{2}(-\hat{\omega})v_a^2 + \beta \hat{h}v_a\right] \nonumber\\
&- b\alpha \tau \int \mathrm{d}\rho \prod_{a=1}^n \left\langle \exp\left[-\frac{\beta}{2}\left(-\frac{W^2}{\sigma}\right)v_a^2 + \beta\left(\frac{W\mu}{\sigma}\right)v_a\right] \right\rangle_W \Biggr)\nonumber \\
&\qquad\times  \prod_{a=1}^n \frac{1}{\mathcal{Z}(\omega,h)} \exp\left[-\frac{\beta}{2}\omega v_a^2 + \beta h v_a\right], \nonumber \\
=&\quad \hat{r} \frac{1}{\mathcal{Z}^n(\omega,h)} \int \mathrm{d}\hat{\pi} \prod_{a=1}^n \int \mathrm{d}v_a \exp\left[-\frac{\beta}{2}(\omega-\hat{\omega})v_a^2 + \beta(h+\hat{h})v_a\right] \nonumber \\
&\quad - b\alpha \tau \frac{1}{\mathcal{Z}^n(\omega,h)} \int \mathrm{d}\rho \prod_{a=1}^n \left\langle \int \mathrm{d}v_a \exp\left[-\frac{\beta}{2}\left(\omega-\frac{W^2}{\sigma}\right)v_a^2 + \beta\left(h+\frac{W\mu}{\sigma}\right)v_a\right] \right\rangle_W.
\end{align}
Performing the $v_a$ integrals we obtain
\begin{equation}
n\gamma_\phi = \hat{r} \; \frac{\int \mathrm{d}\hat{\pi} \; \mathcal{Z}^n(\omega-\hat{\omega}, h+\hat{h})}{\mathcal{Z}^n(\omega,h)} - b\alpha \tau \; \frac{\int \mathrm{d}\rho \left\langle \mathcal{Z}^n\left(\omega-\frac{W^2}{\sigma}, h+\frac{W\mu}{\sigma}\right) \right\rangle_W}{\mathcal{Z}^n(\omega,h)}\ .
\end{equation}
We can now use $x^n\sim1+n\log(x)+o(n)$ to extract the leading $n\to0$ terms
\begin{equation} \label{eq:gamma phi saddle point equation final}
n\gamma_\phi = \hat{r} - b\alpha \tau + \hat{r} \, n \int \mathrm{d}\hat{\pi} \log\left( \frac{\mathcal{Z}(\omega-\hat{\omega}, h+\hat{h})}{\mathcal{Z}(\omega,h)} \right) - b\alpha \tau \, n \int \mathrm{d}\rho \left\langle \log\left( \frac{\mathcal{Z}\left(\omega-\frac{W^2}{\sigma}, h+\frac{W\mu}{\sigma}\right)}{\mathcal{Z}(\omega,h)} \right) \right\rangle_W.
\end{equation}
Thus, by inspection, in order for terms to match we must have that
\begin{equation}
\frac{\hat{r}}{\tau} = b\alpha\ . \label{c hat relation to b alpha v2}
\end{equation}
Furthermore, in order for \eqref{eq:gamma phi saddle point equation final} to hold for all non-integrated variables $\omega$ and $h$, then we must have
\begin{equation} \label{pi hat at the saddle point}
\hat{\pi}(\hat{\omega},\hat{h}) = \int \mathrm{d}\rho \left\langle \delta\left(\hat{\omega} - \frac{W^2}{\sigma}\right) \delta\left(\hat{h} - \frac{W\mu}{\sigma}\right) \right\rangle_W,
\end{equation}
at the saddle point. Taken together, these imply that $\gamma_\phi = 0$. An identical calculation for $\gamma_\psi$ yields
\begin{align}
\frac{\hat{\tau}}{r} &= \frac{b}{\alpha}, \label{t hat and c relation to alpha b v2}\\
\hat{\rho}(\hat{\sigma},\hat{\mu}) &= \int \mathrm{d}\pi \left\langle \delta\left(\hat{\sigma} - \frac{W^2}{\omega}\right) \delta\left(\hat{\mu} - \frac{W h}{\omega}\right) \right\rangle_W, \label{rho hat at the saddle point}\\
\gamma_\psi &= 0.
\end{align}
Next, turning to the stationarity constraint $\frac{\delta \mathcal{A}}{\delta \hat{\rho}}$, we have that:
\begin{align}
n\gamma_{\hat{\psi}} &= \int \mathrm{d}\vec{u} \; \frac{\delta S}{\delta \hat{\psi}(\vec{u})} \prod_{a=1}^n \exp\left[ \frac{\beta}{2}\hat{\sigma}u_a^2 + \beta \hat{\mu} u_a \right] \nonumber \\
&= \int \mathrm{d}\vec{u} \left( \psi(\vec{u}) - \sum_{k=1}^{{k_{\mathrm{max}}}} k p_k(k) \frac{\exp\left[-\frac{\beta}{2}\sum_{a=1}^n u_a^2\right] \left(\mathrm{i}\hat{\psi}(\vec{u})\right)^{k-1}}{\int \mathrm{d}\vec{u}' \exp\left[-\frac{\beta}{2}\sum_{a=1}^n {u'_a}^2\right] \left(\mathrm{i}\hat{\psi}(\vec{u}')\right)^k} \right) \prod_{a=1}^n \exp\left[ \frac{\beta}{2}\hat{\sigma}u_a^2 + \beta \hat{\mu} u_a \right].
\end{align}
Inserting the $\hat\psi$ and $\psi$ versions of \eqref{phi RS Ansatz saddle point} into the RHS and performing the $\mathrm{d}\vec{u}$ integral gives us:
\begin{equation}
\tau \int \mathrm{d}\rho \left( \frac{\mathcal{Z}(\sigma-\hat{\sigma}, \mu+\hat{\mu})}{\mathcal{Z}(\sigma,\mu)} \right)^n - \sum_{k=1}^{k_{\mathrm{max}}} \frac{k p_k(k)}{\hat{\tau}} \frac{\int \{\mathrm{d}\hat{\rho}\}_{k-1} \mathcal{Z}^n\left(1-\{\hat{\sigma}\}_{k-1}-\hat{\sigma}, \{\hat{\mu}\}_{k-1}+\hat{\mu}\right)}{\int \{\mathrm{d}\hat{\rho}\}_k \mathcal{Z}^n\left(1-\{\hat{\sigma}\}_k, \{\hat{\mu}\}_k\right)}\ .
\end{equation}
Thus taking the $n \to 0$ asymptotics gives us:
\begin{align}
\nonumber &\tau + \tau n \int \mathrm{d}\rho \log\left(\mathcal{Z}(\sigma-\hat{\sigma}, \mu+\hat{\mu}) \right) - \tau n \int \mathrm{d}\rho \log\left(\mathcal{Z}(\sigma,\mu)\right)
\\
-& \sum_{k=1}^{{k_{\mathrm{max}}}} \frac{k p_k(k)}{\hat{\tau}} \frac{1 + n \int \{\mathrm{d}\hat{\rho}\}_{k-1} \log\left(\mathcal{Z}\left(1-\{\hat{\sigma}\}_{k-1}-\hat{\sigma}, \{\hat{\mu}\}_{k-1}+\hat{\mu}\right)\right)}{1 + n\int \{\mathrm{d}\hat{\rho}\}_k \log\left(\mathcal{Z}\left(1-\{\hat{\sigma}\}_k, \{\hat{\mu}\}_k\right)\right)} \ .
\end{align}
By inspection, in order for the two sides to match for all non-integrated variables $\hat{\sigma}, \hat{\mu}$, we must have:
\begin{align}
 \tau \hat{\tau} &= \langle k \rangle \label{t anf langle rangle k}\\
\rho(\sigma,\mu) &= \sum_{k=1}^{k_{\mathrm{max}}} \frac{k p_k(k)}{\langle k \rangle} \int \{\mathrm{d}\hat{\rho}\}_{k-1} \delta\left(\sigma - \left(1-\{\hat{\sigma}\}_{k-1}\right)\right) \delta\left(\mu - \{\hat{\mu}\}_{k-1}\right) \label{rho at the saddle point} \\
n\gamma_{\hat{\psi}} &= n \sum_{k=1}^{k_{\mathrm{max}}} \frac{k p_k(k)}{\hat{\tau}} \int \{\mathrm{d}\hat{\rho}\}_k \log\left(\mathcal{Z}\left(1-\{\hat{\sigma}\}_k, \{\hat{\mu}\}_k\right)\right) - \tau n \int \mathrm{d}\rho \log\left(\mathcal{Z}(\sigma,\mu)\right)\ . \label{gamma hat psi saddle point}
\end{align}
Finally, an identical calculation for $\frac{\delta \mathcal{A}}{\delta \hat{\pi}}$ for $t=0$ gives:
\begin{align}
    r\hat r&=\langle s\rangle \label{c hat c to langle rangle s}
    \\
    \pi(\omega,h)&=\sum_{s=1}^{s_\mathrm{max}}\frac{sp_s(s)}{\langle s \rangle}\int \{\mathrm{d}\hat\pi\}_{s-1}\left\langle\delta\left(\omega-(\lambda_{\theta}-\{\hat\omega\}_{s-1})\right)\delta\left(h-(\theta qx+\{\hat h\}_{s-1})\right)\right\rangle_x
    \label{pi at the saddle point}\\
    n\gamma_{\hat\pi}&=n\sum_{s=1}^{s_\mathrm{max}}\frac{sp_s(s)}{\hat r}\int\{\mathrm{d}\hat\pi\}_s\left\langle\log\left(\mathcal{Z}(\lambda_{\theta}-\{\hat\omega\}_s,\theta  qx+\{\hat h \}_s)\right)\right\rangle_x-n\int \mathrm{d}\pi\log\big(\mathcal{Z}(\omega,h)\big)\ . \label{gamma hat pi saddle point}
\end{align}
The values of the Lagrange parameters given by \eqref{gamma hat psi saddle point} and \eqref{gamma hat pi saddle point} have no impact on the value of the action at the saddle point, since by definition the terms multiplying the Lagrange parameters in the action (normalisation conditions for $\hat\pi(\hat\omega,\hat h)$ and $\hat\rho(\hat\sigma,\hat\mu)$ in the form of \eqref{pi normalisation saddle point} and similar) vanish when evaluating at the saddle point.

Finally, we now use \eqref{phi RS Ansatz saddle point} and its equivalent forms to rewrite \eqref{lambda saddle point 1}-\eqref{xi saddle point} (recalling that we write $\lambda(0)=\lambda_{\theta}$ and $q(0)=q$) in terms of our densities.

Setting $t=0$ in \eqref{lambda saddle point 1}  gives the following saddle point equation:
\begin{align}
\nonumber 1 & = \sum_{s=0}^{s_{\mathrm{max}}} p_s(s) \frac{\int \mathrm{d}\vec{v} \left\langle \exp\left[ - \frac{\beta}{2}\lambda_{\theta}\sum_{a=1}^n v_a^2  + \theta\beta x q\sum_{a=1}^n v_a \right] \right\rangle_x v_a^2 \left(\mathrm{i} \hat{\phi}(\vec{v})\right)^s}{\int \mathrm{d}\vec{v}' \left\langle \exp\left[ - \frac{\beta}{2}\lambda_{\theta}\sum_{a=1}^n {v'_a}^2 + \theta\beta x q\sum_{a=1}^n v'_a  \right] \right\rangle_x \left(\mathrm{i} \hat{\phi}
(\vec{v}')\right)^s}
\\
&:= \sum_{s=0}^{s_{\mathrm{max}}} p_s(s) \frac{N}{D}\ ,
\end{align}
where we have used the definition of $q$ given by \eqref{q definition initial}, as well as our ansätze given by \eqref{lambda ansatz}, \eqref{q ansatz}. Inserting the $\hat\phi$ version of \eqref{phi RS Ansatz saddle point} gives us for the denominator
\begin{align}
\nonumber D &= ( \hat{r})^s \int \mathrm{d}\vec{v}' \int \{\mathrm{d}\hat{\pi}\}_s \left\langle\prod_{a=1}^n \exp\left[ -\frac{\beta}{2}(\lambda_{\theta} - \{\hat{\omega}\}_s){v'_a}^2 + \beta(\theta  qx + \{\hat{h}\}_s)v'_a \right]\right\rangle_x \\
&= ( \hat{r})^s \int \{\mathrm{d}\hat{\pi}\}_s \left\langle\mathcal{Z}^n\left(\lambda_{\theta} - \{\hat{\omega}\}_s, \theta qx + \{\hat{h}\}_s\right)\right\rangle_x,
\end{align}
where we identified our expression for $\mathcal{Z}(x,y)$ given by \eqref{Z definition}. Similarly, our numerator term following the insertion of the ansatz becomes:
\begin{equation}
N = (\hat{r})^s \int \{\mathrm{d}\hat{\pi}\}_s \left\langle\int \mathrm{d}\vec{v}\; v_a^2 \exp\left[ -\lambda_{\theta}\frac{\beta}{2}\sum_{a=1}^n v_a^2 + \theta\beta qx \sum_{a=1}^n v_a + \frac{\beta}{2}\{\hat{\omega}\}_s \sum_{a=1}^n v_a^2 + \beta \{\hat{h}\}_s \sum_{a=1}^n v_a \right]\right\rangle_x.
\end{equation}
We note that the $\mathrm{d}\vec{v}$ integral will split up into $n-1$ copies of:
\begin{align}
\int \mathrm{d}v_a \exp\left[ -\frac{\beta}{2}(\lambda_{\theta} - \{\hat{\omega}\}_s)v_a^2 + \beta(\theta qx + \{\hat{h}\}_s)v_a \right] 
= \mathcal{Z}\left(\lambda_{\theta} - \{\hat{\omega}\}_s, \theta qx + \{\hat{h}\}_s\right)\ ,
\end{align}
and one copy of:
\begin{align}
\nonumber &\int \mathrm{d}v_a\; v_a^2 \exp\left[ -\frac{\beta}{2}(\lambda_{\theta} - \{\hat{\omega}\}_s)v_a^2 + \beta(\theta qx + \{\hat{h}\}_s)v_a \right]\\
&= \left(\frac{1}{\beta(\lambda_{\theta}-\{\hat\omega\}_s)}+\left(\frac{\theta  qx+\{\hat h\}_s}{\lambda_{\theta}-\{\hat \omega\}_s}\right)^2\right)\mathcal{Z}(\lambda_{\theta}-\{\hat\omega\}_s,\theta qx+\{\hat h\}_s)\ .\label{Gauss integral for lambda constraint}
\end{align}
Thus, our saddle point equation \eqref{lambda saddle point 1} reduces to
\begin{equation}
    1=\sum_{s=0}^{s_{\mathrm{max}}}p_s(s)\frac{\int\{\mathrm{d}\hat\pi\}_s\left\langle\left(\frac{1}{\beta(\lambda_{\theta}-\{\hat\omega\}_s)}+\left(\frac{\theta qx+\{\hat h\}_s}{\lambda_{\theta}-\{\hat \omega\}_s}\right)^2\right)\mathcal{Z}^n(\lambda_{\theta}-\{\hat\omega\}_s,\theta qx+\{\hat h\}_s)\right\rangle_x}{\int\{\mathrm{d}\hat\pi\}_s\left\langle\mathcal{Z}^n(\lambda_{\theta}-\{\hat\omega\}_s,\theta qx+\{\hat h\}_s)\right\rangle}\ .
\end{equation}
After extracting the leading $n\rightarrow0$ asymptotics, we are left with
\begin{equation} \label{eq:lambda saddle point final pi hat}
    1=\sum_{s=0}^{s_{\mathrm{max}}}p_s(s)\int\{\mathrm{d}\hat\pi\}_s\left\langle\frac{1}{\beta(\lambda_{\theta}-\{\hat\omega\}_s)}+\left(\frac{\theta qx+\{\hat h\}_s}{\lambda_{\theta}-\{\hat \omega\}_s}\right)^2\right\rangle_x\ .
\end{equation}
A very similar calculation also allows us to write our saddle point equation \eqref{q saddle point initial} in the form
\begin{equation}\label{q saddle point condition pi hat}
    q=\sum_{s=0}^{s_{\mathrm{max}}}p_s(s)\int\{\mathrm{d}\hat\pi\}_s\left\langle\frac{\theta qx^2+x\{\hat h\}_s}{\lambda_{\theta}-\{\hat\omega\}_s}\right\rangle_x\ .
\end{equation}
We see that in the $\beta\to \infty$ limit, the first term under the $\{\mathrm{d}\hat\pi\}_s$ integral in \eqref{eq:lambda saddle point final pi hat} vanishes. Hence, after substituting the expression for $\hat\pi$ given by \eqref{pi hat at the saddle point} into \eqref{eq:lambda saddle point final pi hat} and \eqref{q saddle point condition pi hat}, we find that our conditions are given by
\begin{align}
    1 =& \sum_{s=0}^{s_{\mathrm{max}}} p_s(s) \int \left\{ \mathrm{d}\rho \right\}_s \left\langle \left( \frac{\theta qx + \left\{ \frac{\mu W}{\sigma} \right\}_s}{\lambda_\theta - \left\{ \frac{W^2}{\sigma} \right\}_s} \right)^2 \right\rangle_{x,\{W\}_{s}}\  \label{eq:lambda saddle point final}
    \\
    q = &\sum_{s=0}^{s_{\mathrm{max}}} p_s(s) \int \{ \mathrm{d}\rho \}_s \left\langle \frac{\theta  qx^2 + x \left\{ \frac{\mu W}{\sigma} \right\}_s}{\lambda_\theta - \left\{ \frac{W^2}{\sigma} \right\}_s} \right\rangle_{x,\{W\}_{s}}\ . \label{q saddle point condition pi}
\end{align}

We thus have the structure of the densities at the saddle point, as well as constraints that must be met by the Lagrange parameters $\lambda_{\theta}$ and $q$ in terms of the normalised densities. 

Furthermore, we note that the leading terms $\mathcal{A}_0$ and $\tilde{\mathcal{A}}_0$ of the actions for the numerator and the denominator of $\left\langle \mathcal{Z}^n_{\beta} \right\rangle_{\bm A}$ (see \eqref{S num A0} and \eqref{Z dem A0}) turn into each other upon the replacement $r\leftrightarrow \eta$, $\hat r\leftrightarrow\mathrm{i}\hat \eta$, $\tau\leftrightarrow\chi$, and $\hat \tau\leftrightarrow\mathrm{i}\hat\chi$. Moreover, the equations linking corresponding pairs of variables in this mapping [\eqref{c hat relation to t alpha b} and \eqref{c hat relation to b alpha v2}, \eqref{eta to langle rangle s} and \eqref{c hat c to langle rangle s}, \eqref{cb relation to t hat} and \eqref{t hat and c relation to alpha b v2}, as well as \eqref{chi to langle rangle k} and \eqref{t anf langle rangle k}] all coincide. This implies that $\mathcal{A}_0$ and $\tilde{\mathcal{A}_0}$ are equal and thus cancel out in \eqref{eq:partfunasymptwithzeroorder}. The added benefit of this cancellation is that the spurious ER parameter $b$---introduced to assist performing the disorder average---disappears from the final formulae, as expected.
Hence 
\begin{equation}\label{average replication partion function in terms of A1}
     \left\langle \mathcal{Z}^n_{\beta} \right\rangle_{\bm A} \propto \frac{\mathcal{Z}_{\mathrm{num}}}{\mathcal{Z}_{\mathrm{den}}}\approx \exp\left[\sqrt{NM}\left(n \mathcal{A}_1(\beta)+o(n)\right)\right]\ .
\end{equation}
It is now a good moment to summarise what we have achieved, and what remains to be done. Our starting point is the formula \eqref{eq:Lambda_Replica} for the average largest eigenvalue in terms of the average of the replicated partition function at $t=0$
\begin{equation}   \Big\langle\lambda_1\Big\rangle_{\bm{A}} = \lim_{\beta\rightarrow\infty}\lim_{N\to \infty} \frac{2}{\beta N}  \lim_{n\rightarrow 0} \frac{1}{n}\log\Big\langle \mathcal{Z}^n_{\beta} \Big\rangle_{\bm A}\ .
    \label{eq:Lambda_ReplicaLaterOn}
\end{equation}
Assuming that the $N\to\infty$ limit can be exchanged with the replica limit $n\to 0$, and inserting the asymptotics 
\eqref{average replication partion function in terms of A1} into \eqref{eq:Lambda_ReplicaLaterOn}, we get after elementary manipulations
\begin{equation}   \Big\langle\lambda_1\Big\rangle_{\bm{A}} = 2\alpha\lim_{\beta\rightarrow\infty}\frac{\mathcal{A}_1(\beta)}{\beta}\ .   \label{eq:Lambda_ReplicaFinalFormulaLaterOn}
\end{equation}
The next section is devoted to the extraction of the leading $\beta\to\infty$ term of $\mathcal{A}_1(\beta)$ given in \eqref{eq:A1betadef}.

\subsection{Extracting the leading $\beta\to\infty$ term of $\mathcal{A}_1(\beta)$}
We can rewrite $\mathcal{A}_1(\beta)$ as
\begin{equation}\label{A1 decomposed into actions}
    \mathcal{A}_1(\beta) =S_1[\pi,\hat\pi]+S_2[\rho,\hat\rho]+S_3[q,\lambda_{\theta}]+S_4[\pi,\rho]+S_5[\hat\rho]+S_6[\hat\pi,\lambda_{\theta},q]
\end{equation}
where we have written
\begin{align}
    S_1[\pi,\hat\pi]=&- \frac{\langle s\rangle}{\alpha}
\int \mathrm{d}\pi\,\mathrm{d}\hat{\pi}\;\log\!\left[\frac{\mathcal{Z}(\omega-\hat{\omega},\,h+\hat{h})}{\mathcal{Z}(\omega,h)}\right]\label{S1 pre beta limit}
\\
    S_2[\rho,\hat\rho]=& - \alpha\langle k\rangle \int \mathrm{d}\rho \mathrm{d}\hat{\rho} \log\left[ \frac{\mathcal{Z}(\sigma-\hat{\sigma}, \mu+\hat{\mu})}{\mathcal{Z}(\sigma,\mu)} \right] 
\\
    S_3[q,\lambda_{\theta}]=&\frac{\beta}{2\alpha}\,\lambda_{\theta}-\frac{\theta\beta}{2\alpha} q^2
\\
    S_4[\pi,\rho]=&\frac{\langle s\rangle}{\alpha}\int \mathrm{d}{\pi}\,\mathrm{d}\rho\;\left\langle\log \frac{\mathcal{Z}\left(\omega-\frac{W^2}{\sigma},\,h+\frac{\mu W}{\sigma}\right)} {\mathcal{Z}(\omega,h)} \right\rangle_{W}
\\
    S_5[\hat\rho]=&\alpha \sum_{k=0}^{k_{\mathrm{max}}} p_k(k) \int \{\mathrm{d}\hat{\rho}\}_{k}\log\left(\mathcal{Z}\left(1-\{\hat{\sigma}\}_k,\{\hat{\mu}\}_k\right)\right)
    \\
    S_6[\hat\pi,q,\lambda_{\theta}]=&\frac{1}{\alpha} \sum_{s=0}^{s_{\mathrm{max}}} p_s(s) \, \int \{\mathrm{d}\hat{\pi}\}_s \left\langle\log  \biggl( \mathcal{Z}\left(\lambda_{\theta}-\{\hat\omega\}_s,\theta qx+\{\hat h\}_s\right)\biggl)\right\rangle_x\ , \label{S6 in terms of densities}
\end{align}
using \eqref{c hat relation to b alpha v2}, \eqref{c hat c to langle rangle s}, \eqref{t hat and c relation to alpha b v2}, and \eqref{chi to langle rangle k}. We have also set $t=0$ in \eqref{eq:I_s definition}, and identified the integral representation of \eqref{Z definition}, leading to \eqref{S6 in terms of densities}.
\color{black}
Taking the leading order behaviour of $S_1$ using our definition of $\mathcal{Z}(x,y)$ given by \eqref{Z definition}, gives us
\begin{equation}\label{eq:S1 saddle point form}
    S_1 \simeq -\frac{\beta}{2} \frac{\langle s\rangle}{\alpha} \int \mathrm{d}\pi \mathrm{d}\hat{\pi} \left[ \frac{(h + \hat{h})^2}{\omega - \hat{\omega}} - \frac{h^2}{\omega} \right] =: -\frac{\beta}{2}  \frac{\langle s\rangle}{\alpha} I_1\ .
\end{equation}
Similarly, we have that $S_2$'s asymptotic form is given by
\begin{equation} \label{I2 definition}
    S_2  \simeq -\frac{\beta}{2} \alpha \langle k\rangle \int \mathrm{d}\rho \mathrm{d}\hat{\rho} \left[ \frac{(\mu + \hat{\mu})^2}{\sigma - \hat{\sigma}} - \frac{\mu^2}{\sigma} \right] =: -\frac{\beta}{2}  \alpha \langle k\rangle I_2\ .
\end{equation}
Next, we clearly have:
\begin{equation}
    S_3 = \frac{\beta}{2} \left[ \frac{\lambda_{\theta}}{\alpha} - \frac{\theta}{\alpha} q^2 \right]\ .
\end{equation}
Then, the leading order behaviour of $S_4$ is given by:
\begin{equation}
    S_4 \simeq \frac{\beta}{2} \frac{\langle s \rangle}{\alpha} \int \mathrm{d}\pi \mathrm{d}\rho \left\langle \left[ \frac{(h + \frac{\mu W}{\sigma})^2}{\omega - \frac{W^2}{\sigma}} - \frac{h^2}{\omega} \right] \right\rangle_W =: \frac{\beta}{2} \frac{\langle s \rangle}{\alpha} I_3\ .
\end{equation}
Furthermore, in the $\beta\rightarrow\infty$ limit, $S_5$ has leading behaviour
\begin{equation}\label{I4 definition}
    S_5 \simeq \frac{\beta}{2} \alpha \sum_{k=0}^{k_{\mathrm{max}}} p_k(k) \int \{ \mathrm{d}\hat{\rho} \}_k \frac{ \{ \hat{\mu} \}_k^2 }{ 1 - \{ \hat{\sigma} \}_k } =: \frac{\beta}{2}  \alpha \sum_{k=0}^{k_{\mathrm{max}}} p_k(k) {I}_4(k)\ . 
\end{equation}
Finally, our sixth action term is given by:
\begin{align}
    S_6 &\simeq \frac{\beta}{2} \cdot \frac{1}{\alpha} \sum_{s=0}^{s_{\mathrm{max}}} p_s(s) \int \{ \mathrm{d}\hat{\pi} \}_s \left\langle \frac{ (\theta qx + \{ \hat{h} \}_s)^2 }{ \lambda_{\theta} - \{ \hat{\omega} \}_s } \right\rangle_x\ . \label{S6 saddle point form}
\end{align}

We can now insert the saddle point forms of the densities $\pi(\omega,h)$, $\hat\pi(\hat\omega,\hat h)$, $\rho(\sigma,\mu)$, and $\hat\rho(\hat\sigma,\hat\mu)$ into the $\beta\rightarrow\infty $ asymptotic forms of \eqref{S1 pre beta limit}-\eqref{S6 in terms of densities}. This will simplify \eqref{A1 decomposed into actions} allowing us to evaluate the average largest eigenvalue. For instance, we can insert the saddle-point equation for $\hat\pi(\hat\omega,\hat h)$ given by \eqref{pi hat at the saddle point} into \eqref{eq:S1 saddle point form} to find that
\begin{align}
    I_1=&\int \mathrm{d}\pi \mathrm{d}\hat{\pi} \left[ \frac{(h + \hat{h})^2}{\omega - \hat{\omega}} - \frac{h^2}{\omega} \right]  \label{I1 start}
    \\
    =& \int\mathrm{d}\pi\mathrm{d}\rho\left\langle \delta\left(\hat{\omega} - \frac{W^2}{\sigma}\right) \delta\left(\hat{h} - \frac{W\mu}{\sigma}\right) \right\rangle_W\left[ \frac{(h + \hat{h})^2}{\omega - \hat{\omega}} - \frac{h^2}{\omega} \right] 
    \\
    =&\int\mathrm{d}\pi\mathrm{d}\rho\left\langle \left[ \frac{(h + \frac{\mu W}{\sigma})^2}{\omega - \frac{W^2}{\sigma}} - \frac{h^2}{\omega} \right] \right\rangle_W=I_3\ .\label{I1 end}
\end{align}
Thus, the $S_1$ and $S_4$ terms in \eqref{A1 decomposed into actions} cancel out exactly. 

Furthermore, we can split \eqref{S6 saddle point form} into two parts, by expanding the $(\theta qx + \{ \hat{h} \}_s)^2$ term in the numerator. This allows us to write it in terms of two separate $\{\mathrm{d}\hat\pi\}_s$ integrals as follows
\begin{align}
    S_6 &=\frac{\beta}{2} \cdot \frac{1}{\alpha} \sum_{s=0}^{s_{\mathrm{max}}} p_s(s) \left(\int \{ \mathrm{d}\hat{\pi} \}_s \left\langle \frac{ (\theta qx)^2 + \theta qx\{\hat h\}_s }{ \lambda_{\theta} - \{ \hat{\omega} \}_s }\right\rangle_x+\int \{ \mathrm{d}\hat{\pi} \}_s\left\langle\frac{\theta qx\{\hat h \}_s+\{\hat{h} \}_s^2}{\lambda_{\theta} - \{ \hat{\omega} \}_s} \right\rangle_x\right) \nonumber
    \\
    &=: \frac{\beta}{2}\cdot\frac{1}{\alpha}\sum_{s=0}^{s_{\mathrm{max}}} p_s(s)I_5(s)+\frac{\beta}{2}\cdot\frac{1}{\alpha}\sum_{s=0}^{s_{\mathrm{max}}} p_s(s)I_6(s)\ . \label{I6 definition}
\end{align}
By following the same procedure as in \eqref{I1 start}-\eqref{I1 end}, and inserting the saddle point densities for $\hat\pi(\hat\omega,\hat h)$, $\rho(\sigma,\mu)$, and $\hat{\rho}(\hat\sigma,\hat\mu)$ into \eqref{I2 definition}, \eqref{I4 definition}, and the $I_6(s)$ term in \eqref{I6 definition}, then after some algebraic manipulation we find that
\begin{equation}
    \frac{\beta}{2}\cdot\frac{1}{\alpha}\sum_{s=0}^{s_{\mathrm{max}}} p_s(s)I_6(s)-\frac{\beta}{2}\alpha \langle k\rangle I_2+\frac{\beta}{2} \alpha \sum_{k=0}^{k_{\mathrm{max}}} p_k(k) {I}_4(k) = 0\ .
\end{equation}
Hence we can rewrite the leading $\beta\to\infty$ behaviour of the action as
\begin{align}
    \mathcal{A}_1(\beta) \simeq  \frac{\beta}{2} \left( \frac{\lambda_{\theta}}{\alpha} - \frac{\theta}{\alpha} q^2 + \frac{\theta}{\alpha} q  \sum_{s=0}^{s_{\mathrm{max}}} p_s(s) \int \{ \mathrm{d}\hat{\pi} \}_s \left\langle \frac{\theta q x^2 + x \{ \hat{h} \}_s}{\lambda_{\theta} - \{ \hat{\omega} \}_s} \right\rangle_x \right)\ .
\end{align}
Recalling \eqref{q saddle point condition pi hat}, we see that the $q$-dependent terms cancel out exactly, leaving behind the elementary expression
\begin{equation}\label{eq:Action fully evaluated at the saddle point}
    \mathcal{A}_1(\beta) \simeq \frac{\beta}{2\alpha} \lambda_{\theta}\ .
\end{equation}
Inserting \eqref{eq:Action fully evaluated at the saddle point} into \eqref{eq:Lambda_ReplicaFinalFormulaLaterOn} gives eventually the result claimed in the main text, namely
\begin{equation}
    \langle \lambda_1 \rangle_{\bm{A}} = \lambda_{\theta}\ ,
\end{equation}
where $\lambda_{\theta}$ is the parameter that satisfies the set of self-consistency equations given by \crefrange{eq:pi recursive}{eq:q condition}.

\subsection{Further Observables}
Our next goal is to use the formalism introduced in Section \ref{sec:Replica Methodology} to calculate the average density of the entries of the top eigenvector of our matrix $\bm{A}$, as well as the overlap and squared overlap with the signal vector $\bm{x}$. Recalling \crefrange{eq:Function average replics}{eq:Average top eigenvector initial}, we can leverage all the work done so far to write directly
\begin{equation}
    \rho_{\mathrm{top}}(u) = \lim_{\beta \to \infty} \frac{1}{\beta} \frac{\partial}{\partial t} \left. I(\hat{\pi}, \lambda(t), q(t), t) \right|_{t=0}
\end{equation}
where:
\begin{equation} \label{eq:eigenvector I inter 1}
\begin{split}
    \mathcal{I}[\hat{\pi}, \lambda(t), q(t), t] ={} & \left\langle \sum_{s=0}^{s_{\mathrm{max}}} p_s(s) \int \{ \mathrm{d}\hat{\pi} \}_s \right. \log \left( \int \mathrm{d}v \exp \left[ -\frac{\beta}{2} (\lambda(t) - \{ \hat{\omega} \}_s) v^2 \right. \right. \\
    & \left. \left. + \beta (\theta  q(t)x + \{ \hat{h} \}_s) v + \beta tf_{\mathrm{top}}(u,\bm{v}) \right] \right) \bigg\rangle_x 
\end{split}
\end{equation}
(defined in \eqref{eq:I_s definition}) is the only term in the action \eqref{eq:A1betadef} that depends explicitly on the variable $t$. As shown in \cite{Susca2021}, all other terms in the action will not contribute. We can insert \eqref{eq:Top eigenvector F choice} in its factorized form $f_{\mathrm{top}}(u,{v}) = \delta_\epsilon(u - v)$ (where $\delta_\epsilon$ is a smooth regulariser of the delta function \footnote{In the calculation, we take the $\epsilon\to 0^+$ limit of the regulariser.}) into \eqref{eq:eigenvector I inter 1}, and then take the derivative w.r.t. $t$ and evaluate it at $t=0$ to find that
\begin{equation} \label{eq:regulariser}
    \rho_{\mathrm{top}}(u) = \lim_{\beta \to \infty} \sum_{s=0}^{s_{\mathrm{max}}} p_s(s) \int \{ \mathrm{d}\hat{\pi} \}_s \left\langle \frac{ \exp \left[ -\frac{\beta}{2} (\lambda_{\theta} - \{ \hat{\omega} \}_s) u^2 + \beta (\theta  qx + \{ \hat{h} \}_s) u \right] }{ \int \mathrm{d}v \exp \left[ -\frac{\beta}{2} (\lambda_{\theta} - \{ \hat{\omega} \}_s) v^2 + \beta (\theta qx + \{ \hat{h} \}_s) v \right] } \right\rangle_x.
\end{equation}
Taking the $\beta\to\infty$ limit, the normalised exponential collapses to a delta function
\begin{equation}\label{eq:top eigenvector final}
    \rho_{\mathrm{top}}(u) = \sum_{s=0}^{s_{\mathrm{max}}} p_s(s) \int \{ \mathrm{d}\hat{\pi} \}_s \left\langle \delta \left( u - \frac{\theta qx + \{ \hat{h} \}_s}{\lambda_{\theta} - \{ \hat{\omega} \}_s} \right) \right\rangle_x.
\end{equation}
We insert \eqref{pi hat at the saddle point} into \eqref{eq:top eigenvector final} to find 
\begin{equation}
    \rho_{\mathrm{top}}(u)  = \sum_{s=0}^{s_{\mathrm{max}}} p_s(s) \int \{ \mathrm{d}\rho \}_s \left\langle \delta \left( u - \frac{\theta  qx + \left\{ \frac{\mu W}{\sigma} \right\}_s}{\lambda_\theta - \left\{ \frac{W^2}{\sigma} \right\}_s} \right) \right\rangle_{x,\{W\}_s}\ .
\end{equation}
It is important to note that since the partial derivative w.r.t. $t$ only acts on terms containing explicit dependence on $t$, the saddle point equations for the joint probability densities and the order parameters $\lambda_{\theta}$ and $q$ will be the same as for when $t$ was set to zero at the outset. Hence the $\lambda_{\theta}$, $q$, and $\hat{\pi}$ in \eqref{eq:top eigenvector final} satisfy the same saddle-point equations as in \crefrange{eq:pi recursive}{eq:q condition}.

Next, in order to calculate the overlap between the top eigenvector and the signal $\bm{x}$, we follow the same procedure, but with $f$ defined as in \eqref{eq:F choice overlap}, in its factorised form $f(u, v, x) = \delta(u - xv)$. We can insert this into \eqref{eq:eigenvector I inter 1} to obtain
\begin{align}
    \mathcal{I} = \Biggl\langle \sum_{s=0}^{s_{\mathrm{max}}} p_s(s) \int \{ \mathrm{d}\hat{\pi} \}_s \log \Bigg( \int \mathrm{d}v \exp \Big[&-\frac{\beta}{2} (\lambda_{\theta} - \{ \hat{\omega} \}_s) v^2 \nonumber
    \\
    &+\beta (\theta qx + \{ \hat{h} \}_s) v + \beta t \delta_\epsilon(u - xv) \Big] \Bigg) \Biggl\rangle_x\ ,
\end{align}
where, as in \eqref{eq:regulariser}, $\delta_\epsilon$ is a smooth regulariser of the delta function and the $\epsilon\to0$ limit is taken in the calculation. Thus we see the only difference when compared to the eigenvector component density is that the delta will force $v \to \frac{u}{x}$ when we perform the $\mathrm{d}v$ integral in the numerator. Thus we have, recalling the scaling of the $\delta$ function, that the average overlap density is given by:
\begin{equation}
    \rho_{\mathrm{ov}}(u) = \sum_{s=0}^{s_{\mathrm{max}}} p_s(s) \int \{ \mathrm{d}\hat{\pi} \}_s \left\langle \delta \left( u - \frac{\theta q x^2 + x \{ \hat{h} \}_s}{\lambda_{\theta} - \{ \hat{\omega} \}_s} \right) \right\rangle_x\ .
\end{equation}
Again, we insert \eqref{pi hat at the saddle point} into this expression to obtain eventually
\begin{equation}
    \rho_{\mathrm{ov}}(u) = \sum_{s=0}^{s_{\mathrm{max}}} p_s(s) \int \{ \mathrm{d}\rho \}_s \left\langle \delta \left( u - \frac{\theta  qx^2 + x \left\{ \frac{\mu W}{\sigma} \right\}_s}{\lambda_\theta - \left\{ \frac{W^2}{\sigma} \right\}_s} \right) \right\rangle_{x,\{W\}_s}\ .
\end{equation}
Following \eqref{eq:overlap density initial}, we can thus calculate the average overlap between $\bm v_{{\mathrm{top}}}$ and the signal vector $\bm{x}$:
\begin{align}
    \lim_{N \to \infty} \left\langle \frac{\langle \bm{x}, \bm{v}_{\mathrm{top}} \rangle}{N} \right\rangle_{\bm{A}} 
    = \int \mathrm{d}u \, \rho_{\mathrm{ov}}(u) u = \sum_{s=0}^{s_{\mathrm{max}}} p_s(s) \int \{ \mathrm{d}\rho \}_s \left\langle  \frac{\theta  qx^2 + x \left\{ \frac{\mu W}{\sigma} \right\}_s}{\lambda_\theta - \left\{ \frac{W^2}{\sigma} \right\}_s} \right\rangle_{x,\{W\}_s} = q\ , \label{eq:q is overlap}
\end{align}
where we identified the saddle-point equation \eqref{q saddle point condition pi} for $q$ in the last line. Thus, just as the parameter $\lambda_{\theta}$ corresponds to $\langle\lambda_1\rangle_{\bm{A}}$, we can identify the parameter $q$ as being the average overlap between $ \bm v_{{\mathrm{top}}}$ and $\bm{x}$.  Finally, let us assume that the spike is sampled from a distribution such that $\langle x \rangle_x = 0 \quad$ and $\langle x^2 \rangle_x = \sigma_x^2 < \infty$. In this case, Eq. \eqref{eq:q is overlap} simplifies to
\begin{equation} \label{eq:critical threshold inter 1}
    q (1 - \theta \sigma_x^2 Q(\lambda_\theta)) = 0\ ,
\end{equation}
where
\begin{equation}\label{eq:Q_defn}
    Q(\lambda) = \sum_{s=0}^{s_{\mathrm{max}}} p_s(s) \int \{ \mathrm{d}\hat{\pi} \}_s \frac{1}{\lambda - \{ \hat{\omega} \}_s}\ .
\end{equation}
Eq. \eqref{eq:critical threshold inter 1} has two solutions: either $q=0$ (which happens for $\theta \leq \theta_{\mathrm{crit}}$, where $\langle \lambda_{1} \rangle_{\bm{A}} = \lambda_{\theta=0}$, and $\langle \frac{1}{N} \langle \bm{x}, \bm{v}_{\mathrm{top}} \rangle \rangle_{\bm{A}} = 0$) or $\theta \sigma_x^2 Q(\lambda_\theta) = 1$, which corresponds to the $\theta > \theta_{\mathrm{crit}}$ case. The $\theta_{\mathrm{crit}}$ value is therefore determined by the ``marginal'' non-trivial solution at the onset of the transition, i.e. for $\lambda_\theta=\lambda_{\theta=0}$. This gives condition \eqref{eq:recovery threshold}
\begin{equation}\label{eq:theta_crit}
    \theta_{\mathrm{crit}} = \frac{1}{\sigma_x^2 Q(\lambda_{\theta=0})}\ .
\end{equation}

\section{Special degree distribution cases}\label{app:sec:special_cases}
In this section, we consider two special degree distribution cases to obtain more explicit equations for the key results. In particular, we will derive expressions for the recovery threshold value in the signal strength $\theta$, as well as the typical values of the top eigenvalue and squared overlap in the recovery phase $\theta>\theta_{\rm crit}$. We focus on the cases where the marginal distributions $p_k(k)$ and $p_s(s)$ are truncated Poisson (in \ref{app:sec:poisson}) and random biregular (in \ref{app:sec:random_regular}), with means $\langle k\rangle$ and $\langle s\rangle$ respectively, related through $\langle k\rangle=\frac{1}{\alpha^2}\langle s\rangle$ as per the handshake constraint \eqref{handshake constraint math}. In the following, we assume the spike components to be zero-centered $\langle x\rangle_x=0$ with finite variance $\langle x^2\rangle_x=\sigma_x^2<\infty$. 
The first step towards a more explicit expression is to define the following inverse moments
\begin{align}\label{app:eq:m_v_cavityfn}
m_{\omega,\tau}(\lambda_{\theta}) &:= \int \mathrm{d}\pi(\omega,h)\, \omega^{-\tau}  , 
\end{align}
and
\begin{align}\label{app:eq:m_u_cavityfn}
m_{\sigma,\tau}(\lambda_{\theta}) &:= \int \mathrm{d}\rho(\sigma,\mu)\, \sigma^{-\tau}  , 
\end{align}
with $m_{\sigma,0} = m_{\omega,0} = 1$.
Before specifying to particular cases, it is useful to express the relationship between them. 

We insert the definition of $\rho$ in \eqref{eq:rho recursive} into $m_{\sigma,\tau}$ defined in \eqref{app:eq:m_u_cavityfn} to write
\begin{align}
m_{\sigma,\tau}(\lambda_{\theta}) &= \sum_{k=1}^{k_{\rm max}} \frac{k p_k(k)}{\langle k \rangle} \int \prod_{\ell=1}^{k-1}\left[\mathrm{d}\pi(\omega_\ell,h_\ell)\,\mathrm{d}\varrho_W(W_\ell)\right]\frac{1}{(1-\sum_{\ell=1}^{k-1}W_\ell^2/\omega_\ell)^{\tau}} \ .
\end{align}
Similarly, we use the RDE for $\pi$ (defined in \eqref{eq:pi recursive}) to write  $m_{\omega,\tau}$ (defined in \eqref{app:eq:m_v_cavityfn}) as
\begin{align}
m_{\omega,\tau} (\lambda_{\theta}) &= \sum_{s=1}^{s_{\rm max}} \frac{s p_s(s)}{\langle s \rangle} \int \prod_{\ell=1}^{s-1}\left[\mathrm{d}\rho(\sigma_\ell,\mu_\ell)\,\mathrm{d}\varrho_W(W_\ell)\right]\frac{1}{(\lambda_\theta - \sum_{\ell=1}^{s-1}W_\ell^2/\sigma_\ell)^{\tau}}  \ .
\end{align}
For populations where $\pi$ and $\rho$ are non-degenerate neither of the two expressions above is closed, as the denominators couple $m_{\sigma,\tau}$ and $m_{\omega,\tau}$ to the inverse moments of all orders $\tau \ge 1$. 

Denote the sums by
\begin{align}\label{app:eq:zy}
z_k := \sum_{\ell=1}^{k-1}\frac{W_\ell^2}{\omega_\ell} \ , \qquad
y_s := \sum_{\ell=1}^{s-1}\frac{W_\ell^2}{\sigma_\ell} \ ,
\end{align}
so that $\sigma = 1 - z_k$ and $\omega = \lambda_{\theta} - y_s$ by
\eqref{eq:rho recursive} and \eqref{eq:pi recursive}.
Since the measure factorises over $\ell$, we can evaluate any integer power of $z_k$ by the multinomial theorem
\begin{align}\label{app:eq:multinomial}
\int \prod_{\ell=1}^{k-1}\left[\mathrm{d}\pi(\omega_\ell,h_\ell)\,
\mathrm{d}\varrho_W(W_\ell)\right] z_k^{\,p}
= \sum_{\substack{r_1+\dots+r_{k-1}=p\\ r_\ell \ge 0}}
\frac{p!}{r_1!\cdots r_{k-1}!}
\prod_{\ell=1}^{k-1}\left[\langle W^{2r_\ell}\rangle_W\,
m_{\omega,r_\ell}\right] \ ,
\end{align}
where $\langle W^{2r}\rangle_W := \int \mathrm{d}\varrho_W(W)\,W^{2r}$, and where $r_\ell = 0$ contribute a factor one by the convention $m_{\omega,0} = 1$.
For $|z_k| < 1$ we have
\begin{align}\label{app:eq:geometric}
\sigma^{-\tau} = (1-z_k)^{-\tau}
= \sum_{p=0}^{\infty} \binom{\tau+p-1}{p} z_k^{\,p} \ ,
\end{align}
and we truncate this series at order $p = P$.
Together with
\eqref{app:eq:multinomial} this gives, for every $\tau = 1,\dots,P$, the following approximation for \eqref{app:eq:m_u_cavityfn}
\begin{align}\label{app:eq:hierarchy_sigma}
m_{\sigma,\tau}^{(P)}(\lambda_{\theta})
= \sum_{k=1}^{k_{\rm max}} \frac{k p_k(k)}{\langle k \rangle}
\sum_{p=0}^{P} \binom{\tau+p-1}{p}
\sum_{\substack{r_1+\dots+r_{k-1}=p\\ r_\ell \ge 0}}
\frac{p!}{r_1!\cdots r_{k-1}!}
\prod_{\ell=1}^{k-1}\left[\langle W^{2r_\ell}\rangle_W\,
m_{\omega,r_\ell}^{(P)}\right] \ .
\end{align}
The approximated equation for $m^{(P)}_{\omega,\tau}$ of \eqref{app:eq:m_v_cavityfn} is obtained in the same way, once the
explicit dependence on $\lambda_{\theta}$ has been isolated by writing
$\omega^{-\tau} = \lambda_{\theta}^{-\tau}(1-y_s/\lambda_{\theta})^{-\tau}$,
\begin{align}\label{app:eq:hierarchy_omega}
m_{\omega,\tau}^{(P)}(\lambda_{\theta})
= \sum_{s=1}^{s_{\rm max}} \frac{s p_s(s)}{\langle s \rangle}
\sum_{p=0}^{P} \binom{\tau+p-1}{p}
\frac{1}{\lambda_{\theta}^{\,\tau+p}}
\sum_{\substack{r_1+\dots+r_{s-1}=p\\ r_\ell \ge 0}}
\frac{p!}{r_1!\cdots r_{s-1}!}
\prod_{\ell=1}^{s-1}\left[\langle W^{2r_\ell}\rangle_W\,
m_{\sigma,r_\ell}^{(P)}\right] \ .
\end{align}
We have thus defined a system of $2P$ coupled equations with $2P$ unknowns, namely $m_{\omega,1},\dots,m_{\omega,P}$ and $m_{\sigma,1},\dots,m_{\sigma,P}$.

\subsection{Truncated Poisson}\label{app:sec:poisson}
In this section, we specialise to the truncated Poisson degree distribution, namely
\begin{align}
p_s^{(\rm Poi)}(s) &= \frac{\bar c^s}{\Gamma_s s!} \ , \quad s = 0,\ldots,s_{\max} \ ,
\end{align}
where the normalisation factor is $\Gamma_s = \sum_{s=0}^{s_{\max}}\frac{\bar c^s}{s!}$ and the average degree is $\langle s \rangle = \sum_{s=0}^{s_{\max}} s p_s^{(\rm Poi)}(s)$. We also specialise the following
\begin{align}\label{app:eq:rs_poisson}
\frac{s p_s^{(\rm Poi)}(s)}{\langle s \rangle} &= \frac{\bar c}{\langle s \rangle}\frac{\bar c^{s-1}}{\Gamma_s (s-1)!} \ , \quad s = 1,\ldots,s_{\max} \ ,
\end{align}
where $\bar c$ satisfies
\[ \langle s\rangle = \frac{
\sum_{s=0}^{s_{\max}} s\,\dfrac{\bar c^s}{s!} }{ \sum_{s=0}^{s_{\max}} \dfrac{\bar c^s}{s!}
}\ . \]
Starting from the expression in \eqref{eq:Big Q definition}, namely 
\begin{equation} \label{Q unspecified}
    Q(\lambda_\theta) = \sum_{s=0}^{s_{\mathrm{max}}} p_{s}^{(\rm Poi)}(s) \int \{ \mathrm{d}\rho \}_s \left\langle\frac{1}{\lambda_\theta - \left\{ W^2/\sigma \right\}_s}\right\rangle_{\{W\}_s} \ ,
\end{equation}
the goal is to express the right-hand side in terms of $m_\omega$ and $m_\sigma$, defined in \eqref{app:eq:m_v_cavityfn} and \eqref{app:eq:m_u_cavityfn}, respectively.

Multiplying the RDE for $\pi$ in \eqref{eq:pi recursive} on both sides by $1/\omega$ and integrating, we denote by $\tilde{Q}(\lambda_\theta)$ the function $Q(\lambda_\theta)$ given by \eqref{Q unspecified} specialised to the Poissonian degree distribution and obtain 

\begin{align}
m_{\omega,1}(\lambda_{\theta}) &= \sum_{s=1}^{s_{\max}} \frac{\bar c}{\langle s \rangle}\frac{\bar c^{s-1}}{\Gamma_s (s-1)!} \int \{ \mathrm{d}\rho \}_{s-1} \left\langle \frac{1}{\lambda_{\theta}-\{ W^2/\sigma \}_{s-1}} \right\rangle_{\{W\}_{s-1}} \\ 
&= \frac{\bar c}{\langle s \rangle} \sum_{s=0}^{s_{\max}-1} \frac{\bar c^{s}}{\Gamma_s s!} \int \{ \mathrm{d}\rho \}_{s} \left\langle \frac{1}{\lambda_{\theta}-\{ W^2/\sigma \}_{s}} \right\rangle_{\{W\}_{s}} \\
&= \frac{\bar c}{\langle s \rangle} [\tilde Q(\lambda_{\theta}) - R(\lambda_{\theta},\theta, s_{\rm max})]\ ,
\end{align}
where we have performed the shift \(s\to s-1\), and introduced the final remainder term corresponding to $s_{\rm max}$ as 
\begin{align}
R(\lambda_{\theta},\theta,s_{\rm max}) &= \frac{\bar c^{s_{\max}}}{\Gamma_s s_{\max}!} \int \{ \mathrm{d}\rho \}_{s_{\max}} \left\langle \frac{1}{\lambda_{\theta}-\{ W^2/\sigma \}_{s_{\max}}} \right\rangle_{\{W\}_{s_{\rm max}}} \\
&= \frac{\bar c^{s_{\max}}}{\Gamma_s s_{\max}!}
\sum_{p=0}^{\infty} \frac{1}{\lambda_{\theta}^{\,1+p}}
\sum_{\substack{r_1+\dots+r_{s_{\max}}=p\\ r_\ell \ge 0}}
\frac{p!}{r_1!\cdots r_{s_{\max}}!}
\prod_{\ell=1}^{s_{\max}}\left[\langle W^{2r_\ell}\rangle_W\, m_{\sigma,r_\ell}\right] \\
&\simeq \frac{\bar c^{s_{\max}}}{\Gamma_s s_{\max}!}
\sum_{p=0}^{P} \frac{1}{\lambda_{\theta}^{\,1+p}}
\sum_{\substack{r_1+\dots+r_{s_{\max}}=p\\ r_\ell \ge 0}}
\frac{p!}{r_1!\cdots r_{s_{\max}}!}
\prod_{\ell=1}^{s_{\max}}\left[\langle W^{2r_\ell}\rangle_W\, m_{\sigma,r_\ell}^{(P)}\right] \\
&=: R^{(P)}(\lambda_{\theta},s_{\rm max}) \ ,
\end{align}
Truncating the above expression for $m_{\omega,1}^{(P)}(\lambda_{\theta})$, we obtain
\begin{align}\label{app:eq:QW}
\tilde Q(\lambda_{\theta}) \simeq \frac{\langle s \rangle}{\bar c}m_{\omega,1}^{(P)}(\lambda_{\theta})+ R^{(P)}(\lambda_{\theta},s_{\rm max}) =: \tilde Q^{(P)}(\lambda_{\theta}) \ .
\end{align}
As $P\to\infty$, we have that $\tilde Q^{(P)}(\lambda_{\theta}) \to \tilde Q(\lambda_{\theta})$. Moreover, in the non-truncated Poisson limit \(s_{\max}\to\infty\), \(\bar c \to \langle s \rangle \), and we recover $\tilde Q^{(P)}(\lambda_{\theta}) \to m_{\omega,1}^{(P)}(\lambda_{\theta})$. 
\subsubsection*{Transition value, top eigenvalue and squared overlap.}
We now show that the expression \(\tilde Q(\lambda_{\theta})\) in \eqref{app:eq:QW} is sufficient to determine the transition value, the typical top eigenvalue and the squared overlap between the signal and the \textit{top} eigenvector. 
We then use \eqref{eq:theta_crit} to write down where the transition between non-recovery and partial recovery via the top eigenvector occurs at
\begin{align} \label{theta crit poisson for invert}
\theta_{\rm crit} &\simeq \frac{1}{\sigma_x^2 \tilde Q^{(P)}(\lambda_{\theta=0})} \ ,
\end{align}
which is exact as $P\to\infty$.
Moreover, as \(\theta>\theta_{\rm crit}\), the signal-associated eigenvalue is the top eigenvalue and is computed by  rearranging \eqref{theta crit poisson for invert} and using the inverse function of $\tilde Q^{(P)}$ which we denote by $(\tilde Q^{(P)})^{-1}$. This gives the following expression for the signal-associated eigenvalue
\begin{align}
\langle\lambda_{1}\rangle_{\bm{A}} &= \lambda_\theta \simeq (\tilde Q^{(P)})^{-1}\left(\frac{1}{\theta \sigma^2_x}\right)  \ .
\end{align}
\begin{figure}
    \centering
    \includegraphics[width=0.5\linewidth]{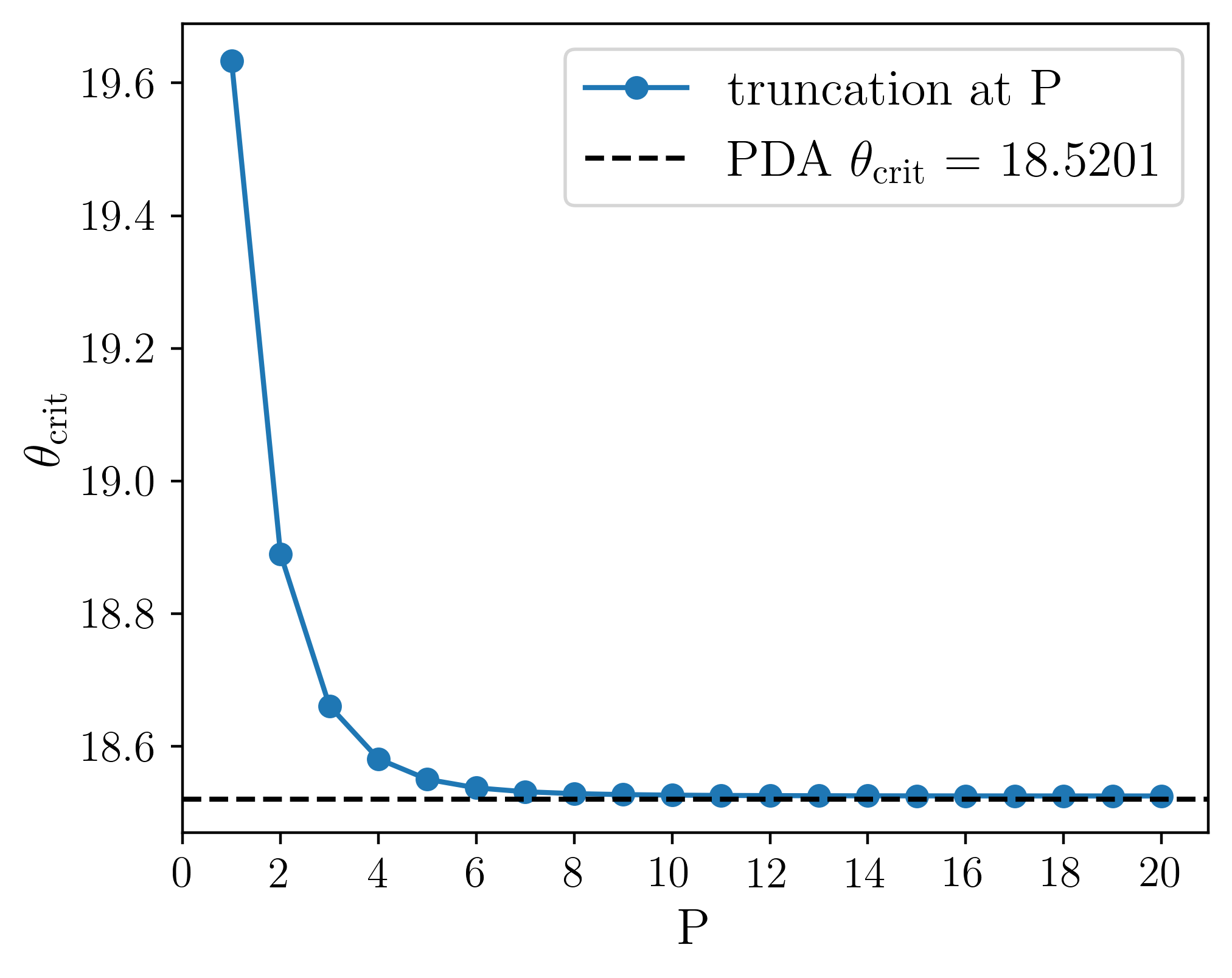}
    \caption{The truncation estimate of $\theta_{\rm crit}$ as the truncation order $P$ grows converges to the value obtained via PDA, for truncated Poisson with $\langle s \rangle=3, \langle k \rangle=4$ and $k_{\rm max}=s_{\rm max}=20$.}
    \label{app:fig:trunc}
\end{figure}

To compute the squared overlap, we recall that in section \ref{sec:The Model} we showed that
\begin{equation}
    \left\langle \frac{\langle\bm x,\bm v_{\mathrm{top}}\rangle^2}{N^2} \right\rangle_{\bm A} =\frac{\mathrm{d}\langle\lambda_1\rangle_{\bm{A}}}{\mathrm{d}\theta} \ .
\end{equation}
Differentiating the implicit equation $\theta\sigma_x^2 \tilde Q(\lambda_\theta) = 1$ with respect to \(\theta\), we obtain
\begin{align}
\sigma_x^2 \tilde Q(\lambda_\theta)+\theta\sigma_x^2 \tilde Q'(\lambda_\theta)\frac{\rm d \lambda_\theta}{\rm d \theta} &= 0 \ .
\end{align}
Therefore, we obtain the final expression in \eqref{eq:Poisson_overlap} for $\theta>\theta_{\rm crit}$
\begin{align}
\lim_{N \to \infty} \left\langle\frac{\langle \bm x,\bm v_{\rm top}\rangle^2}{N^2}\right\rangle_{\bm A, \bm x} &= \frac{\rm d \lambda_\theta}{\rm d \theta} = -\frac{\tilde Q(\lambda_\theta)}{\theta \tilde Q'(\lambda_\theta)} = -\frac{1}{\theta^2\sigma_x^2 \tilde Q'(\lambda_\theta)} \simeq -\frac{1}{\theta^2\sigma_x^2 \tilde Q{^{(P)}}'(\lambda_\theta) } \ ,
\end{align}
where we have used $\tilde Q(\lambda_\theta) = \frac{1}{\theta\sigma_x^2}$ in the last step.
Finally, as \(\theta\leq \theta_{\rm crit}\), the typical top eigenvalue is the structural noise eigenvalue and the typical squared overlap between the top eigenvector and the signal is zero
\begin{align}
\langle\lambda_{1}\rangle_{\bm{A}} &= \lambda_{\theta=0} \ , \qquad \lim_{N \to \infty} \left\langle\frac{\langle \bm x,\bm v_{\rm top}\rangle^2}{N^2}\right\rangle_{\bm{A}} = 0 \ .
\end{align}

\subsection{Random Biregular}\label{app:sec:random_regular}
We now consider the case for $p_s^{(R)}(s) = \delta_{s,\langle s \rangle}$ and $p_{k}^{(R)}(k) = \delta_{k,\langle k \rangle}$, where $\langle s \rangle = \alpha^2 \langle k \rangle$. For simplicity, we set $W_{ij}=1$. The saddle point equations, upon insertion of the `hat' into the `non-hat' densities, acquire the simplified form
\begin{align} \label{eq:main_saddle_point_RR_short}
    \pi(\omega, h) &= \int \{ \mathrm{d}\rho \}_{\langle s \rangle-1} \left\langle \delta\left( \omega - \left(\lambda_{\theta} - \left\{ \frac{1}{\sigma} \right\}_{\langle s \rangle-1}\right) \right) \delta\left( h - \left(\theta qx + \left\{ \frac{\mu }{\sigma} \right\}_{\langle s \rangle-1}\right) \right) \right\rangle_x \\
    \rho(\sigma, \mu) &= \int \left\{ \mathrm{d}\pi \right\}_{\langle k \rangle-1} \delta\left( \sigma - \left(1 - \left\{ \frac{1}{\omega} \right\}_{\langle k \rangle-1}\right) \right) \delta\left( \mu - \left\{ \frac{h}{\omega} \right\}_{\langle k \rangle-1} \right) \ .
\end{align}

Using the following ans\"atze
\begin{align}
    \pi(\omega,h)
    &= \delta(\omega-\bar{\omega})
    \left\langle \pi_h(h\mid\bm{x})\right\rangle_{\bm{x}}\ ,
    \tag{B.29}\\
    \rho(\sigma,\mu)
    &= \delta(\sigma-\bar{\sigma})\,\rho_\mu(\mu)\ ,
    \tag{B.30}
\end{align}

Eq. \eqref{eq:Q_defn} specialises to
\begin{equation}
Q^{RR}(\lambda_\theta) = \sum_{s=0}^{s_{\mathrm{max}}} p_{s}^{(R)}(s) \int \{ \mathrm{d}\hat{\pi} \}_s \frac{1}{\lambda_\theta - \{ \hat{\omega} \}_s} = \int \{ \mathrm{d} \rho \}_{\langle s \rangle} \frac{1}{\lambda_\theta - \{ 1/\sigma \}_{\langle s \rangle}}  = \left( \lambda_\theta - \frac{\langle s \rangle}{\bar \sigma} \right)^{-1} \ , 
\end{equation}
using the definition of $\hat\pi(\hat\omega,\hat h)$ given by \eqref{pi hat at the saddle point} in the second step, and the above ansatz for $\rho$ in the third step. Note that only the $(\omega,\sigma)$ pair is required, since $Q$ depends on
$\hat{\omega}$ alone and is independent of the field distributions
$\pi_h$ and $\rho_\mu$. Moreover, no truncation is required because the distribution of $\sigma$
is a point mass and we use a constant bond weight $W_{ij}=1$.

Inserting $\bar \omega = \lambda_{\theta} - (\langle s \rangle-1)/\bar \sigma$ into $\bar \sigma = 1 - (\langle k \rangle-1) / \bar \omega$, we obtain
\begin{equation}\label{eq:RR:sigma}
\bar \sigma = \frac{\langle s \rangle-\langle k \rangle+\lambda_{\theta} + \sqrt{(\langle k \rangle-\langle s \rangle-\lambda_{\theta})^2 - 4\lambda_{\theta}(\langle s \rangle-1)}}{2\lambda_{\theta}} \ ,
\end{equation}
where we choose the physical $\bar \sigma>0$ branch because $\bar \sigma$ is an inverse variance in the Gaussian replica ansatz, and we have $\bar \sigma \to 1$ as $\lambda_\theta \to \infty$.
We thus have an explicit expression
\begin{equation}
Q^{RR}(\lambda_{\theta}) = \left( \lambda_{\theta} - \frac{2\lambda_{\theta} \langle s \rangle}{\langle s \rangle-\langle k \rangle+\lambda_{\theta} + \sqrt{(\langle k \rangle-\langle s \rangle-\lambda_{\theta})^2 - 4\lambda_{\theta}(\langle s \rangle-1)}} \right)^{-1} \ .
\end{equation}
As a result, we can obtain the relevant observables explicitly in closed form: the critical threshold $\theta_{\rm crit}$, and both the typical top eigenvalue and squared overlap in the recovery phase as $\theta>\theta_{\rm crit}$. 
First, the critical signal strength is obtained as
\begin{equation}
\theta_{\mathrm{crit}} = \frac{1}{\sigma_x^2 Q^{RR}(\lambda_{\theta=0})} = \frac{1}{\sigma_x^2 Q^{RR}(\langle s \rangle \langle k \rangle)}= \frac{\langle s \rangle \big(\langle s \rangle \langle k \rangle-\langle s \rangle-\langle k \rangle \big)}{\sigma_x^2 \big(\langle s \rangle-1\big)} \ ,
\end{equation}
where we have used the value $\lambda_{\theta=0}=\langle s \rangle \langle k \rangle$ of the noise outlier eigenvalue. In the random biregular case every row of $\bm X$ as defined in \eqref{X definition} sums to $\langle k\rangle$ and every column to $\langle s\rangle$, so we have that $\bm J \bm 1_N = \bm X^\top \left( \bm X\mathbf{1}_N \right) = \langle k\rangle\, \bm X^\top \mathbf{1}_M = \langle s \rangle \langle k \rangle \bm 1_N$, where $\bm 1_N = (1,\dots,1)^{\intercal}\in \mathbb R^N$. Since $\bm J$ is entry-wise non-negative, we have by Perron-Frobenius that $\lambda_{\theta=0}=\langle s \rangle \langle k \rangle$, for more details see Chapter 2 of Ref.~\cite{GraphSpectraBook2012}. 

Second, the typical top eigenvalue in the recovery phase as $\theta>\theta_{\mathrm{crit}}$ is explicitly determined using $\theta \sigma_x^2 Q^{RR}(\lambda_\theta)=1$, that is $\theta \sigma_x^2 = \lambda_\theta - \frac{\langle s \rangle}{\bar \sigma}$, where $\bar \sigma>0$ is defined in \eqref{eq:RR:sigma} (positive branch). Rearranging, it reads
\begin{equation}
\langle\lambda_{1}\rangle_{\bm A} = \lambda_\theta = \frac{1}{2} \left( \langle s \rangle \langle k \rangle + (2-\langle s \rangle)\,\theta \sigma_x^2
+\langle s \rangle \sqrt{(\langle k \rangle -\theta\sigma_x^2)^2 + 4\theta\sigma_x^2}
 \right) \ . 
\end{equation}
Setting $\theta=0$ we indeed recover the top outlier eigenvalue of the noise matrix $\bm J$ of this particular case, $\lambda_{\theta=0}=\langle s \rangle \langle k \rangle$.
Finally, we obtain the typical squared overlap via
\begin{equation}
 \lim_{N \to \infty} \left\langle\frac{\langle \bm x,\bm v_{\rm top}\rangle^2}{N^2}\right\rangle = \frac{\mathrm{d} \lambda_\theta}{\mathrm{d} \theta} = \frac{\sigma_x^2}{2}
\left[2-\langle s \rangle+\langle s \rangle \ \frac{\theta \sigma_x^2 + 2 - \langle k \rangle} {\sqrt{(\langle k \rangle - \theta\sigma_x^2)^2 + 4\theta\sigma_x^2}} \right].
\end{equation}
Moreover, the right bulk edge of the spectrum is 
\begin{equation} \label{bulk lambda_b random regular}
    \lambda_{\rm b}=\left(\sqrt{\langle s \rangle -1}+\sqrt{\langle k \rangle -1}\right)^2 \ ,
\end{equation}
which is known and derived independently in the mathematical literature, for $\langle s \rangle, \langle k \rangle \geq 2$. Namely, for the random biregular case with $W=1$, $\bm X$ defined in \eqref{X definition} is precisely the biadjacency matrix of a $(\langle k \rangle, \langle s \rangle)$-biregular bipartite graph with $M$ row nodes with degree $
\langle k \rangle$, and $N$ column nodes with degree $\langle s \rangle$.
The full adjacency matrix of this bipartite graph is
$$\bm B = \begin{pmatrix} \bm 0 & \bm X\\ \bm X^\top & \bm 0\end{pmatrix} \ , \qquad \bm B^2 = \begin{pmatrix} \bm X\bm X^\top & \bm 0\\ \bm 0 & \bm X^\top\bm X\end{pmatrix} \ .$$
The non-zero eigenvalues of $\bm J=\bm X^\top\bm X$ are the squared singular values of $\bm X$, equivalently the squares of the positive eigenvalues of the bipartite matrix $\bm B$, whose non-zero spectrum occurs in $\pm$ pairs. Moreover, $\bm J$ has at least $\max(0,N-M)$ zero eigenvalues, with equality when $\bm X$ has full rank.

The trivial eigenvalues $\pm\sqrt{\langle s\rangle\langle k\rangle}$ of a biregular
adjacency matrix square to the noise outlier $\lambda_{\theta=0}=\langle s\rangle\langle k\rangle$, while the edge \eqref{bulk lambda_b random regular} of the non-trivial spectrum is inherited from the infinite $(\langle k\rangle,\langle s\rangle)$-biregular tree (the local limit of the graph) and was first computed by Godsil and Mohar~\cite{GODSIL1988191}. Evaluating at \(\lambda_{\rm b}\) and rearranging gives us the transition value
\[
\theta_{\mathrm b} =  \frac{1}{\sigma_x^2 Q^{RR}(\lambda_{\rm b})} = 
\frac{\langle k \rangle-2+ \sqrt{(\langle s \rangle-1)(\langle k \rangle-1)}-\sqrt{\frac{\langle k \rangle-1}{\langle s \rangle-1}}}{\sigma_x^2} \ .
\]

\section{The Dense Limit} \label{sec:dense limit}

We now take the degree distributions $p_s(s)$ and $p_k(k)$ to be Poissonian
and consider the high-connectivity limit
$\langle s\rangle,\langle k\rangle\to\infty$ at fixed
\begin{equation}
    \langle k\rangle=\alpha^{-2}\langle s\rangle \ ,
\end{equation}
as imposed by \eqref{handshake constraint math}. More precisely, we consider
the high-connectivity limit of the thermodynamic saddle-point equations
derived above. With the centered connectivity-dependent scaling introduced
below, this limit recovers the additive BBP-type transition for dense
Wishart noise.

We first insert \eqref{pi hat at the saddle point} into
\eqref{pi at the saddle point}, and \eqref{rho hat at the saddle point}
into \eqref{rho at the saddle point}, obtaining
\begin{equation}\label{eq: pi in terms of rho}
    \pi(\omega,h)
    =
    \sum_{s=1}^{\infty}
    \frac{s p_s(s)}{\langle s\rangle}
    \int \{\mathrm{d}\rho\}_{s-1}
    \left\langle
    \delta\left(
        \omega-
        \left[
            \lambda_\theta-
            \left\{\frac{W^2}{\sigma}\right\}_{s-1}
        \right]
    \right)
    \delta\left(
        h-
        \left[
            \theta qx+
            \left\{\frac{\mu W}{\sigma}\right\}_{s-1}
        \right]
    \right)
    \right\rangle_{x,\{W\}_{s-1}},
\end{equation}
and
\begin{equation}\label{eq:rho in terms of pi}
    \rho(\sigma,\mu)
    =
    \sum_{k=1}^{\infty}
    \frac{k p_k(k)}{\langle k\rangle}
    \int \{\mathrm{d}\pi\}_{k-1}
    \left\langle
    \delta\left(
        \sigma-
        \left[
            1-
            \left\{\frac{W^2}{\omega}\right\}_{k-1}
        \right]
    \right)
    \delta\left(
        \mu-
        \left\{\frac{Wh}{\omega}\right\}_{k-1}
    \right)
    \right\rangle_{\{W\}_{k-1}}.
\end{equation}

To obtain the standard dense Wishart limit, we now introduce a centered,
connectivity-dependent scaling of the bond weights, departing from the
nonzero-mean weights considered in the main body. Namely, we set
\begin{equation}\label{eq:dense:rescaling_weigths}
    W=\frac{Z}{\sqrt{\langle s\rangle}} \ ,
\end{equation}
where $Z$ is a random variable satisfying
\begin{equation}
    \langle Z\rangle_Z=0,
    \qquad
    \langle Z^2\rangle_Z=1.
\end{equation}
This scaling keeps the spectrum of the noise matrix of order one as the
connectivity diverges. In particular, $\langle W\rangle_W=0$.

For Poisson degree distributions, the size-biased laws appearing in
\eqref{eq: pi in terms of rho} and \eqref{eq:rho in terms of pi} satisfy
\begin{equation}
    s-1\sim\mathrm{Poisson}(\langle s\rangle),
    \qquad
    k-1\sim\mathrm{Poisson}(\langle k\rangle).
\end{equation}
Their relative fluctuations therefore vanish in the high-connectivity
limit. Using \eqref{eq:dense:rescaling_weigths}, the law of large numbers
gives
\begin{equation}
    \frac{1}{\langle s\rangle}
    \sum_{l=1}^{s-1}\frac{Z_l^2}{\sigma_l}
    \longrightarrow
    \int \mathrm{d}\sigma\,\mathrm{d}\mu\,
    \rho(\sigma,\mu)\,
    \frac{\langle Z^2\rangle_Z}{\sigma} \ ,
\end{equation}
and
\begin{equation}
    \frac{1}{\langle k\rangle}
    \sum_{l=1}^{k-1}\frac{Z_l^2}{\omega_l}
    \longrightarrow
    \int \mathrm{d}\omega\,\mathrm{d}h\,
    \pi(\omega,h)\,
    \frac{\langle Z^2\rangle_Z}{\omega} \ .
\end{equation}
Hence $\omega$ and $\sigma$ become non-fluctuating in the limit
$\langle s\rangle,\langle k\rangle\to\infty$, and the saddle-point
distributions take the form
\begin{equation}
    \pi(\omega,h)
    =
    \delta(\omega-\bar\omega)\pi_h(h) \ ,
\end{equation}
and
\begin{equation}
    \rho(\sigma,\mu)
    =
    \delta(\sigma-\bar\sigma)\rho_\mu(\mu) \ .
\end{equation}
The resulting self-consistency equations for the deterministic curvature
variables are
\begin{equation}\label{appc:eq:baromegaLLN}
    \bar\omega
    =
    \lambda_\theta
    -\frac{1}{\bar\sigma}\langle Z^2\rangle_Z
    =
    \lambda_\theta-\frac{1}{\bar\sigma} \ ,
\end{equation}
and
\begin{equation}\label{appc:eq:barsigmaLLN}
    \bar\sigma
    =
    1-\frac{1}{\alpha^2\bar\omega}\langle Z^2\rangle_Z
    =
    1-\frac{1}{\alpha^2\bar\omega} \ .
\end{equation}
Inserting \eqref{appc:eq:barsigmaLLN} into
\eqref{appc:eq:baromegaLLN} gives
\begin{equation}\label{eq:omega bar}
    \bar\omega
    =
    \lambda_\theta
    -
    \frac{\alpha^2\bar\omega}
         {\alpha^2\bar\omega-1} \ ,
\end{equation}
while the converse substitution gives
\begin{equation}\label{eq:sigma bar}
    \bar\sigma
    =
    1-
    \frac{\bar\sigma}
         {\alpha^2(\bar\sigma\lambda_\theta-1)} \ .
\end{equation}
The field variables satisfy
\begin{align}
    h
    &=
    \theta qx
    +
    \frac{1}{\bar\sigma\sqrt{\langle s\rangle}}
    \sum_{l=1}^{s-1}\mu_l Z_l ,
    \label{appc:eq:hdense_1}\\
    \mu
    &=
    \frac{1}{\alpha\bar\omega\sqrt{\langle k\rangle}}
    \sum_{l=1}^{k-1}h_l Z_l .
    \label{appc:eq:mudense_1}
\end{align}
Under the usual finite-moment conditions, the random sums in
\eqref{appc:eq:hdense_1}--\eqref{appc:eq:mudense_1} become Gaussian by the
central limit theorem. The term $\theta qx$ in
\eqref{appc:eq:hdense_1}, however, contains a single sample
$x\sim\varrho_x$ and is not averaged over. Consequently, the conditional
distribution of $h$ at fixed $x$ is Gaussian, whereas its marginal
distribution is generally a mixture of Gaussians. The dense-limit
saddle-point distributions therefore read
\begin{align}
    \pi(\omega,h)
    &=
    \delta(\omega-\bar\omega)\,\pi_h(h),
    &
    \pi_h(h)
    &=
    \int\mathrm{d}x\,\varrho_x(x)\,
    \phi\!\left(h;\theta qx,\Delta_h\right),
    \label{eq:C-pi-dense}\\
    \rho(\sigma,\mu)
    &=
    \delta(\sigma-\bar\sigma)\,
    \phi(\mu;0,\Delta_\mu),
    &&
    \label{eq:C-rho-dense}
\end{align}
where $\phi(z;m,v)$ denotes the Gaussian density with mean $m$ and variance
$v$. The corresponding Gaussian variances satisfy
\begin{equation}
    \Delta_h
    =
    \frac{\mathbb{E}[\mu^2]}{\bar\sigma^2},
    \qquad
    \Delta_\mu
    =
    \frac{\mathbb{E}[h^2]}{\alpha^2\bar\omega^2}\ .
\end{equation}
We next determine the recovery threshold. Recall from
\eqref{eq:theta_crit} that
\begin{equation}
    \theta_{\mathrm{crit}}
    =
    \frac{1}{\sigma_x^2 Q(\lambda_{\theta=0})} \ ,
\end{equation}
where
\begin{equation}
    Q(\lambda_{\theta=0})
    =
    \sum_{s=0}^{\infty}
    p_s(s)
    \int \{\mathrm{d}\rho\}_s
    \frac{1}{
        \lambda_{\theta=0}
        -
        \left\{\frac{W^2}{\sigma}\right\}_s
    } \ .
\end{equation}
Using the dense-limit concentration of $\sigma$ and the law of large
numbers, we obtain
\begin{equation}\label{eq:dense:intermediate_step1}
    \theta_{\mathrm{crit}}
    =
    \frac{1}{\sigma_x^2}
    \left(
        \lambda_{\theta=0}
        -
        \frac{1}{\bar\sigma}
    \right)\ .
\end{equation}
The field variable $\mu$ does not enter $Q$, so the corresponding
$\mathrm{d}\mu\,\phi(\mu;0,\Delta_\mu)$ integral contributes unity.

The null-model value $\lambda_{\theta=0}$ is the upper edge of the
Marchenko--Pastur spectrum for the present scaling. Although this value is
known directly from the Marchenko--Pastur law, it can also be recovered
from the saddle-point equations above. In the null problem, $q=0$, so
\eqref{eq:C-pi-dense}--\eqref{eq:C-rho-dense} imply
\begin{equation}
    \Delta_h=\mathbb{E}[h^2],
    \qquad
    \Delta_\mu=\mathbb{E}[\mu^2]\ .
\end{equation}
Equations \eqref{appc:eq:hdense_1}--\eqref{appc:eq:mudense_1} then give
\begin{equation}
    \Delta_h
    =
    \frac{\Delta_\mu}{\bar\sigma^2},
    \qquad
    \Delta_\mu
    =
    \frac{\Delta_h}{\alpha^2\bar\omega^2}\ .
\end{equation}
At the spectral edge, the zero-field solution becomes marginal. A
non-trivial solution for the second moments therefore requires
$\Delta_h\neq0$, yielding
\begin{equation}
    1
    =
    \frac{1}{
        \alpha^2\bar\omega^2\bar\sigma^2
    }\ .
\end{equation}
Choosing the physical positive branch gives
\begin{equation}
    \alpha\bar\omega\bar\sigma=1\ .
\end{equation}
Combining this condition with \eqref{appc:eq:barsigmaLLN} gives
\begin{equation}
    \bar\omega
    =
    \frac{1+\alpha}{\alpha^2},
    \qquad
    \bar\sigma
    =
    \frac{\alpha}{1+\alpha}\ .
\end{equation}
Finally, \eqref{appc:eq:baromegaLLN} gives
\begin{equation}
    \lambda
    =
    \bar\omega+\frac{1}{\bar\sigma}\ ,
\end{equation}
and hence the upper edge of the dense noise spectrum is
\begin{equation}
    \lambda_{\theta=0}
    =
    \frac{1+\alpha}{\alpha^2}
    +
    \frac{1+\alpha}{\alpha}
    =
    \frac{(1+\alpha)^2}{\alpha^2}\ .
\end{equation}
Substituting this result into
\eqref{eq:dense:intermediate_step1} yields
\begin{equation}
    \theta_{\rm crit}
    =
    \frac{1+\alpha}{\sigma_x^2\alpha^2}\ .
\end{equation}
This is the critical threshold for the additive rank-one deformation of
the dense Wishart (Marchenko--Pastur) ensemble, i.e.\ the corresponding
BBP-type transition for the present scaling \cite{Forrester2023}.

We now determine the typical signal-associated outlier eigenvalue and the
squared overlap with the signal in the recovery phase
$\theta>\theta_{\rm crit}$. For a zero-mean signal distribution and
$q\neq0$, we may divide \eqref{eq:q condition} by $q$. Using the
dense-limit form of $\rho(\sigma,\mu)$ gives
\begin{equation}
    1
    =
    \frac{\theta\sigma_x^2}{
        \lambda_\theta-\frac{1}{\bar\sigma}
    }\ .
\end{equation}
Equivalently,
\begin{equation}
    \bar\sigma(\lambda_\theta)
    =
    \frac{1}{
        \lambda_\theta-\theta\sigma_x^2
    }\ .
\end{equation}
Substituting this expression into \eqref{eq:sigma bar} and solving for
$\lambda_\theta$ gives the outlier eigenvalue in the recovery phase:
\begin{equation}
    \lambda_\theta
    =
    \theta\sigma_x^2
    \left(
        1+
        \frac{\alpha^2}{
            \alpha^2\theta\sigma_x^2-1
        }
    \right)\ .
\end{equation}
This recovers the outlier branch of the additive dense-Wishart BBP-type
transition for the present scaling \cite{Forrester2023}.

Finally, we compute the typical squared overlap between the top eigenvector
and the signal. For $\theta>\theta_{\rm crit}$, where the outlier
eigenvalue is simple, the relation \eqref{eq:HF} gives
\begin{equation}
    \lim_{N\to\infty}
    \left\langle
        \frac{
            \langle \bm{x},\bm{v}_{\mathrm{top}}\rangle^2
        }{N^2}
    \right\rangle_{\bm A}
    =
    \frac{\mathrm{d}\lambda_\theta}{\mathrm{d}\theta}
    =
    \sigma_x^2
    -
    \frac{
        \alpha^2\sigma_x^2
    }{
        \left(
            \alpha^2\sigma_x^2\theta-1
        \right)^2
    }\ .
\end{equation}
This overlap reflects the normalization used throughout this work,
$\|\bm v_{\mathrm{top}}\|^2=N$ and
$\|\bm x\|^2/N\to\sigma_x^2$, and is therefore not normalized to unity.
Indeed, the corresponding squared cosine overlap is
\begin{equation}
    \frac{
        \langle \bm x,\bm v_{\mathrm{top}}\rangle^2
    }{
        \|\bm x\|^2\|\bm v_{\mathrm{top}}\|^2
    }
    \xrightarrow[N\to\infty]{}
    \frac{1}{\sigma_x^2}
    \frac{
        \langle \bm x,\bm v_{\mathrm{top}}\rangle^2
    }{N^2}\ .
\end{equation}
Thus, dividing the overlap obtained above by $\sigma_x^2$ gives the
dimensionless squared cosine overlap, which lies between $0$ and $1$ and approaches $1$ for perfect signal recovery. This is the normalisation typically used when comparing sample and population eigenvectors in multiplicative spiked-covariance models \cite{Paul2007}.


\newpage


\begin{thebibliography}{100}
\bibitem{Jolliffe2002} Jolliffe, I. T. (2002). Principal Component Analysis. \textit{Springer}, https://doi.org/10.1007/b98835.
\bibitem{Candes2011}Candès, E. J., Li, X., Ma, Y., \& Wright, J. (2011). Robust Principal Component Analysis? \textit{J. ACM}, Volume \textbf{58}, Number 3, https://doi.org/10.1145/1970392.1970395.
\bibitem{Bishop2006}Bishop, C. M. (2006). Pattern recognition and machine learning. \textit{Springer}.
\bibitem{Montanari2021}Montanari, A., \& Venkataramanan, R. (2021). Estimation of Low-Rank Matrices via Approximate Message Passing. \textit{The Annals of Statistics}, Volume \textbf{49}, Number 1, Pages 321-345, https://doi.org/10.1214/20-AOS1958.
\bibitem{Liu2022a} Liu, J., Liu, R. W., Sun, J., \& Zeng, T. (2023). Rank-One Prior: Real-Time Scene Recovery. \textit{IEEE Transactions on Pattern Analysis and Machine Intelligence}, Volume \textbf{45}, Number 7, Pages 8845-8860, https://doi.org/10.1109/TPAMI.2022.3226276.
\bibitem{Liu2022b}Liu, T., Li, Y., Zhou, E., \& Zhao, T. (2022). Noise Regularizes Over-parameterized Rank One Matrix Recovery, Provably. Proceedings of the 25th International Conference on Artificial Intelligence and Statistics (AISTATS) 2022, PMLR: Volume \textbf{151}, Pages 2784-2802.
\bibitem{Coeurdoux2026} Coeurdoux, F., Ferré, G., \& Bouchaud, J.-P. (2026). Random Matrix Theory of Early-Stopped Gradient Flow: A Transient BBP Scenario. \textit{arXiv preprint}, arXiv:2604.18450.
\bibitem{Johnstone2001} Johnstone, I. M. (2001). On the distribution of the largest eigenvalue in principal components analysis. \textit{The Annals of Statistics}, Volume \textbf{29}, Number 2, Pages 295-327, https://doi.org/10.1214/aos/1009210544.
\bibitem{Baik2005} Baik, J., Ben Arous, G., \& Péché, S. (2005). Phase transition of the largest eigenvalue for nonnull complex sample covariance matrices. \textit{The Annals of Probability}, Volume \textbf{33}, Number 5, Pages 1643-1697, https://doi.org/10.1214/009117905000000233.
\bibitem{Forrester2023} Forrester, P. J. (2023). Rank 1 perturbations in random matrix theory---a review of exact results. \textit{Random Matrices: Theory and Applications}, Volume \textbf{12}, Number 4, 2330001, https://doi.org/10.1142/S2010326323300012.
\bibitem{Benaych-Georges2011} Benaych-Georges, F., \& Nadakuditi, R. R. (2011). The eigenvalues and eigenvectors of finite, low rank perturbations of large random matrices. \textit{Advances in Mathematics}, Volume \textbf{227}, Issue 1, Pages 494-521, https://doi.org/10.1016/j.aim.2011.02.007.
\bibitem{UrtePaper} Adomaityte, U., Sicuro, G., \& Vivo, P. (2025). PCA recovery thresholds in low-rank matrix inference with sparse noise. \textit{arXiv preprint}, arXiv:2511.11927.
\bibitem{Mantegna2012} Mantegna, R. N. (1999). Hierarchical structure in financial markets. \textit{The European Physical Journal B - Condensed Matter and Complex Systems}, Volume \textbf{11}, Pages 193-197, https://doi.org/10.1007/s100510050929.
\bibitem{Fan2016} Fan, J., Liao, Y., \& Liu, H. (2016). An overview of the estimation of large covariance and precision matrices. \textit{The Econometrics Journal}, Volume \textbf{19}, Number 1, Pages C1-C32, https://doi.org/10.1111/ectj.12061.
\bibitem{Laloux1999} Laloux, L., Cizeau, P., Bouchaud, J.-P., \& Potters, M. (1999). Noise Dressing of Financial Correlation Matrices. \textit{Physical Review Letters}, Volume \textbf{83}, Number 1467,  https://doi.org/10.1103/PhysRevLett.83.1467.
\bibitem{Kenett2010} Kenett, D. Y., Tumminello, M., Madi, A., Gur-Gershgoren, G., Mantegna, R. N., \& Ben-Jacob, E. (2010). Dominating Clasp of the Financial Sector Revealed by Partial Correlation Analysis of the Stock Market. \textit{PLoS ONE}, {\bf 5}(12): e15032, https://doi.org/10.1371/journal.pone.0015032.
\bibitem{Bouchaud2018} Bouchaud, J.-P., \& Potters, M. (2018). Financial applications of random matrix theory: a short review. \textit{The Oxford Handbook of Random Matrix Theory}, Pages 824-850, https://doi.org/10.1093/oxfordhb/9780198744191.013.40.
\bibitem{Borghesi2007} Borghesi, C., Marsili, M., \& Miccichè, S. (2007). Emergence of time-horizon invariant correlation structure in financial returns by subtraction of the market mode. \textit{Physical Review E}, Volume \textbf{76}, Number 026104, https://doi.org/10.1103/PhysRevE.76.026104.
\bibitem{Fan2013} Fan, J., Liao, Y., \& Mincheva, M. (2013). Large Covariance Estimation by Thresholding Principal Orthogonal Complements. \textit{Journal of the Royal Statistical Society Series B: Statistical Methodology}, Volume \textbf{75}, Issue 4, Pages 603-680, https://doi.org/10.1111/rssb.12016.
\bibitem{Deng2013} Deng, Y., Jiang, Y.-H., Yang, Y., He, Z., Luo, F. \& Zhou, J. (2012). Molecular ecological network analyses. \textit{BMC Bioinformatics}, Volume \textbf{13}, Number 113, https://doi.org/10.1186/1471-2105-13-113.
\bibitem{Kurtz2015} Kurtz, Z. D., Müller, C. L., Miraldi, E. R., Littman, D. R., Blaser, M. J., \& Bonneau, R. A. (2015). Sparse and Compositionally Robust Inference of Microbial Ecological Networks. \textit{PLoS Computational Biology}, {\bf 11}(5): e1004226,  https://doi.org/10.1371/journal.pcbi.1004226.
\bibitem{Tikhonov2017} Tikhonov, G., Abrego, N., Dunson, D., \& Ovaskainen, O. (2017). Using joint species distribution models for evaluating how species-to-species associations depend on the environmental context. \textit{Methods in Ecology and Evolution}, Volume \textbf{8}, Issue 4, Pages 443-452, https://doi.org/10.1111/2041-210X.12723.
\bibitem{BookParisi} Mézard, M., Parisi, G., \& Virasoro, M. (1987). Spin Glass Theory and Beyond. \textit{World Scientific Lecture Notes in Physics Series}, Volume \textbf{9}, https://doi.org/10.1142/0271.
\bibitem{Kuhn2024} Kühn, R. (2008). Spectra of sparse random matrices. \textit{Journal of Physics A: Mathematical and Theoretical}, Volume \textbf{41}(29), 295002, https://doi.org/10.1088/1751-8113/41/29/295002.
\bibitem{Nagao2007} Nagao, T., \& Tanaka, T. (2007). Spectral density of sparse sample covariance matrices. \textit{Journal of Physics A: Mathematical and Theoretical}, Volume \textbf{40}(19), 4973,  https://doi.org/10.1088/1751-8113/40/19/003.
\bibitem{Kabashima2012} Kabashima, Y., \& Takahashi, H. (2012). First eigenvalue/eigenvector in sparse random symmetric matrices: influences of degree fluctuation. \textit{Journal of Physics A: Mathematical and Theoretical}, Volume \textbf{45}(32), 325001,  https://doi.org/10.1088/1751-8113/45/32/325001.
\bibitem{Susca2019} Susca, V. A. R., Vivo, P., \& Kühn, R. (2019). Top eigenpair statistics for weighted sparse graphs. \textit{Journal of Physics A: Mathematical and Theoretical}, Volume \textbf{52}(48), 485002,  https://doi.org/10.1088/1751-8121/ab4d63.
\bibitem{Susca2021} Susca, V. A. R., Vivo, P., \& Kühn, R. (2021). Cavity and replica methods for the spectral density of sparse symmetric random matrices. \textit{SciPost Physics Lecture Notes}, Volume \textbf{33}, https://doi.org/10.21468/SciPostPhysLectNotes.33.
\bibitem{Budnick2025} Budnick, B., Forer, P., Vivo, P., Aufiero, S., Bartolucci, S., \& Caccioli, F. (2025). Top eigenpair statistics of diluted Wishart matrices. \textit{Journal of Physics A: Mathematical and Theoretical}, Volume \textbf{58}(32), 325001,  https://doi.org/10.1088/1751-8121/add821.
\bibitem{Decelle2011} Decelle, A., Krzakala, F., Moore, C., \& Zdeborová, L. (2011). Asymptotic analysis of the stochastic block model for modular networks and its algorithmic applications. \textit{Physical Review E}, Volume \textbf{84}, Number 066106, https://doi.org/10.1103/PhysRevE.84.066106.
\bibitem{Lelarge2018} Lelarge, M., \& Miolane, L. (2019). Fundamental limits of symmetric low-rank matrix estimation. \textit{Probability Theory and Related Fields}, Volume \textbf{173}, Pages 859-929, https://doi.org/10.1007/s00440-018-0845-x.
\bibitem{Abbe2018} Abbe, E. (2018). Community Detection and Stochastic Block Models: Recent Developments. \textit{Journal of Machine Learning Research}, Volume \textbf{18}, Number 177, Pages 1-86.
\bibitem{Rajan2006} Rajan, K., \& Abbott, L. F. (2006). Eigenvalue Spectra of Random Matrices for Neural Networks. \textit{Physical Review Letters}, Volume \textbf{97}, Number 188104, https://doi.org/10.1103/PhysRevLett.97.188104.
\bibitem{Mastrogiuseppe2018} Mastrogiuseppe, F., \& Ostojic, S. (2018). Linking Connectivity, Dynamics, and Computations in Low-Rank Recurrent Neural Networks. \textit{Neuron}, Volume \textbf{99}, Issue 3, Pages 609-623, https://doi.org/10.1016/j.neuron.2018.07.003.
\bibitem{Park2026} Park, C., Bocchi, D., D'Amico, F., Lucini, B., \& Aarts, G. (2026). Spectral phase transitions and trainability in neural network learning dynamics. \textit{arXiv preprint}, arXiv:2606.28486.
\bibitem{Krzakala2013} Krzakala, F., Moore, C., Mossel, E., Neeman, J., Sly, A., Zdeborová, L. \& Zhang, P. (2013). Spectral redemption in clustering sparse networks. \textit{Proceedings of the National Academy of Sciences}, Volume \textbf{110}, Number 52, Pages 20935-20940, https://doi.org/10.1073/pnas.1312486110.
\bibitem{Abbe2015}Abbe, E., \& Sandon, C. (2015). Community Detection in General Stochastic Block models: Fundamental Limits and Efficient Algorithms for Recovery. \textit{2015 IEEE 56th Annual Symposium on Foundations of Computer Science}, Pages 670-688, https://doi.org/10.1109/FOCS.2015.47.
\bibitem{Vreeswijk1998} van Vreeswijk, C., \& Sompolinsky, H. (1998). Chaotic Balanced State in a Model of Cortical Circuits. \textit{Neural Computation}, Volume \textbf{10}, Number 6, Pages 1321-1371, https://doi.org/10.1162/089976698300017214.
\bibitem{Brunel2000} Brunel, N. (2000). Dynamics of Sparsely Connected Networks of Excitatory and Inhibitory Spiking Neurons. \textit{Journal of Computational Neuroscience}, Volume \textbf{8}, Pages 183-208, https://doi.org/10.1023/A:1008925309027.
\bibitem{Herbert2022} Herbert, E., \& Ostojic, S. (2022). The impact of sparsity in low-rank recurrent neural networks. \textit{PLoS Computational Biology}, {\bf 18}(8): e1010426,  https://doi.org/10.1371/journal.pcbi.1010426.
\bibitem{Peche2006} Péché, S. (2006). The largest eigenvalue of small rank perturbations of Hermitian random matrices. \textit{Probability Theory and Related Fields}, Volume \textbf{134}, Pages 127-173, https://doi.org/10.1007/s00440-005-0466-z.
\bibitem{Peche2007} Féral, D., \& Péché, S. (2007). The Largest Eigenvalue of Rank One Deformation of Large Wigner Matrices. \textit{Communications in Mathematical Physics}, Volume \textbf{272}, Pages 185-228, https://doi.org/10.1007/s00220-007-0209-3.
\bibitem{Capitaine2009} Capitaine, M., Donati-Martin, C., \& Féral, D. (2009). The largest eigenvalues of finite rank deformation of large Wigner matrices: Convergence and nonuniversality of the fluctuations. \textit{Annals of Probability}, Volume \textbf{37}, Number 1, Pages 1-47, https://doi.org/10.1214/08-AOP394.
\bibitem{Bloemendal2013} Bloemendal, A., \& Virág, B. (2013). Limits of spiked random matrices I. \textit{Probability Theory and Related Fields}, Volume \textbf{156}, Pages 795-825, https://doi.org/10.1007/s00440-012-0443-2.
\bibitem{Bocchi2026} Bocchi, D., Biroli, G., Cammarota, C. \& Ricci-Tersenghi, F. (2026). Discontinuous BBP transitions. \textit{arXiv preprint}, arXiv:2604.27992.
\bibitem{Ferreira2026} Ferreira, L. S., \& Metz, F. L. (2026). BBP transition and the leading eigenvector of the spiked Wigner model with inhomogeneous noise. \textit{arXiv preprint}, arXiv:2604.18523.
\bibitem{Baik2006} Baik, J., \& Silverstein, J. W. (2006). Eigenvalues of large sample covariance matrices of spiked population models. \textit{Journal of Multivariate Analysis}, Volume \textbf{97}, Issue 6, Pages 1382-1408, https://doi.org/10.1016/j.jmva.2005.08.003.
\bibitem{Paul2007} Paul, D. (2007). Asymptotics of Sample Eigenstructure for a Large Dimensional Spiked Covariance Model. \textit{Statistica Sinica}, Volume \textbf{17}, Number 4, Pages 1617-1642.
\bibitem{Onatski2013} Onatski, A., Moreira, M. J., \& Hallin, M. (2013). Asymptotic power of sphericity tests for high-dimensional data. \textit{The Annals of Statistics}, Volume \textbf{41}, Number 3, Pages 1204-1231, http://doi.org/10.1214/13-AOS1100.
\bibitem{Mezard2000} Mézard, M., \& Parisi, G. (2001). The Bethe lattice spin glass revisited. \textit{European Physical Journal B - Condensed Matter and Complex Systems}, Volume \textbf{20}, Pages 217-233, https://doi.org/10.1007/PL00011099.
\bibitem{Nishimori2001} Nishimori, H. (2001). Mean-Field Theory of Spin Glasses. \textit{Statistical Physics of Spin Glasses and Information Processing: An Introduction}, Chapter 2 https://doi.org/10.1093/acprof:oso/9780198509417.003.0002.
\bibitem{Kuhn2007} Kühn, R., v. Mourik, J., Weigt, M., \& Zippelius, A. (2007). Finitely coordinated models for low-temperature phases of amorphous systems. \textit{Journal of Physics A: Mathematical and Theoretical}, Volume \textbf{40}, Number 31, Pages 9227–9252, https://doi.org/10.1088/1751-8113/40/31/004.
\bibitem{Edwards1976} Edwards, S. F., \& Jones, R. C. (1976). The eigenvalue spectrum of a large symmetric random matrix. \textit{Journal of Physics A: Mathematical and General}, Volume \textbf{9}(10), 1595,  https://doi.org/10.1088/0305-4470/9/10/011.
\bibitem{BrayRodgers} Rodgers, G. J., \& Bray, A. J. (1988). Density of states of a sparse random matrix. \textit{Physical Review B}, Volume \textbf{37}, Number 3557, https://doi.org/10.1103/PhysRevB.37.3557.
\bibitem{Rodgers1990} Rodgers, G. J., \& De Dominicis, C. (1990). Density of states of sparse random matrices. \textit{Journal of Physics A: Mathematical and General}, Volume \textbf{23}, Number 1567, https://doi.org/10.1088/0305-4470/23/9/019.
\bibitem{Akara2025} Akara-pipattana, P., \& Evnin, O. (2025). Hammerstein equations for sparse random matrices. Journal of Physics A: Mathematical and Theoretical, Volume \textbf{58}(3), 035006,  https://doi.org/10.1088/1751-8121/ada8ea.
\bibitem{Bianconi2008} Bianconi, G. (2008). Spectral properties of complex networks. \textit{arXiv preprint}, arXiv:0804.1744.
\bibitem{Annibale2017}
Coolen, A. C. C., Annibale, A., \& Roberts, E. S. (2017).
\textit{Generating Random Networks and Graphs}.
First Edition, Oxford University Press, Oxford.
ISBN 9780198709893.
https://doi.org/10.1093/oso/9780198709893.001.0001.
\bibitem{Zdeborov2016} Zdeborová, L., \& Krzakala, F. (2016). Statistical physics of inference: Thresholds and algorithms. \textit{Advances in Physics}, Volume \textbf{65}, Number 5, Pages 453-552, https://doi.org/10.1080/00018732.2016.1211393.
\bibitem{Kuhn2017} Kühn, R., \& Rogers, T. (2017). Heterogeneous micro-structure of percolation in sparse networks. \textit{Europhysics Letters}, Volume \textbf{118}, Number 6, https://doi.org/10.1209/0295-5075/118/68003
\bibitem{Bartolucci2024} Bartolucci, S., Caccioli, F., Caravelli, F., \& Vivo, P. (2024). Distribution of centrality measures on undirected random networks via the cavity method. \textit{Proceedings of the National Academy of Sciences}, Volume \textbf{121}, Number 40, https://doi.org/10.1073/pnas.2403682121.
\bibitem{Dumitriu2026} Dumitriu, I., Flynn, J. D., \& Wang, Z. (2026). BBP Phase Transition for a Doubly Sparse Deformed Model. \textit{arXiv preprint}, arXiv:2603.04832.
\bibitem{Susca2020b} Susca, V. A. R., Vivo, P., \& Kühn, R. (2021). Second largest eigenpair statistics for sparse graphs. \textit{Journal of Physics A: Mathematical and Theoretical}, Volume \textbf{54}(1), 015004, https://doi.org/10.1088/1751-8121/abcbad.
\bibitem{GraphSpectraBook2012} Brouwer, A. E., \& Haemers, W. H. (2012). Spectra of graphs. \textit{New York: Springer.}
\bibitem{GODSIL1988191} Godsil, C. D., \& Mohar, B. (1988). Walk generating functions and spectral measures of infinite graphs. \textit{Linear Algebra and its Applications}, Volume \textbf{107}, Pages 191-206, https://doi.org/10.1016/0024-3795(88)90245-5.
\bibitem{Brito_2021} Brito, G., Dumitriu, I., \& Harris, K. D. (2022). Spectral gap in random bipartite biregular graphs and applications. \textit{Combinatorics, Probability and Computing}, Volume \textbf{31}, Number 2, Pages 229-267, https://doi.org/10.1017/S0963548321000249.


\end{thebibliography}
\end{document}